%% file: sn-article.tex
\ifdefined\CSUR
  \documentclass[acmsmall,review]{acmart}
\else
  \documentclass[acmsmall,nonacm]{acmart}
\fi

\ifdefined\CSUR\else
  \setcopyright{none}
  \renewcommand\footnotetextcopyrightpermission[1]{}
\fi

\input{sn-preamble}

\begin{document}

\title{Projection and Quantisation: A Unifying View of Learning to Hash, from Random Projections to the RAG Era}

\author{Sean Moran}
\orcid{0000-0001-5377-1608}
\email{sean.j.moran@gmail.com}
\affiliation{%
  \institution{Independent Researcher}
  \city{London}
  \country{United Kingdom}
}
\renewcommand{\shortauthors}{Moran}

\begin{abstract}
Approximate nearest-neighbour search underpins large-scale retrieval and retrieval-augmented generation, yet its methods are studied in communities that seldom read one another. We argue that they form one field with three design choices. We develop the projection--quantisation--organisation lens: every method places its projections, places its quantisation thresholds, and organises the resulting codes for search. We test the lens with a reproducible measurement, released as the open \textsc{BitBudget} benchmark, and report three findings. First, the quantisation axis delivers the largest memory savings: a one-bit code with full-precision re-ranking matches uncompressed quality for six of seven embedders, the scanned code one thirty-second of the float's size. Second, the orderings the lens anticipates, including a learned-embedding regime where binary codes overtake an inverted-file product quantiser at a matched byte budget, recur as the embedding is enlarged. Third, given class labels, an eight-byte supervised code more than doubles the retrieval quality of the two-kilobyte task-agnostic float it replaces. We also recast the semantic identifiers of generative retrieval as quantisation codes. The main contribution is a single, tested account of compact-code search, from random projections to the retrieval-augmented era.

\end{abstract}


%% --- CSUR submission front matter: journal metadata + CCS concepts (preprint build omits it) ---
\ifdefined\CSUR
  \acmJournal{CSUR}
  \acmYear{2026}\acmVolume{1}\acmNumber{1}\acmArticle{1}\acmMonth{1}
  \setcopyright{acmlicensed}
  \acmDOI{10.1145/nnnnnnn.nnnnnnn}     % placeholder; ACM assigns a DOI on acceptance
\fi

\maketitle
\section{Motivation}\label{sec:ch1_motivation}

Approximate nearest-neighbour (ANN) search has become foundational infrastructure for modern artificial intelligence. It is the retrieval substrate beneath retrieval-augmented generation (RAG), semantic search, recommendation, and cross-modal retrieval. In RAG, a large language model is grounded by fetching semantically related passages from a billion-item corpus. In each of these applications, content is represented as a high-dimensional embedding vector, and the central operation is finding the vectors most similar to a query. As corpora have grown from millions to billions of items, the binding constraint has shifted. Query latency was for years the primary concern, and mature index structures now routinely meet it. The dominant cost is \emph{memory}: the footprint of storing billions of high-dimensional vectors. This has renewed interest in \emph{compact codes}, representations that shrink each vector by an order of magnitude or more while still answering similarity queries accurately. Compact binary codes are the most aggressive such compression, and they are the central object of this survey. After a decade in which dense embeddings and graph indexes dominated, binary codes are resurgent: production vector databases now offer them as a compression layer beneath a graph or a re-ranking stage. The need for algorithms whose overhead does not scale with data volume is therefore acute.

This article studies nearest-neighbour (NN) search: finding the data-points most similar to a query in a large database. Similarity is judged by representing data-points as fixed-dimensional vectors and computing a distance such as the Euclidean or cosine metric. The problem recurs across science, from information retrieval to genomic assembly. The na\"ive solution, \emph{brute-force search}, compares the query against every database point, so its query time is linear in the database size. This is tractable only at small scale. Exhaustively searching a modern corpus of millions to billions of points is infeasible, and streaming settings compound the difficulty. Efficient retrieval therefore demands algorithms whose query time grows sub-linearly with dataset size.

Hashing-based methods are a widely used family that delivers such sub-linear retrieval. They map both database entries and queries into compact binary representations, called hashcodes or fingerprints. The codes are designed so that nearby points in the original space receive closely related codes under the Hamming metric, and code similarity can be compared by fast bitwise operations. Note that this inverts the usual goal of hashing: a cryptographic hash collides only on identical inputs, whereas a similarity-preserving hash deliberately assigns similar codes to related inputs. The search then retrieves the records whose meaning is closest to the query, rather than those whose exact content matches.

\iflong
\begin{figure}[!t]
\centering
\noindent\includegraphics[width=\textwidth, height=100mm, keepaspectratio]{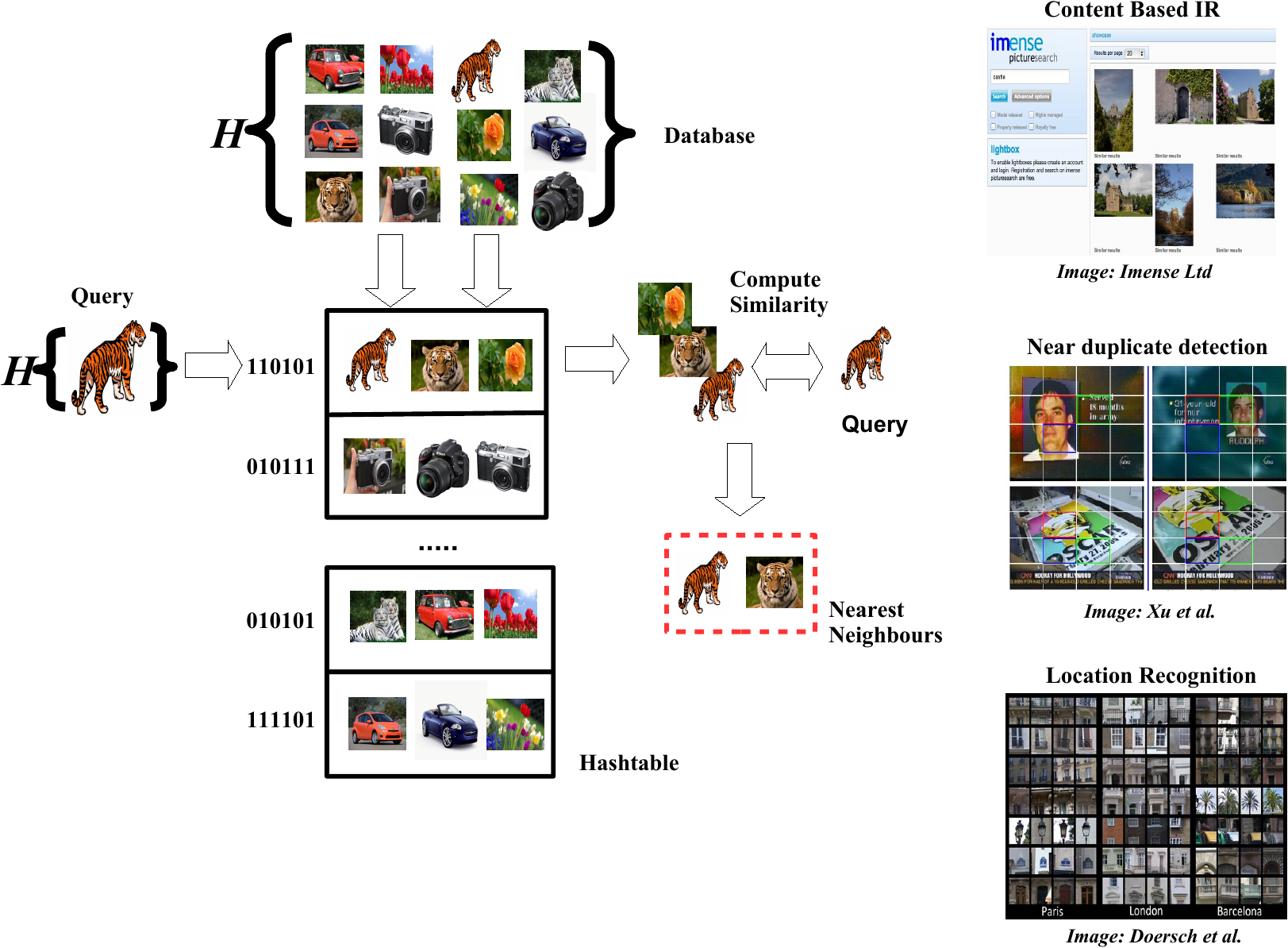}
\caption[Using hashcodes as the indices into the buckets of a hashtable for constant time search.]{Nearest neighbour search with hashcodes. Similarity-preserving binary codes generated by a hash function $\mathcal{H}$ serve as indices into the buckets of a hashtable for constant-time search: only those images sharing the query's bucket need be compared, reducing the search space. The focus of this review is learning $\mathcal{H}$ so as to maximise the similarity of hashcodes for similar data-points. On the right are tasks for which nearest neighbour search has proved fundamental, from content-based information retrieval (IR) to near-duplicate detection and location recognition. The three images on the right are taken from Imense Ltd and~\cite{Doersch12, Xu10, Grauman13}.}
\label{fig:ch1_hash_table}
\end{figure}
\fi

\longshort{This property lets the hashcodes serve as keys into the buckets of a hashtable. Similar but not necessarily identical data-points then collide in the same bucket\longonly{ (Figure~\ref{fig:ch1_hash_table})}. This is the opposite of the usual use of hashtables, where collisions between non-identical items are to be avoided. Search overhead falls because at query time we compare the query only to the data-points sharing its bucket. The drawback is a non-zero probability of missing a true nearest neighbour that falls in a different bucket. This false-negative rate is quantifiable, and in many application areas it is an acceptable trade-off, since sub-optimal neighbours are often almost as good as the exact ones~\cite{Dean13,Petrovic10}.}{This property lets the hashcodes serve as keys into the buckets of a hashtable: similar but not necessarily identical data-points collide in the same bucket, so a query is compared only against the points sharing its bucket. The drawback is a quantifiable probability of missing a true neighbour that falls elsewhere. This proves acceptable in many applications, since sub-optimal neighbours are almost as good as exact ones~\cite{Dean13,Petrovic10}.}

Locality-Sensitive Hashing (LSH)~\cite{Indyk98} is the foundational randomised algorithm of the area. It found early large-scale application in image retrieval and streaming detection. LSH drove real-time event detection over Twitter streams~\cite{Petrovic10}, and it scaled object recognition to a hundred thousand categories on a single workstation~\cite{Dean13}. Each system replaced the worst-case $\mathcal{O}(N)$ cost of an exhaustive scan with sub-linear query time over millions of items.\longonly{\footnote{This holds when the cost of computing hashcodes is excluded and each data-point maps to a unique bucket. In practice, for a single table the per-query cost comprises hashing ($\mathcal{O}(KD)$, with $K$ the hashcode length), lookup ($\mathcal{O}(1)$), and candidate verification, whose expected cost depends on bucket occupancy. The data-oblivious reading of LSH is the practitioner's view; data-dependent LSH constructions achieve strictly better exponents in theory~\cite{Andoni15}. The formal guarantee of Indyk and Motwani~\cite{Indyk98} is that the randomised $(c,R)$-near-neighbour problem is answered in $\mathcal{O}(N^{\rho})$ time with $\rho = \ln P_{1}/\ln P_{2} < 1$, where $P_{1} > P_{2}$ are the collision probabilities of near and far points, so the query time is sub-linear in the database size $N$.}}\shortonly{\footnote{Formally, Indyk and Motwani~\cite{Indyk98} answer the randomised $(c,R)$-near-neighbour problem in $\mathcal{O}(N^{\rho})$ time, $\rho = \ln P_{1}/\ln P_{2} < 1$, with $P_{1} > P_{2}$ the collision probabilities of near and far points.}} Since then the representation has changed: the hand-crafted descriptors of that era, such as GIST~\cite{Oliva01}, have given way to learned neural embeddings, and corpora have grown to billions of items. The underlying problem is unchanged and has grown in importance: compressing high-dimensional vectors so that similarity survives. The present survey organises this problem.

\iflong\input{supp/pipeline_figure}\fi

Hashing-based methods partition the feature space into non-overlapping regions using separating hypersurfaces, either linear (hyperplanes) or non-linear. The resulting cells serve as the buckets of a hashtable\longonly{ (Figure~\ref{fig:ch1_pipeline})}. A data-point is encoded by recording on which side of every surface it lies, which yields a binary string as long as the number of surfaces. Algorithmically this is a two-stage pipeline of projection and quantisation. Each vector is first projected onto the hyperplane normals, which are chosen randomly or learned so that similar points are more likely to share a region and a code. The resulting real-valued projections are then converted into bits, usually by thresholding each projected dimension at zero for centred data.\footnote{A projected dimension refers to the set of scalar projection values obtained by taking dot products between all data-points and the normal vector of a particular hyperplane.}

Despite their success, both steps leave considerable room for improvement, because LSH sets the hyperplanes and the thresholds independently of the data. In the cosine-similarity variant the normals are sampled from a zero-mean unit-variance Gaussian. This data-oblivious mechanism risks slicing through dense regions and partitioning related points into different buckets\longonly{ (e.g.\ points $a$ and $b$ in Figure~\ref{fig:ch1_pipeline})}. The usual remedy is multiple independent hashtables whose colliding buckets are unioned into a candidate set. This raises recall, but at a commensurate cost in memory; multi-probe variants reduce the cost by additionally probing the neighbouring buckets of each table~\cite{Lv07}. The thresholds are likewise fixed at zero. Note that zero is typically the densest part of a projected dimension, so related points frequently fall on opposite sides and receive different bits.

Substantial opportunity therefore remains, and this survey emphasises two challenges in particular. The first is preserving the neighbourhood structure of the original space at each stage of the pipeline, so that similar points yield similar codes. The second is doing so with the shortest possible hashcode, since length drives both storage and computational overhead.

No existing survey reads the classical, deep, quantisation, graph, and retrieval-augmented literatures under a single geometric account; Section~\ref{sec:contributions} contrasts the present work with the principal incumbents\longshort{ (Table~\ref{tab:surveys})}{ (Table~S15 of the supplementary material)}. That projection and quantisation are the two stages of a hash function is uncontroversial and implicit in several of these surveys; this article adds a third, co-equal stage, the \emph{organisation} of the codes for search, and the claim that this three-axis factorisation runs unbroken across all of these eras. This factorisation presents two decades of progress, from sign-random-projection LSH through spectral, supervised, and deep hashing to product quantisation and the binary-embedding resurgence within production vector databases, as successively more data-driven answers to three coupled questions: \emph{where to place the projections}, \emph{where to place the quantisation thresholds}, and \emph{how to organise the resulting codes}. This PQO lens unifies methods the literature treats as unrelated, clarifies why graph-based indexes, which act wholly on the organisation axis, outperform hashing in regimes it makes explicit, and reframes the learned ``semantic identifiers'' of generative retrieval as quantisation codes rather than a rival paradigm; a reproducible measurement, released as \textsc{BitBudget}, an open, extensible benchmark, tests the orderings the lens anticipates.\footnote{Source at \url{https://github.com/sjmoran/bitbudget} (\texttt{pip install bitbudget}); a live, community-submittable leaderboard at \url{https://sjmoran.github.io/bitbudget/}.}

\section{Survey Contributions}\label{sec:contributions}

Every system that searches a large collection of embeddings faces the same three decisions: how to project each vector, how to quantise that projection into a compact code, and how to organise the codes so that a query can be answered without scanning them all. This article surveys the methods that make these decisions, from the classical hashing literature to the embedding-compression layers of contemporary vector databases. We organise the survey around a single unifying idea, introduced formally in Section~\ref{sec:lens} as the \emph{PQO lens}: the methods of the field, however varied their motivation and machinery, are particular settings of these three coupled choices. A natural point of departure is Locality-Sensitive Hashing (LSH), the data-oblivious baseline against which much of the field has defined itself. LSH selects a projection at random and applies a fixed threshold at zero to quantise each data-point into a single bit. It repeats the operation $K$ times to yield a $K$-bit hashcode. In doing so it rests on three assumptions:

\begin{itemize}
\item $A_{1}$: A single, static threshold placed at zero (assuming mean-centred data)
\item $A_{2}$: Uniform allocation of thresholds across all dimensions
\item $A_{3}$: Linear hyperplanes positioned randomly
\end{itemize}

These assumptions are common in the literature~\cite{Indyk98, Weiss08, Liu12, Gong11, Raginsky09, Kulis09b}. They simplify the design of the hash function, but they frequently lead to suboptimal retrieval. A great deal of subsequent work, surveyed here, may be read as the progressive relaxation of each assumption: learning the quantisation thresholds from the data, allocating them non-uniformly across projected dimensions, and learning the projection itself in a data-dependent or, latterly, deeply non-linear manner.

The contributions of this survey are correspondingly threefold, each tested where stated.
\begin{enumerate}
\item[C1.] We take the established projection--quantisation decomposition of a hash function and add the \emph{organisation} of the codes as a third, co-equal axis. We use the resulting three-stage lens as a consistent vocabulary throughout: each method is introduced as an answer to where the projections and thresholds are placed, how tightly the two are coupled, and how the codes are organised for search. The lens is defined in Section~\ref{sec:lens} and instantiated per family in Table~\ref{tab:lens}.
\item[C2.] We apply the lens beyond the classical binary-hashing literature to the families that now dominate large-scale retrieval: deep end-to-end hashing, product quantisation, and graph-based indexes. We show that these too are settings of the same axes, and we identify where the lens reaches its limit in the case of graph indexes (Sections~\ref{sec:deep}--\ref{sec:graph}). The lens's predicted orderings are tested in Tables~\ref{tab:sift_measured} and~\ref{tab:cifar_measured}.
\item[C3.] We trace the recent resurgence of compact codes within vector databases and retrieval-augmented generation (Section~\ref{sec:resurgence}). We argue that this development reassembles the classical projection--quantisation pipeline from contemporary, learned components, and that it returns the questions with which the field began to the centre of large-scale retrieval. The resurgence's compression claims are measured in Table~\ref{tab:emb_quant}.
\end{enumerate}

\shortonly{Table~S15 of the supplementary material contrasts the present work with the principal prior surveys. These are organised by mechanism family~\cite{Wang14a, Chi17}, by level of supervision~\cite{Wang16}, by similarity-preserving objective~\cite{Wang18, Luo23}, or by empirical benchmark~\cite{Li20}. Each is individually strong, but none spans the classical, deep, quantisation, graph, and RAG-era literatures under a single organising principle, and none offers the practitioner guidance of Section~\ref{sec:practitioner}. This survey adds both by carrying the organisation axis across the eras they predate.}
\iflong\input{supp/surveys_table}\fi

\iflong\section{Survey Overview}\fi

\shortonly{\paragraph{Overview.} }This survey first motivates efficient similarity search and the limitations of data-oblivious methods such as Locality-Sensitive Hashing (LSH). It then introduces the background needed to read the field consistently: the nearest-neighbour problem and the projection and sign-random-projection operations on which the later sections build (Section~\ref{sec:ch2_lsh}). Section~\ref{sec:lens} develops the PQO lens that organises the remainder of the article. The lens factorises a similarity-search system into a projection, a quantisation, and an organisation stage, with three coupled questions: where to place the projections, where to place the quantisation thresholds, and how to organise the resulting codes. The survey then proceeds axis by axis. We review the quantisation axis, from single-bit thresholding to the multi-threshold schemes that relax it (Section~\ref{sec:ch2_quantisation}). We review the projection axis across data-independent, unsupervised, supervised, and cross-modal settings (Section~\ref{sec:ch2_projection}), and we summarise the classical material (Section~\ref{sec:classical_summary}). We then turn to the families that dominate large-scale retrieval: deep hashing, in which the projection becomes a learned network optimised jointly with the quantiser (Section~\ref{sec:deep}); the product-quantisation family, which pursues the quantisation axis through vector codebooks (Section~\ref{sec:pq}); graph-based indexes, which act entirely on the organisation axis and so mark the limit of the lens (Section~\ref{sec:graph}); and the resurgence of compact codes within vector databases and retrieval-augmented generation (Section~\ref{sec:resurgence}). Section~\ref{ch:chapter4} reviews evaluation methodology. The article closes with a synthesis and a discussion of open problems.

\section{Background on Hashing-Based Approximate Nearest Neighbour Search}
\input{body/background}
\section{The Projection--Quantisation Lens}\label{sec:lens}

Locality-Sensitive Hashing is best read as one particular setting of a small number of reusable design decisions, rather than as a single fixed algorithm. Its decomposition into a random \emph{projection} followed by a fixed \emph{quantisation} at zero\longonly{ (Figure~\ref{fig:ch1_pipeline})} generalises: essentially every technique surveyed here makes a choice along the same small set of axes, and we will argue that the same holds for every dominant method in the contemporary approximate nearest neighbour (ANN) literature. The embed-and-compress reading is long established for the binary-hashing literature. We add the \emph{organisation} of the codes for search as a third, co-equal stage. The resulting \emph{projection--quantisation--organisation} lens, the PQO lens for short, factorises a similarity-search system into an \emph{embed}, a \emph{compress}, and an \emph{organise} stage. It poses three coupled questions: \emph{where to place the projections}, \emph{where to place the quantisation thresholds}, and \emph{how to organise the resulting codes}. Their progressively more data-driven answers structure the remainder of this survey and organise the field as a whole.

\subsection{A factorisation of similarity search}\label{sec:lens_factorisation}

Let $\mathbf{x} \in \mathbb{R}^{D}$ denote a database item under a $D$-dimensional feature representation. We describe a similarity-search system as the composition of three maps.

\begin{definition}[The projection--quantisation factorisation]\label{def:lens}
A similarity-search index is characterised by a triple $(p, q, \mathcal{I})$:
\begin{enumerate}
  \item a \emph{projection} (embed) map $p : \mathbb{R}^{D} \rightarrow \mathbb{R}^{M}$, producing a real-valued representation $\mathbf{y} = p(\mathbf{x})$;
  \item a \emph{quantisation} (compress) map $q : \mathbb{R}^{M} \rightarrow \mathcal{C}$, mapping $\mathbf{y}$ to a code $\cc = q(\mathbf{y})$ drawn from a finite alphabet $\mathcal{C}$, with an associated code-space distance $d_{\mathcal{C}}$;
  \item an \emph{organisation} (search) structure $\mathcal{I}$ over the codes $\{q(p(\mathbf{x}_i))\}_{i=1}^{N}$ that, given a query, returns a candidate set without an exhaustive scan.
\end{enumerate}
The composite code map is $g = q \circ p$. This recovers the binary embedding $g(\mathbf{x}_i) = \bb_i$ of Section~\ref{sec:ch2_lsh} when $\mathcal{C} = \{0,1\}^{K}$ and $d_{\mathcal{C}}$ is the Hamming distance. Throughout, $M$ denotes the dimensionality of the real-valued projection and $K$ the length of the resulting code. In the single-bit hashing instance each projected dimension contributes one bit, so the two coincide, $M = K$.
\end{definition}

\input{supp/lens_figure}

The value of the factorisation is that the three stages are largely independent design choices (Figure~\ref{fig:lens}). Named methods correspond to characteristic settings of each stage. Binary hashing fixes $\mathcal{C} = \{0,1\}^{K}$ with Hamming distance and organises codes in one or more hashtables. Vector and product quantisation (Section~\ref{sec:pq}) instead take $\mathcal{C}$ to be a codebook of centroids with an asymmetric lookup distance. As discussed in Section~\ref{sec:lens_limits}, graph indexes act almost entirely on $\mathcal{I}$ and leave $p$ and $q$ untouched. Reading methods through $(p, q, \mathcal{I})$ exposes which stage a contribution actually advances. The taxonomies of supervision level or similarity-preserving objective~\cite{Wang18, Luo23} tend to obscure this distinction.

\subsection{Two design questions and the data-dependence spectrum}\label{sec:lens_questions}

Within this factorisation, the learning-to-hash literature has overwhelmingly targeted $p$ and $q$. Each poses a single design question.

\paragraph{Where to place the projections.} The projection $p$ determines the geometry of the partition before any bits are assigned. In the canonical hyperplane setting, $p(\mathbf{x}) = \mathbf{W}^{\intercal}\mathbf{x}$ for a matrix of normals $\mathbf{W} = [\mathbf{w}_1,\ldots,\mathbf{w}_M]$, and the question is how to choose $\mathbf{W}$. Answers span a spectrum of increasing data dependence. \emph{Data-independent} methods sample $\mathbf{w}_k$ at random, as in LSH~\cite{Indyk98, Charikar02, Datar04}. \emph{Unsupervised data-dependent} methods fit $\mathbf{W}$ to the second-order or manifold structure of the data, as in spectral~\cite{Weiss08} and rotation-based~\cite{Gong11} methods. \emph{Supervised} methods shape $\mathbf{W}$ with label or pairwise-similarity information. \emph{Nonlinear or deep} methods realise $p$ by a kernel map or a neural network. Section~\ref{sec:ch2_projection} traces this progression for linear and kernelised projections. The deep methods that later subsume it~\cite{Salakhutdinov09, Li16, Cao17} differ only in making $p$ a learned, high-capacity function.

\paragraph{Where to place the thresholds.} Given a projected dimension $\mathbf{y}^{k}$, the quantiser $q$ decides how its real-valued range is partitioned into codes. LSH answers with a single, static threshold at zero applied uniformly to every dimension. The methods of Section~\ref{sec:ch2_quantisation} relax this in three ways: they learn the threshold positions from the projected density; they allocate multiple thresholds, and hence multiple bits, per dimension; and they allocate those bits non-uniformly, so that the more informative projections are quantised more finely. At the furthest relaxation, $q$ replaces per-dimension scalar thresholds with a vector codebook. This yields the product-quantisation family~\cite{Jegou11, Ge14} and its score-aware descendants~\cite{Guo20}.

\paragraph{Coupling.} A third, orthogonal dimension is whether $p$ and $q$ are learned \emph{separately} or \emph{jointly}. Classical pipelines almost always fix the projection first and quantise afterwards, so quantisation error is incurred against a geometry chosen without reference to it. Joint optimisation trains $p$ and $q$ against a single criterion, so that each is shaped in anticipation of the other. End-to-end deep hashing is the most complete form of this coupling~\cite{Li16, Cao17, Su18}. The central hypothesis of this survey is that relaxing the static, uniform and data-oblivious choices of LSH along these axes improves retrieval effectiveness. In these terms, the claim is that a data-dependent $p$, a data-dependent $q$, and the coupling of the two each contribute measurable gains.

\subsection{The lens applied to the modern families}\label{sec:lens_modern}

\shortonly{These axes are worth formalising because they extend cleanly beyond the binary-hashing methods that motivated them. Table~\ref{tab:lens} reads the dominant contemporary families through $(p, q, \mathcal{I})$. It shows the deep, product-quantisation, graph, and binary-embedding eras to be successive settings of the same three axes rather than rival paradigms: the deep transition escalates $p$ and $q$ jointly; product quantisation and binary hashing are two settings of the compression axis; and the binary-embedding resurgence in production retrieval is the lens reassembled from modern components, a deep projection, an aggressive binary quantiser, and a cheap re-ranking pass.}\iflong
These axes are worth formalising because they extend cleanly beyond the binary-hashing methods that motivated them. Table~\ref{tab:lens} reads the dominant contemporary families through $(p, q, \mathcal{I})$, and three observations are worth drawing out. First, under the lens the deep-learning transition is the simultaneous escalation of two existing axes: $p$ becomes a learned network and $q$ is trained jointly with it, often by relaxing or annealing the binary constraint~\cite{Cao17, Su18}. Second, the product-quantisation family and binary hashing are two settings of the compression axis, scalar thresholds against vector codebooks. Methods that interpolate between them make the duality explicit. Third, the binary-embedding ``resurgence'' in production retrieval~\cite{Yamada21, Wang21} is the lens reassembled from modern components: a deep, semantically rich projection $p$, an aggressive binary quantiser $q$, and a cheap re-ranking pass that recovers the precision lost to compression. Matryoshka-style adaptive embeddings~\cite{Kusupati22} act on the same projection axis. They learn representations that remain useful after truncation, itself a learned analogue of non-uniform bit allocation across projected dimensions.
\fi

\begin{table}[!t]
\centering
\small
\begin{tabular}{@{}p{2.7cm} p{3.1cm} p{3.0cm} p{2.7cm}@{}}
\toprule
\textbf{Method family} & \textbf{Projection ($p$)} & \textbf{Quantisation ($q$)} & \textbf{Organisation ($\mathcal{I}$)} \\
\midrule
LSH~\cite{Indyk98, Charikar02} & random hyperplanes & single static threshold at $0$ & multiple hashtables \\
Learned binary hashing~\cite{Weiss08, Gong11} & unsupervised / supervised linear or kernel & learned, multi-threshold, non-uniform bits & hashtable / Hamming ranking \\
Product quantisation~\cite{Jegou11, Ge14} & rotation / subspace split & vector codebook (asymmetric distance) & inverted lists / multi-index \\
Deep hashing~\cite{Li16, Cao17} & deep network (\emph{jointly} with $q$) & relaxed/annealed binary, learned jointly & Hamming ranking \\
Graph indexes~\cite{Malkov20, Fu19} & \emph{(identity / pretrained)} & \emph{(none or PQ side-codes)} & navigable proximity graph \\
Binary-embedding retrieval~\cite{Yamada21, Kusupati22} & deep embedding (+ truncation) & binary, with real re-ranking & flat / graph + rescore \\
Generative semantic IDs~\cite{Tay22, Rajput23} & encoder to latent & hierarchical or residual-quantised (RQ-VAE) codebook & autoregressive decoder \\
\bottomrule
\end{tabular}
\caption[The PQO lens applied to dominant ANN families]{Dominant families of approximate nearest neighbour search read through the PQO lens (Definition~\ref{def:lens}). Each family is a characteristic setting of the embed ($p$), compress ($q$), and organise ($\mathcal{I}$) stages; italicised entries mark stages a family leaves essentially untouched.}
\label{tab:lens}
\end{table}

\subsection{Limits of the lens: the organisation stage}\label{sec:lens_limits}

The lens is a claim about $p$ and $q$, and its limit is the third stage, $\mathcal{I}$. Graph-based indexes such as HNSW~\cite{Malkov20}, NSG~\cite{Fu19}, and DiskANN~\cite{Subramanya19} make essentially no commitment about projection or quantisation. They operate on raw or independently compressed vectors and place their design effort entirely in the organisation structure: a navigable proximity graph supporting greedy routing to a query's neighbourhood. On the recall--latency trade-offs that dominate empirical benchmarking~\cite{Aumuller20} they frequently outperform hashing. The lens explains why: hashing assumes a good $p$ and $q$ render organisation a mere bucket lookup, whereas graph methods assume a sufficiently sophisticated $\mathcal{I}$ renders embedding and compression unnecessary. This delineation is one of the framework's uses. It locates which axis a method competes on, and it frames the open question of how the three stages should be \emph{co-designed} rather than optimised in isolation; we return to this question in Section~\ref{sec:graph} and the conclusion. The lens also accommodates the generative-retrieval methods often positioned as an alternative to indexing~\cite{Tay22}. Their learned ``semantic identifiers'' are residual-quantisation codes~\cite{Rajput23}, a particular $q$, with $\mathcal{I}$ an autoregressive decoder rather than a lookup. They therefore instantiate the factorisation.

\iflong
\subsection{Roadmap}\label{sec:lens_roadmap}

The remainder of the survey is organised by axis. Section~\ref{sec:ch2_quantisation} examines the quantisation axis in depth, from single-bit thresholding to multi-threshold and multi-bit schemes. Section~\ref{sec:ch2_projection} examines the projection axis across data-independent, unsupervised, supervised, and cross-modal settings. Throughout, $(p, q, \mathcal{I})$ provides the shared vocabulary: each method is introduced as a specific answer to where the projections and thresholds should be placed, and how tightly the two should be coupled.
\fi

\section{Quantisation for Nearest Neighbour Search}\label{sec:ch2_quantisation}
\input{body/quant}
\section{Projection for Nearest Neighbour Search}\label{sec:ch2_projection}
\input{body/projection}
\section{Summary: The Classical Era of Learning to Hash}\label{sec:classical_summary}

This part of the survey has set out the background to classical learning to hash. We motivated scalable alternatives to brute-force nearest-neighbour search and the seminal framework of Locality-Sensitive Hashing (LSH) (Sections~\ref{sec:ch2_ann_search}--\ref{sec:ch2_lsh}), whose similarity-preserving hashcodes give near-constant-time retrieval but, by sampling both the hyperplanes and the thresholds at random, rest on asymptotic guarantees that often fail at the limited code lengths used in practice, scattering related data-points across buckets. A central line of classical research therefore relaxed this data-independence assumption, learning the projection (Section~\ref{sec:ch2_projection}) and the quantisation thresholds (Section~\ref{sec:ch2_quantisation}) directly from the data, across unsupervised (Section~\ref{sec:ch2_dependent_projection}), supervised (Section~\ref{sec:ch2_supervised_projection}), and cross-modal (\longshort{Section~\ref{sec:ch2_crossmodal_projection}}{Section~S3 of the supplementary material}) settings, which formed the comparative baselines of the period.

Four limitations nonetheless defined the boundaries of this pre-deep era: the multi-threshold quantisers exploited at most pairwise constraints rather than full label information; thresholds were allocated uniformly across projected dimensions, ignoring their differing discriminative power; supervised projections often relied on computationally expensive eigendecomposition or kernel methods; and, fourth and most consequentially, the projection and the quantisation thresholds were almost never learned in a single unified framework, the few exceptions including minimal loss hashing~\cite{Norouzi11mlh} and a shallow scheme that learned the two jointly~\cite{Moran16}. Subsequent work advanced on each of these limitations, most of all the fourth: the unification of projection and quantisation learning is resolved at scale by the deep era of learning to hash, to which we now turn before returning to evaluation methodology.

\section{Deep Hashing: Jointly Learning Projection and Quantisation}\label{sec:deep}
\input{body/deep}
\section{The Product-Quantisation Family: Vector Codebooks on the Quantisation Axis}\label{sec:pq}
\input{body/pq}
\section{The Organisation Axis: Graph-Based Indexes}\label{sec:graph}
\input{body/graph}
\section{The Resurgence of Compact Codes in Large-Scale Retrieval}\label{sec:resurgence}
\input{body/resurgence}
\section{Experimental Methodology}\label{ch:chapter4}
\shortonly{\label{sec:ch4_datasets}\label{sec:ch4_dataset_unimodal}\label{sec:ch4_dataset_crossmodal}\label{sec:ch4_groundtruth}\label{sec:ch4_epsilon_ground}\label{sec:ch4_class_ground}\label{sec:ch4_paradigms}\label{sec:ch4_splits}\label{sec:ch4_improved_splits}}

\subsection{Introduction}

\shortonly{Two complementary strands of evaluation bear on the methods we compare. The classical retrieval-effectiveness protocol scores ranking quality under metrics such as mean average precision (Section~\ref{sec:ch4_metrics}). The recall--latency--memory protocol evaluates the dominant modern methods at scale (Section~\ref{sec:ch4_scale}). The former isolates the discriminative quality of the codes. The latter isolates the system-level trade-offs among recall, speed, and memory discussed in Section~\ref{sec:graph}. A complete account of a modern method reports both.}

\iflong
This section outlines the experimental methodology adopted in the nearest-neighbour search literature: \longshort{the datasets used as testbeds (Section~\ref{sec:ch4_datasets}), the prevailing definitions of ground truth (Section~\ref{sec:ch4_groundtruth}), and the metrics used to quantify retrieval effectiveness (Section~\ref{sec:ch4_metrics})}{the standard datasets and ground-truth definitions, and the metrics used to quantify retrieval effectiveness (Section~\ref{sec:ch4_metrics})}. We note where practice varies and where evaluation may be placed on a more consistent footing. The protocol described here is that of the classical retrieval-effectiveness literature, in which methods are compared by ranking quality under metrics such as mean average precision. It isolates the discriminative quality of the codes. It should be read alongside the complementary and now equally important strand of evaluation, exemplified by the ANN-Benchmarks and billion-scale efforts~\cite{Aumuller20, Simhadri22} and taken up in Section~\ref{sec:ch4_scale}. That strand compares methods by recall against query latency and by memory footprint per item rather than by ranking quality alone. It isolates the system-level trade-offs among recall, speed, and memory discussed in Section~\ref{sec:graph}.
\fi

\iflong
\subsection{Datasets}\label{sec:ch4_datasets}

We are concerned chiefly with large-scale image retrieval, in three settings: image-to-image, text-to-image, and image-to-annotation. We therefore describe unimodal datasets, in which query and database share a visual representation, and cross-modal datasets, in which a text query is run over an image database. To preserve comparability with prior art, we catalogue the widely used, publicly available collections below (Sections~\ref{sec:ch4_dataset_unimodal}--\ref{sec:ch4_dataset_crossmodal}). The reproducible measurement of Section~\ref{sec:ch4_measured} also draws on modern text-retrieval and large-scale collections that these classical image datasets predate. We catalogue those in Section~\ref{sec:ch4_dataset_modern}, so that every testbed the measurement names is described here in one place.

\subsubsection{Unimodal Retrieval Datasets}\label{sec:ch4_dataset_unimodal}

Four image datasets appear frequently in unimodal hashing studies: LabelMe, CIFAR-10, NUS-WIDE, and SIFT1M. They span sizes from 22{,}019 to 1M images, use diverse descriptors (GIST, SIFT, BoW), and cover varied content (natural scenes, personal photos, logos, drawings). Together they form a challenging evaluation suite. Table~\ref{tab:ch4_datasets} summarises their salient statistics. The configurations follow widely cited work (\cite{Kong12b, Shen15, Liu12}) and are publicly available.

\begin{table}[t]
\centering
\caption[Table showing the salient statistics of the image datasets used in this survey]{Salient statistics of four datasets frequently used in unimodal hashing evaluations. Labels/image and Images/label are means over the full dataset.}
\label{tab:ch4_datasets}
\begin{tabular}{@{}llllll@{}}
\toprule
\textbf{Dataset} & \textbf{\# images} & \textbf{\# labels} & \textbf{Labels/image} & \textbf{Images/label} & \textbf{Descriptor} \\
\midrule
LABELME & 22,019 & -- & -- & -- & 512-D GIST \\
CIFAR-10 & 60,000 & 10 & 1 & 6,000 & 512-D GIST \\
NUS-WIDE & 269,648 & 81 & 1.87 & 6,220 & 500-D BoW \\
SIFT1M & 1,000,000 & -- & -- & -- & 128-D SIFT \\
\bottomrule
\end{tabular}
\end{table}

\subsubsection{Cross-modal Retrieval Datasets}\label{sec:ch4_dataset_crossmodal}

Cross-modal evaluations in the learning-to-hash literature most commonly use the \emph{Wiki} dataset~\cite{Rasiwasia10} and a cross-modal version of NUS-WIDE~\cite{Chua09}. Both provide paired image--text descriptions~\cite{kumar11, Zhen12, Song13, Rastegari13, Bronstein10}. For Wiki, images are represented by a SIFT bag-of-words and the paired text by LDA topic proportions~\cite{Blei03}. For NUS-WIDE, the user tags form the textual modality alongside the visual bag-of-words used in the unimodal setting.

\subsubsection{Text-Retrieval and Large-Scale Datasets}\label{sec:ch4_dataset_modern}

The classical image collections above predate the embedding-retrieval and billion-scale literatures. The reproducible measurement of Section~\ref{sec:ch4_measured} therefore appeals to a second group of testbeds, summarised in Table~\ref{tab:ch4_datasets_modern}. For text retrieval under graded relevance, we use four corpora from the BEIR benchmark~\cite{Thakur21}: scifact, nfcorpus, arguana and fiqa. These span corpus sizes from a few thousand to some fifty-seven thousand passages. We add the half-million Quora question embeddings used for the recall--latency--memory study on a modern corpus. For metric search at scale, we use the SIFT1M collection of Table~\ref{tab:ch4_datasets} and, two orders of magnitude larger, a ninety-million-vector slice of the BigANN \texttt{SIFT1B} collection~\cite{Jegou11a}. For retrieval at the scale of retrieval-augmented generation, we use the full MS~MARCO passage corpus~\cite{Nguyen16} of some $8.84$ million passages. The neural-retrieval corpora carry graded relevance judgements and are scored by nDCG@$10$ (Section~\ref{sec:ch4_metrics}). The metric collections define ground truth by exact Euclidean nearest neighbours and are scored by recall against that exact ranking (Section~\ref{sec:ch4_scale}).

\begin{table}[t]
\centering
\caption[Text-retrieval and large-scale datasets used in the measurement]{Datasets used by the reproducible measurement (Section~\ref{sec:ch4_measured}) beyond the classical image collections of Table~\ref{tab:ch4_datasets}: the modern text-retrieval and large-scale ANN testbeds. Counts are database items; the BEIR and MS~MARCO corpora are embedded with sentence-transformers, and the metric collections are searched in their native descriptor space.}
\label{tab:ch4_datasets_modern}
\begin{tabular}{@{}l r l l l@{}}
\toprule
\textbf{Dataset} & \textbf{\# items} & \textbf{Content} & \textbf{Ground truth} & \textbf{Measured in} \\
\midrule
scifact (BEIR) & $5{,}183$ & text passages & graded qrels & \S\ref{sec:ch4_measured} (compression) \\
nfcorpus (BEIR) & $3{,}633$ & text passages & graded qrels & \S\ref{sec:ch4_measured} (compression) \\
arguana (BEIR) & $8{,}674$ & text passages & graded qrels & \S\ref{sec:ch4_measured} (compression) \\
fiqa (BEIR) & $57{,}638$ & text passages & graded qrels & \S\ref{sec:ch4_measured} (compression) \\
Quora & $522{,}931$ & text embeddings & exact $k$-NN & \S\ref{sec:ch4_measured} (recall--latency) \\
BigANN \texttt{SIFT1B} slice & $90{,}000{,}000$ & 128-D SIFT & exact $k$-NN & \S\ref{sec:ch4_scale_measured} (at scale) \\
MS~MARCO passages & $8{,}841{,}823$ & text passages & graded qrels & \S\ref{sec:ch4_scale_measured} (at scale) \\
\bottomrule
\end{tabular}
\end{table}

\fi
\iflong
\subsection{Nearest Neighbour Ground-truth Definition}\label{sec:ch4_groundtruth}

Evaluation requires a notion of \emph{true} nearest neighbours for each query. Two definitions are prevalent: an $\epsilon$-ball in feature space (Section \ref{sec:ch4_epsilon_ground}) and class/label-based relevance (Section \ref{sec:ch4_class_ground}). To date, little evidence connects the $\epsilon$-ball definition to user satisfaction.

\subsubsection{$\epsilon$-Ball Nearest Neighbours}\label{sec:ch4_epsilon_ground}

A common choice is the $\epsilon$-NN definition (Figure~\ref{fig:ch4_groundtruth_1}). Following~\cite{Kong12a, Kong12b, Kulis09b, Gong11}, we set $\epsilon$ to the Euclidean distance at which a point has, on average, $R$ neighbours (commonly $R{=}50$). The ground-truth matrix $\mathbf{S} \in \{0,1\}^{N_{trd}\times N_{trd}}$ is then obtained by thresholding the pairwise distances at $\epsilon$ (Equation~\ref{eqn:ch4_groundtruth}).

\begin{equation}
\mathbf{S} = \begin{cases}
S_{ij}=1, & \text{if } D_{ij}\le \epsilon,\\
S_{ij}=0, & \text{if } D_{ij} > \epsilon.
\end{cases}
\label{eqn:ch4_groundtruth}
\end{equation}

\begin{figure}[!t]
\centering
\hspace{-0.3in}\subfloat[\textbf{$\epsilon$-nearest neighbour ground truth}]{\includegraphics[width=0.42\textwidth, height=55mm, keepaspectratio]{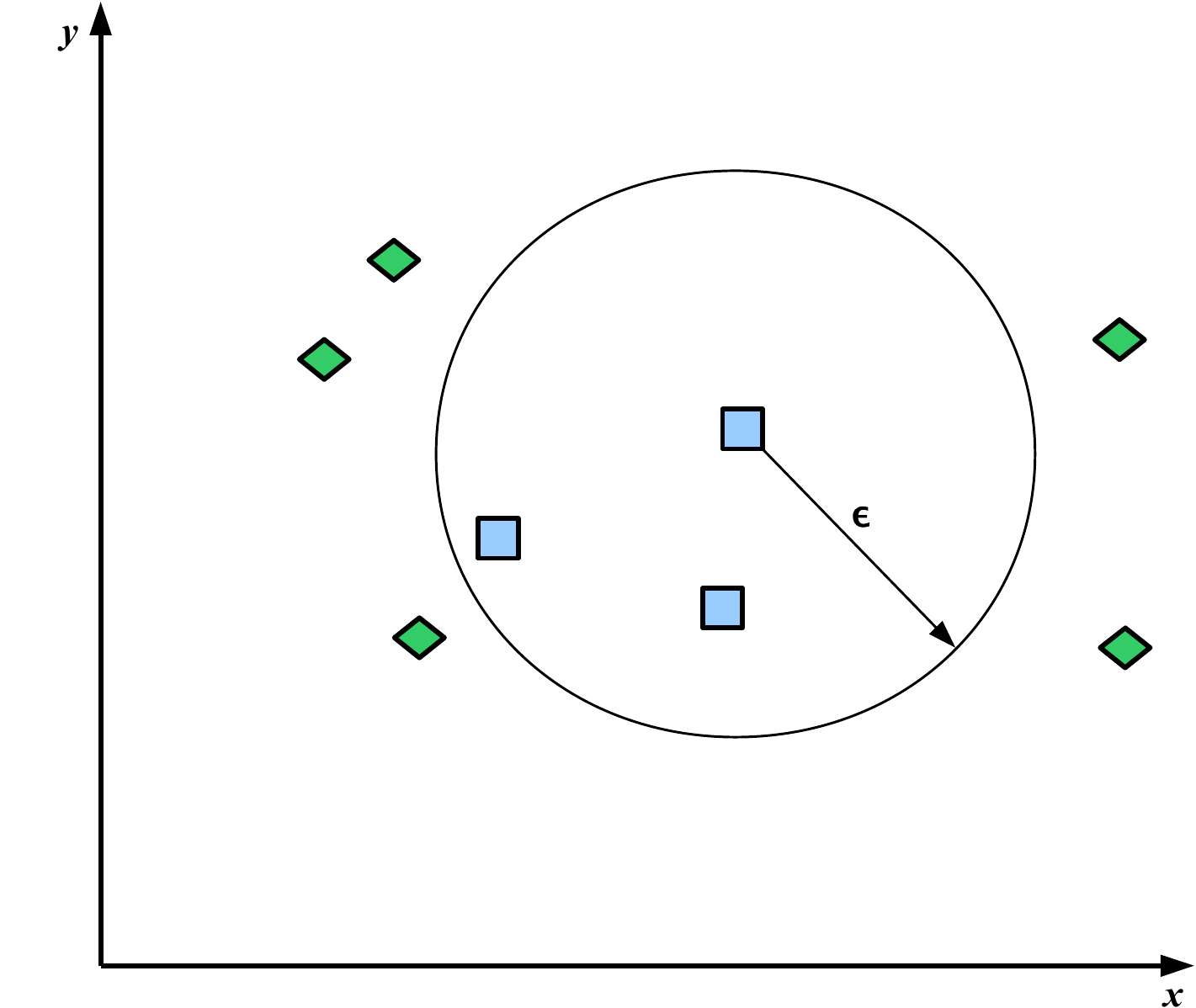}\label{fig:ch4_groundtruth_1}}
\subfloat[\textbf{Class-based ground truth}]{\includegraphics[width=0.42\textwidth, height=55mm, keepaspectratio]{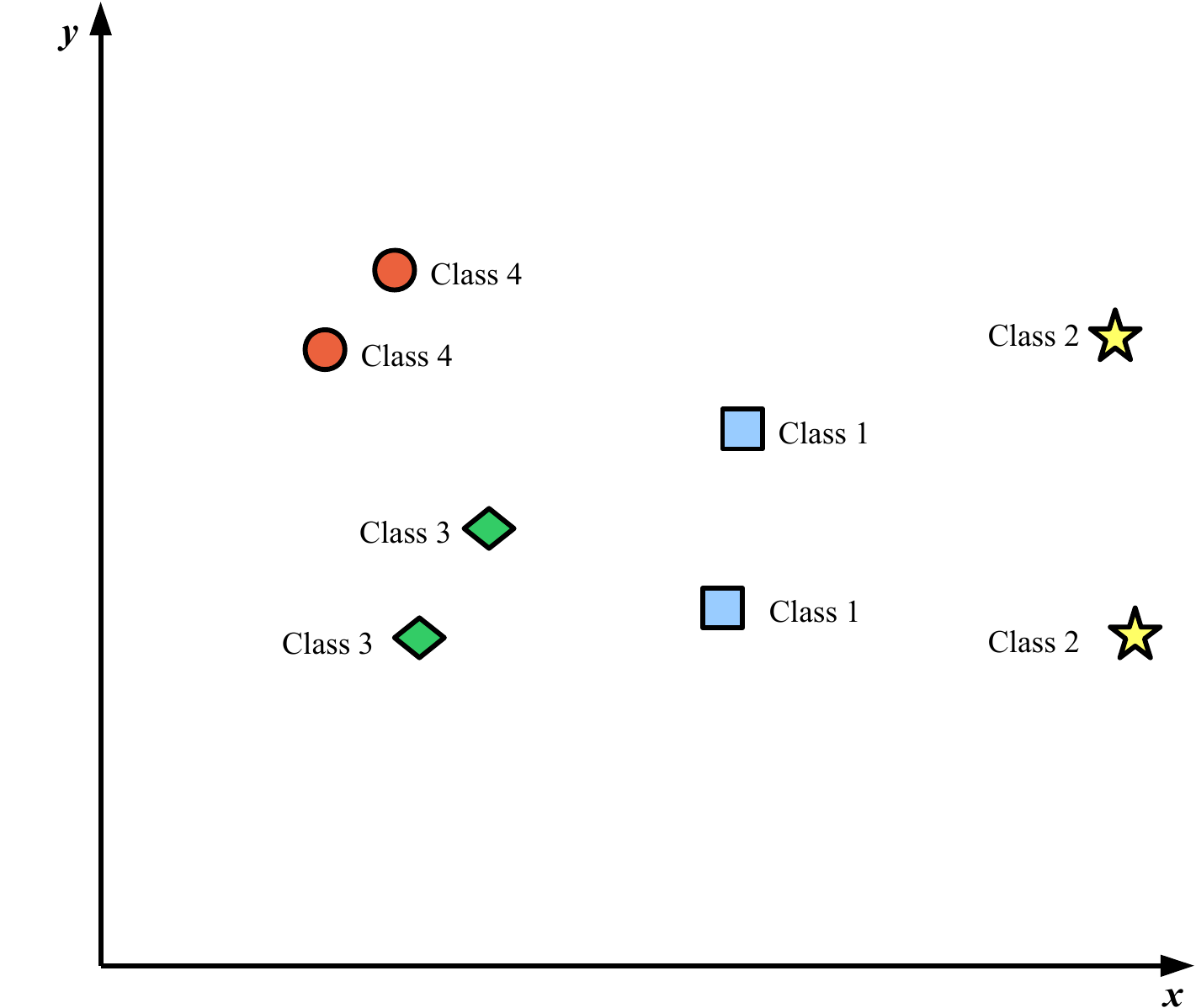}\label{fig:ch4_groundtruth_2}}
\caption[$\epsilon$-NN and class-based ground-truth definitions]{Two common ground-truth definitions. (a) $\epsilon$-ball neighbours; (b) class/label overlap.}
\label{fig:ch4_groundtruth}
\end{figure}

\subsubsection{Class-Based Nearest Neighbours}\label{sec:ch4_class_ground}

Class- or tag-based definitions are widely used when modalities differ (e.g., text vs. image) or to align with streams of prior work (\cite{Gong11,Liu12}). Here, $S_{ij}=1$ if $\mathbf{x}_{i}$ and $\mathbf{x}_{j}$ share at least one label/annotation, and $S_{ij}=0$ otherwise (Figure \ref{fig:ch4_groundtruth_2}).

\subsection{Evaluation Paradigms}\label{sec:ch4_paradigms}

Two paradigms dominate. In \emph{Hamming ranking}, the standard in the learning-to-hash literature, database items are ranked by increasing Hamming distance from the query. The ranking is scored by AUPRC (Section~\ref{sec:ch4_auprc}) or mAP (Section~\ref{sec:ch4_map}). By abstracting away the particular $(K,L)$ of a hashtable, this paradigm summarises expected behaviour across many configurations~\cite{Kong12a, Liu12, Gong11, Zhang10}. \emph{Hash-table bucket} evaluation mirrors the operational $\mathcal{O}(1)$-lookup setting. The items colliding with the query across $L$ tables form an unranked retrieved set, scored by precision, recall, and $F_{\beta}$. Bucket effectiveness depends on implementation choices such as $(K,L)$ and chaining. A single setting therefore binds the result to one design point, and a sweep multiplies the reported volume. Hamming ranking provides an implementation-agnostic proxy, which is why aggregate ranking metrics predominate in comparative studies.

\subsection{Constructing Random Dataset Splits}\label{sec:ch4_splits}

The literature-standard protocol uses repeated random subsampling, commonly ten runs~\cite{Liu12, Kong12a, Kong12b, Liu14, Wang12}. The data $\mathbf{X} \in \mathbb{R}^{N\times D}$ is split into held-out test queries $\mathbf{X}_{teq}$ and a database $\mathbf{X}_{db}$. The database further supplies a training split $\mathbf{X}_{trd}$, together with validation queries and a validation database for tuning. The test queries are scored once against $\mathbf{X}_{db}$. A known concern is overfitting, since the same database both trains the model and serves as the index against which the test queries are scored. An improved strategy\label{sec:ch4_improved_splits} therefore holds out a separate \emph{test database} $\mathbf{X}_{ted}$. As Figure~\ref{fig:ch4_improved_splits} shows, the test queries and test database are used once for final scoring. A disjoint development pool is partitioned into training, validation-query, and validation-database splits. Generalisation to an unseen database is therefore assessed rather than assumed.

\begin{figure}
\centering
\includegraphics[width=110mm, height=50mm]{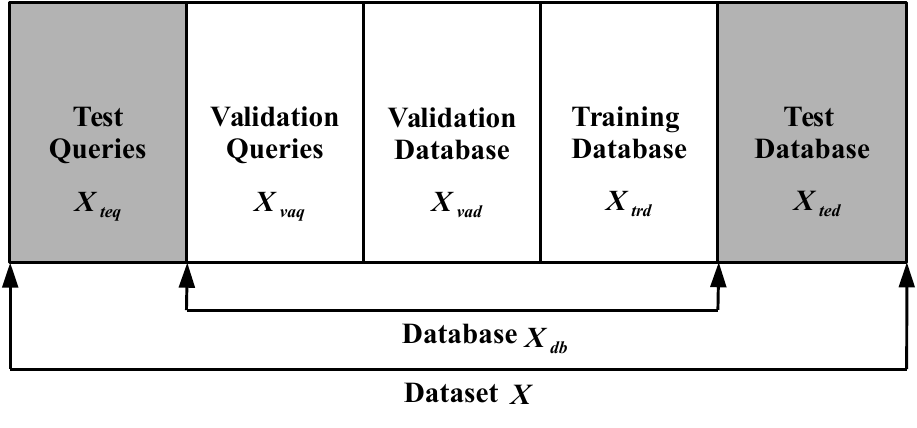}
\caption[Improved dataset splitting procedure]{An evaluation-oriented split that holds out both test queries and a test database (grey), mitigating coupling between training and final scoring.}
\label{fig:ch4_improved_splits}
\end{figure}

\fi
\shortonly{We follow standard evaluation practice. Nearest-neighbour ground truth is defined either by a metric $\varepsilon$-ball or by shared class labels. Methods are compared by Hamming ranking rather than by hashtable lookup, and a held-out database split assesses generalisation on unseen data. The measured study draws on the classical SIFT1M and CIFAR-10 collections, four BEIR text corpora (scifact, nfcorpus, arguana and fiqa)~\cite{Thakur21}, half a million Quora question embeddings, a ninety-million-vector slice of BigANN \texttt{SIFT1B}~\cite{Jegou11a}, and the full MS~MARCO passage corpus~\cite{Nguyen16}. We next define the metrics and the at-scale protocol on which the measured results rely.}

\subsection{Evaluation Metrics}\label{sec:ch4_metrics}

\shortonly{We summarise retrieval effectiveness with standard information-retrieval measures, defined in full in Section~S8 of the supplementary material.\label{sec:ch4_auprc}\label{sec:ch4_map} For an unranked retrieved set we report precision, recall, and their $F_{\beta}$ combination~\cite{Rijsbergen79}. For a ranking under binary relevance we report the area under the precision--recall curve (AUPRC) and mean average precision (mAP). The two coincide when the relevant items per query are balanced and diverge under skew; AUPRC favours queries with many relevant items, whereas mAP weights queries equally~\cite{Turpin06, Sebastiani02}. Where relevance is graded, as in the neural-retrieval benchmarks, we report the normalised discounted cumulative gain (nDCG@$k$), the standard measure on the BEIR and MTEB suites~\cite{Thakur21, Muennighoff23}. For the supervised CIFAR-10 study of Section~\ref{sec:ch4_measured} we report mean average precision under the class-based ground truth of Section~\ref{sec:ch4_class_ground}; for its embedding-compression study we report nDCG@$10$.}
\iflong\input{supp/methodology_metrics}\fi

\subsection{Evaluation at Scale: Recall, Latency, and Memory}\label{sec:ch4_scale}

\shortonly{The methods that dominate large-scale retrieval are most usefully compared along three axes rather than at a fixed code length. The first is \emph{$k$-recall@$k$}, the fraction of the true $k$ nearest neighbours recovered. The second is the throughput, or queries answered per second (QPS), on fixed hardware, against which recall is traced out as a curve. The third is the \emph{memory footprint}, the bytes stored per vector. These map cleanly onto the lens: recall reflects projection-and-quantiser quality, the recall--throughput curve reflects the organisation structure, and bytes-per-vector is the compression axis. The classical ranking metrics and these operating-point metrics are therefore projections of one system onto different axes rather than competitors. The formal definitions, the standard datasets, and the ANN-Benchmarks and billion-scale protocols are given in Section~S8 of the supplementary material.}

\iflong\input{supp/scale_metrics}\fi

\iflong\input{supp/meas_empirical}\fi

\subsection{A Reproducible Measurement}\label{sec:ch4_measured}

The published benchmarks supply representative per-family figures\longshort{ (Table~\ref{tab:empirical})}{ (Table~S8 of the supplementary material)}. We complement them with a study of our own, run under the protocol just described and released in full, so that every number may be reproduced from a single command.\footnote{The benchmark, \textsc{BitBudget}, exposes two reproducible leaderboards: one for \emph{compression} (retrieval quality retained per byte) and one for \emph{indexing} (recall against query throughput). Source and leaderboard locations are given where the benchmark is introduced in Section~\ref{sec:ch1_motivation}. Each leaderboard is extensible with a one-line plugin decorator, and the harness downloads the standard datasets automatically. The measured study includes a method due to the author (GRH); we release the harness, configurations, and data splits so that this selection can be audited and extended.} The study aims to make the lens's trade-offs concrete on real data, rather than to rank the methods overall. It spans the two regimes the survey distinguishes: the unsupervised regime, where a code must preserve a \emph{metric} neighbourhood, and the supervised regime, where it must preserve a \emph{semantic} one.

\paragraph{The unsupervised regime.} Table~\ref{tab:sift_measured} reports the three axes of Definition~\ref{def:lens} for one representative method per family on the one-million-vector SIFT collection, under the metric ground truth of the ANN-Benchmarks suite~\cite{Aumuller20}. The axes are recall against an exact ranking, single-thread throughput in queries per second, and bytes per vector. We observe the pattern the lens anticipates. The graph~\cite{Malkov20} attains the highest throughput and the largest memory footprint. The product quantiser~\cite{Jegou11} reduces memory twentyfold at a loss of recall. Within the smallest-footprint corner, the measurement isolates the effect with which the survey began. At an identical sixteen bytes, the learned code of iterative quantisation~\cite{Gong11} recovers a recall of $0.77$, where data-oblivious LSH reaches only $0.55$. The gain is attributable entirely to learning where to place the projections.

\begin{table}[!t]
\centering
\small
\caption[Reproducible per-family measurements on SIFT1M and Quora]{A reproducible measurement of one representative method per family, on the one-million-vector SIFT collection (metric ground truth of the ANN-Benchmarks suite) and on $522{,}931$ Quora question embeddings under two embedders ($10{,}000$ held-out queries; recall@$10$ against the exact ranking throughout). FAISS and the underlying BLAS are pinned to one thread for every timed search (environment in the scope note); every method is reported at its accuracy-parameter sweep point nearest recall $0.9$ (the scope note lists the sweeps), the compact codes at the fixed byte budget of their panel; the IVF-PQ$\,+\,$rerank rows grant the product quantiser the identical full-precision re-ranking pass the binary pipelines receive, over the matched candidate shortlist; and the SIFT exact row's residual $0.001$ stems from tied distances at the rank-ten cut-off in the published ground truth ($145$ of the $10{,}000$ queries), not from search error. Bytes per vector are measured from the serialised index; the binary methods report the scanned compact code, while re-ranking fetches the full-precision vectors, held in memory during this measurement and relegated to secondary storage in the deployments of Section~\ref{sec:res_rerank}. The exact rows' throughput benefits from the platform BLAS's matrix coprocessor, which the re-ranking gathers cannot use (scope note).}
\label{tab:sift_measured}\label{tab:quora_pareto}
\begin{tabular}{@{}p{3.2cm} p{2.0cm} p{1.9cm} p{2.0cm}@{}}
\toprule
\textbf{Family} & \textbf{Bytes / vec} & \textbf{Recall@10} & \textbf{QPS} \\
\midrule
\multicolumn{4}{@{}l}{\emph{SIFT1M ($128$-d SIFT descriptors)}}\\
Exact (flat) & $512$ & $0.999$ & $1{,}397$ \\
HNSW~\cite{Malkov20} & $784$ & $0.943$ & $11{,}332$ \\
IVF-PQ~\cite{Jegou11} & $26$ & $0.588$ & $3{,}121$ \\
IVF-PQ $+$ rerank & $26$ & $0.926$ & $5{,}157$ \\
LSH $+$ rerank & $16$ & $0.554$ & $697$ \\
ITQ $+$ rerank~\cite{Gong11} & $16$ & $0.768$ & $686$ \\
\midrule
\multicolumn{4}{@{}l}{\emph{Quora, MiniLM-L6 ($384$-d)}}\\
Exact (flat) & $1536$ & $1.000$ & $1{,}783$ \\
HNSW~\cite{Malkov20} & $1808$ & $0.987$ & $13{,}503$ \\
IVF-PQ~\cite{Jegou11} & $65$ & $0.755$ & $2{,}465$ \\
IVF-PQ $+$ rerank & $65$ & $0.927$ & $8{,}462$ \\
LSH $+$ rerank & $48$ & $0.943$ & $561$ \\
ITQ $+$ rerank~\cite{Gong11} & $48$ & $0.967$ & $560$ \\
\midrule
\multicolumn{4}{@{}l}{\emph{Quora, mxbai-embed-large ($1024$-d)}}\\
Exact (flat) & $4096$ & $1.000$ & $905$ \\
HNSW~\cite{Malkov20} & $4368$ & $0.987$ & $7{,}782$ \\
IVF-PQ~\cite{Jegou11} & $161$ & $0.779$ & $932$ \\
IVF-PQ $+$ rerank & $161$ & $0.911$ & $3{,}598$ \\
LSH $+$ rerank & $128$ & $0.977$ & $421$ \\
ITQ $+$ rerank~\cite{Gong11} & $128$ & $0.989$ & $413$ \\
\bottomrule
\end{tabular}
\end{table}

\paragraph{The supervised regime.} Table~\ref{tab:cifar_measured} turns to CIFAR-10~\cite{Krizhevsky09}, where relevance is a shared class label and effectiveness is mean average precision over a Hamming ranking. This is the standard supervised-hashing protocol, here run over features from an ImageNet-pretrained network. Supervision produces the largest gains in this regime. Graph regularised hashing~\cite{Moran15c} roughly doubles the precision of unsupervised ITQ at every code length, and roughly triples that of LSH. Supervised hashing with kernels~\cite{Liu12} raises ITQ's precision by between sixty and eighty per cent. The gap reflects class information the features alone do not carry, and both supervised methods improve as the code lengthens. A deep hash head trained on the same frozen features and label budget with the pairwise DPSH objective of Section~\ref{sec:deep}\longonly{ (Equation~\ref{eqn:deep_dpsh_loss})} improves further still, to $0.598$ at sixty-four bits. Coupling projection and quantisation in a learned network yields a further margin over the two-stage supervised codes.\longonly{\footnote{This is a two-stage deep hash: the DPSH objective is trained over the fixed ResNet-18 features, so the comparison isolates the hashing objective at a matched feature and label budget. The fully end-to-end variants, which also fine-tune the backbone, are surveyed in Section~\ref{sec:deep} but not re-run here, for the reasons set out in the scope note below.}} \longonly{A per-bit diagnostic in the supplementary material traces that margin against the effective-hashcode properties of Section~\ref{sec:ch2_effective_hashcodes}. The deep code is no more redundant ($E_{4}$) than the shallow supervised ones, yet it attains the largest class separation and so the most precision per bit. A stronger projection achieves greater class separation without increased redundancy.}\shortonly{A per-bit diagnostic (Figure~S13 of the supplementary material) traces that margin to a stronger projection achieving greater class separation ($E_{4}$, Section~\ref{sec:ch2_effective_hashcodes}) without increased redundancy.} The supervisory signal improves the per-byte gain as much as the per-bit gain. At a matched eight-byte budget ($64$ bits), the same supervised codes reach a class-mAP of roughly $0.5$, against $0.223$ for the floating-point ranking of the same features ($512$-d single precision, two kilobytes). They therefore more than double the quality at a two-hundred-and-fifty-six-fold compression. The gain measures the value of the labels the task-agnostic float cannot exploit. The full comparison against float and product quantisation is given in Table~S9 of the supplementary material.

\begin{table}[!t]
\centering
\small
\caption[A reproducible supervised measurement on CIFAR-10]{A reproducible supervised measurement on CIFAR-10 ($512$-dimensional ImageNet-pretrained ResNet-18 features, class-based ground truth, mAP over a Hamming ranking; $1{,}000$ queries against a $59{,}000$-image database). The ordering is stable across code lengths and is the result of interest.}
\label{tab:cifar_measured}
\begin{tabular}{@{}p{3.8cm} ccc@{}}
\toprule
\textbf{Method} & \textbf{mAP ($16$ bits)} & \textbf{mAP ($32$)} & \textbf{mAP ($64$)} \\
\midrule
LSH (unsupervised) & $0.145$ & $0.172$ & $0.190$ \\
ITQ (unsupervised)~\cite{Gong11} & $0.249$ & $0.256$ & $0.264$ \\
GRH (supervised)~\cite{Moran15c} & $0.482$ & $0.504$ & $0.505$ \\
KSH (supervised)~\cite{Liu12} & $0.400$ & $0.446$ & $0.467$ \\
DPSH (deep)~\cite{Li16} & $0.560$ & $0.585$ & $0.598$ \\
\bottomrule
\end{tabular}
\end{table}

\iflong\input{supp/meas_cifar_diag}\fi

\iflong\input{supp/meas_cifar_pareto}\fi

\paragraph{The resurgence regime.} The compression claims of Section~\ref{sec:resurgence} rest, in the literature, largely on industrial white papers; we substantiate them here on real sentence embeddings (Table~\ref{tab:emb_quant}). We embed four BEIR corpora~\cite{Thakur21}, scifact, nfcorpus, arguana and fiqa, with two sentence-transformers~\cite{Reimers19}: a $384$-dimensional MiniLM-L6~\cite{Wang20minilm} and a $1024$-dimensional Matryoshka-trained model (mxbai-embed-large). Table~\ref{tab:emb_quant} reports the retrieval quality (nDCG@$10$) retained under each compression of the document embeddings, with the spread taken across corpora. The same pattern holds for both embedders. Scalar quantisation to eight bits is indistinguishable from full precision, and is marginally better in both cases. One-bit binary codes, a thirty-two-fold reduction, retain $88$ to $92\%$ of the floating-point quality alone; the larger embedding loses less because its discriminative information is spread across more coordinates. A single full-precision re-ranking pass over the binary candidates recovers the floating-point quality entirely. The rotated one-bit code of RaBitQ~\cite{Gao24} recovers a little more, $94$ to $96\%$, by isotropising the quantisation error. The central claim of the resurgence is that an embedding may be compressed thirty-two-fold at no measurable loss, provided a cheap re-ranking pass is retained. We observe that the measurement bears this claim out, so it no longer rests on vendor reports alone. The thirty-two-fold figure is the size of the code scanned in the first pass. A deployment either keeps the re-ranking vectors on cheaper secondary storage, in the manner of DiskANN (Section~\ref{sec:graph_disk}), reading only the few it must re-rank, or forgoes them and scores asymmetrically against the binary codes. RaBitQ takes the latter route, which is why it recovers somewhat less.

\begin{table}[!t]
\centering
\small
\caption[Measured embedding-compression quality on BEIR]{Retrieval quality (nDCG@$10$, mean over four BEIR corpora~\cite{Thakur21}) retained under each compression of the document embeddings, for two sentence-transformers; $\pm$ is the standard deviation of the retained percentage across corpora. Binary codes re-ranked over their top-$100$ Hamming candidates match full precision for both embedders, with the scanned code occupying one thirty-second of the memory; the re-ranking vectors are held as Section~\ref{sec:res_rerank} describes.}
\label{tab:emb_quant}
\begin{tabular}{@{}l r r c@{}}
\toprule
\textbf{Scheme} & \textbf{Bytes/vec} & \textbf{nDCG@10} & \textbf{\% of float} \\
\midrule
\multicolumn{4}{@{}l}{\emph{MiniLM-L6 ($384$-d)}}\\
float32 & $1536$ & $0.425$ & $100 \pm 0$ \\
int8 (scalar) & $384$ & $0.426$ & $100 \pm 0$ \\
binary (1-bit) & $48$ & $0.375$ & $88 \pm 3$ \\
binary $+$ re-rank & $48$ & $0.424$ & $100 \pm 0$ \\
RaBitQ (1-bit)~\cite{Gao24} & $48$ & $0.400$ & $94 \pm 2$ \\
\midrule
\multicolumn{4}{@{}l}{\emph{mxbai-embed-large ($1024$-d, Matryoshka)}}\\
float32 & $4096$ & $0.508$ & $100 \pm 0$ \\
int8 (scalar) & $1024$ & $0.509$ & $100 \pm 0$ \\
binary (1-bit) & $128$ & $0.469$ & $92 \pm 2$ \\
binary $+$ re-rank & $128$ & $0.509$ & $100 \pm 0$ \\
RaBitQ (1-bit)~\cite{Gao24} & $128$ & $0.487$ & $96 \pm 1$ \\
\bottomrule
\end{tabular}
\end{table}

\paragraph{Generality across embedders.} \longshort{Table~\ref{tab:emb_generality} shows that the thirty-two-fold result is not an artefact of the two embedders above. Through the released benchmark (Section~\ref{sec:ch4_measured}) we repeat the one-bit compression across seven embedders, including the two commercial OpenAI \texttt{text-embedding-3} models; the table reports the quality retained over three of the four corpora. The pattern is consistent. For six of the seven embedders, the two OpenAI models among them, a one-bit code with a re-ranking pass recovers essentially all of the floating-point quality at one thirty-second of the memory. The resurgence result therefore holds on the representations most widely deployed in practice as well as on open research models. The exception is instructive, and on inspection it lies in the quantiser's threshold rather than the embedding. The e5-base-v2 representation~\cite{Wang22e5} is strongly off-centre: some sixty per cent of its coordinates carry the same sign for almost every document, against a quarter for the embedders that compress cleanly. A one-bit code thresholded at zero therefore allocates most of its bits to coordinates of near-constant sign, and it retains only fifty-three per cent of the ranking. This is the failure of the static, zero-placed threshold that presumes mean-centred data, the first of the assumptions with which we characterised LSH at the outset. Placing the threshold instead at each coordinate's mean, the elementary decision of the quantisation axis (Section~\ref{sec:ch2_quantisation}), recovers the code to eighty-six per cent of the floating-point ranking. Re-ranking then behaves as for the others. The fault lies with the data-oblivious threshold rather than with e5 itself, this case shows that threshold placement can matter as much as the number of bits.}{We verify through the released benchmark that the thirty-two-fold result holds across seven embedders (Table~S14 of the supplementary material), including the two commercial OpenAI \texttt{text-embedding-3} models. The single exception, e5-base-v2~\cite{Wang22e5}, proves to lie in the quantiser rather than the embedding. Its strongly off-centre representation is mishandled by a one-bit code thresholded at zero, which retains only $53$ per cent of the ranking. Re-centring the threshold at each coordinate's mean, the elementary decision of the quantisation axis, recovers it to $86$ per cent.}

\iflong\input{supp/emb_generality}\fi

\shortonly{The same embeddings also answer a question the lens poses but the literature rarely tests directly. At a fixed memory budget, is it better to spend the bytes on the projection axis, by reducing dimensions, or on the quantisation axis, by reducing bits? A matched-byte comparison across three budgets per embedder, reported in Table~S10 of the supplementary material, is decisive: the quantisation axis dominates at every budget. At a matched $96$-byte budget, full-dimensional coarse quantisation of MiniLM retains an nDCG@$10$ of $0.40$, against $0.18$ for the best dimension-reduced float, more than doubling it. The reason is that the discriminative information is spread thinly across all the coordinates rather than concentrated in a few. Even the fair Matryoshka prefix-truncation baseline, the operation that embedding is trained to support, does not overturn the conclusion; it reaches $0.22$ against quantisation's $0.48$ at $256$ bytes on the larger embedder. For these representations, reducing precision preserves more retrieval quality than reducing dimensionality at the same byte budget.}
\iflong\input{supp/emb_axis}\fi

\paragraph{The trade-off as the lens anticipates it.} The Quora panels of Table~\ref{tab:quora_pareto} set the three families against one another on a modern corpus of half a million question embeddings, in the manner of the ANN benchmarks~\cite{Aumuller20}. The graph index again attains the highest throughput at near-exact recall, now at the largest footprint of all. The product quantiser reduces memory more than twentyfold at a controlled loss of recall. The learned binary code with a re-ranking pass attains a recall of $0.97$ with a scanned code one thirty-eighth of the graph's bytes. The lens anticipates two features of this comparison. First, the learned binary code (ITQ) again outranks the random one (LSH) at the same size, $0.97$ against $0.94$. Second, both one-bit codes with a re-ranking pass outrank the inverted-file product quantiser at a comparable footprint. This ordering survives granting the quantiser the identical pass. The IVF-PQ$\,+\,$rerank rows close most of its recall deficit yet still trail the learned binary code at every matched shortlist length of fifty candidates or more, $0.960$ against $0.967$ at fifty and $0.986$ against $0.999$ at five hundred \longshort{(Table~\ref{tab:rerank_control})}{(Table~S11 of the supplementary material)}, while leading on throughput at matched recall; at a shortlist of ten, where re-ranking is a no-op, the quantiser is ahead. On the raw-feature SIFT collection of Table~\ref{tab:sift_measured} the comparison runs the other way: the re-ranked quantiser dominates both axes ($0.93$ at five thousand queries per second against ITQ's $0.77$). On the learned embedding the binary code overtakes it. The binarised learned embedding reverses the ordering because, as Table~S10 of the supplementary material showed, its discriminative information is spread across the coordinates and survives one-bit quantisation. The crossover is visible directly in the operating curves of Figure~\ref{fig:quora_curves}. The graph dominates when memory is abundant, and the compact code with re-ranking dominates when memory is the binding constraint, the regime of retrieval-augmented serving at scale.\longonly{ Note that these orderings recur on a second, larger embedder. On the $1024$-dimensional mxbai embedding the learned binary code again reaches recall $0.99$ against the random code's $0.98$. Both again overtake the inverted-file quantiser at $0.78$ and its re-ranked control at the matched shortlist ($0.989$ against $0.958$ at fifty candidates). The lens therefore anticipates the design crossovers across embedding scale rather than naming them after the fact.}\shortonly{ As the mxbai panel of Table~\ref{tab:quora_pareto} shows, each of these orderings recurs on the larger embedder. The lens therefore anticipates the design crossovers across embedding scale rather than naming them after the fact.}

\input{supp/meas_quora_curves}\iflong\input{supp/rerank_control}\fi

\longshort{Taken together, the four regimes trace the survey's argument in miniature on measured data. On the unsupervised metric task, a learned projection improves upon the data-oblivious baseline. On the supervised semantic task, supervision improves upon both, and a coupled deep objective improves further still. In the resurgence, a learned embedding is compressed thirty-two-fold without loss, and the quantisation axis proves the more byte-efficient. In the system-level trade-off, the preferred structure shifts from a graph to a compact code as available memory decreases. Each step yields a quantifiable gain, and the supplementary diagnostic analyses its per-bit encoding.}{Taken together, the four regimes trace the survey's argument in miniature. A learned projection improves on a random one, and supervision and deep coupling improve further. A learned embedding compresses thirty-two-fold without loss, and the preferred structure shifts from a graph to a compact code as available memory decreases.}

\subsection{At scale: ninety million vectors and an $8.8$-million-passage corpus}\label{sec:ch4_scale_measured}

\shortonly{We repeated two of these measurements two orders of magnitude larger on a single cloud machine: on a ninety-million-vector slice of BigANN \texttt{SIFT1B} and on the full $8.84$-million-passage MS~MARCO corpus (Tables~S12--S13 of the supplementary material). The lens's crossovers hold at scale. A one-bit code with a re-ranking pass remains lossless at thirty-two-fold compression at retrieval-augmented scale. At $10^{8}$ vectors, the recall of a flat-shortlist binary code falls away. This is a failure of organisation rather than of compression, since the same bits searched by a better structure need not lose those neighbours. Organisation structures for binary codes, from the exact sub-linear Hamming search of multi-index hashing~\cite{Norouzi12mih} to indexes built natively on compact codes, are intended to close this gap.}
\iflong\input{supp/at_scale}\fi

\paragraph{Scope and limitations of the measurement.} \shortonly{These results are a controlled illustration of the lens rather than a competition leaderboard. Throughput is single-thread, and it indicates relative ordering rather than absolute speed. The deep regime is represented by a single two-stage DPSH point; the end-to-end variants are surveyed but not re-run. The product quantiser is reported both through its customary inverted file and, in the IVF-PQ$\,+\,$rerank rows, granted the identical full-precision re-ranking pass the binary pipelines receive. The operating parameters are pinned in the released harness (HNSW $M{=}32$, $\mathit{efConstruction}{=}200$, $\mathit{efSearch}$ swept $16$--$256$; IVF-PQ ${\sim}D/8$ sub-quantisers of eight bits, $4\sqrt{N}$ lists, $\mathit{nprobe}$ swept $1$--$128$; re-ranking over a shortlist swept to $500$ candidates; the at-scale runs of Section~\ref{sec:ch4_scale_measured} use scale-adjusted settings recorded in the released harness). Each table reports the sweep point nearest recall $0.9$ under single-batch timing (one batched call over all $10{,}000$ queries, best of three repetitions) on an Apple M2, with FAISS ($1.13.2$) and BLAS pinned to one thread for timed searches (builds multi-threaded; dense scans dispatch to Accelerate's matrix coprocessor, hence the flat scan's lead over the re-ranking gathers). The embedding ground truth is computed in the search metric (vectors normalised before the exact pass) over held-out queries. Retention percentages are reported to the rounding shown without significance testing (per-query distributions ship with the harness), and a deployment that keeps its float store resident holds $33/32$ of the original bytes: the compression figures describe the scanned code. The released benchmark (Section~\ref{sec:ch4_measured}) is designed so that others may extend the measurement along any of these axes.}
\iflong
These results are a controlled illustration of the lens rather than a competition leaderboard, and three boundaries should be kept in view. First, the throughput figures are single-thread queries per second on one machine with untuned index parameters. They indicate the relative ordering the lens anticipates rather than the absolute speeds a tuned, multi-threaded, hardware-specific deployment would reach; for those, the standardised system benchmarks~\cite{Aumuller20, Simhadri22} remain the authority. Second, most corpora here are of moderate scale, chosen so that every number can be reproduced on a single workstation from the released harness. The ninety-million-vector and eight-million-passage measurements of Section~\ref{sec:ch4_scale_measured} confirm that the crossovers sharpen rather than soften as the corpus grows and memory binds. The billion-vector regime that motivates the largest deployments is reviewed in Sections~\ref{sec:graph} and~\ref{sec:resurgence} rather than re-measured here. Third, the deep hashing of Section~\ref{sec:deep} is represented in the supervised probe by a single method: the pairwise DPSH objective\longonly{ of Equation~\ref{eqn:deep_dpsh_loss}} trained as a hash head over the same frozen features and label budget as the other supervised codes. The comparison therefore isolates the hashing objective rather than the representation, and it concerns a two-stage rather than a fully end-to-end deep hash. The end-to-end variants that fine-tune the backbone, and the continuation schedule of HashNet~\cite{Cao17}, are surveyed but not re-run. A faithful comparison would require each method's original training budget, and it would couple the projection and quantisation axes in a manner that resists the clean, axis-by-axis attribution the other regimes permit. The single deep point we report suffices to place the coupled, learned objective above the two-stage supervised codes, without settling the wider deep-hashing literature. The supervised CIFAR probe is likewise deliberately compact: a transparent demonstration that supervision can be encoded in a handful of bytes rather than a claim about any production system. These boundaries follow from requiring a reproducible, single-workstation measurement, and the released benchmark (Section~\ref{sec:ch4_measured}) is designed so that others may extend it along any of these axes. Fourth, the product quantiser is reported both through its customary inverted file and, in the IVF-PQ$\,+\,$rerank rows, granted the identical full-precision re-ranking pass the binary pipelines receive over the matched candidate shortlist. The binary-versus-PQ comparison therefore isolates the codes rather than coupling each quantiser to its customary organisation alone. The operating parameters are pinned in the released harness (HNSW $M{=}32$, $\mathit{efConstruction}{=}200$, $\mathit{efSearch}$ swept $16$--$256$; IVF-PQ ${\sim}D/8$ sub-quantisers of eight bits, $4\sqrt{N}$ lists, $\mathit{nprobe}$ swept $1$--$128$; re-ranking over a shortlist swept to $500$ candidates; the at-scale runs of Section~\ref{sec:ch4_scale_measured} use scale-adjusted settings recorded in the released harness). Each table reports the sweep point nearest recall $0.9$ under single-batch timing (one batched call over all $10{,}000$ queries, best of three repetitions) on an Apple M2, with FAISS ($1.13.2$) and the underlying BLAS pinned to one thread for every timed search; index construction is multi-threaded. FAISS's dense scans dispatch to the platform BLAS (Apple Accelerate), whose matrix coprocessor lets a single thread exceed scalar-core arithmetic peak. This is why the batched flat scan outpaces the per-candidate re-ranking gathers. The embedding ground truth is computed in the search metric, with the vectors normalised before the exact pass, over the $10{,}000$ queries held out of the corpus. Fewer than half a per cent of those queries have a near-duplicate in the corpus. Retention percentages are reported to the rounding shown without significance testing (per-query distributions ship with the harness), and a deployment that keeps its float store resident holds $33/32$ of the original bytes: the compression figures describe the scanned code.
\fi

\subsection{Summary}

\shortonly{This section has consolidated evaluation practice for hashing-based ANN retrieval. We defined the standard IR metrics (AUPRC, mAP) (Section~\ref{sec:ch4_metrics}) and the complementary recall--latency--memory protocol by which the dominant modern methods are evaluated at scale (Section~\ref{sec:ch4_scale}). We grounded both in a reproducible measurement that traces the survey's argument across four regimes on real data.}
\iflong
This section has consolidated evaluation practice for hashing-based ANN retrieval. \longshort{We catalogued commonly used unimodal and cross-modal datasets (Section \ref{sec:ch4_datasets}). We described two prevailing ground-truth definitions (Section \ref{sec:ch4_groundtruth}). We contrasted the standard evaluation paradigms, namely Hamming ranking and hash-table lookup, and motivated ranking-based summaries in comparative studies (Section \ref{sec:ch4_paradigms}). We reviewed dataset split strategies, including an improved approach that isolates a held-out test database to better assess generalisation (Section \ref{sec:ch4_splits}). We defined and compared the standard IR metrics (AUPRC, mAP) (Section \ref{sec:ch4_metrics}). We introduced the complementary recall--latency--memory protocol by which the dominant modern methods are evaluated at scale (Section \ref{sec:ch4_scale}).}{We defined the standard IR metrics (AUPRC, mAP) (Section \ref{sec:ch4_metrics}). We introduced the complementary recall--latency--memory protocol by which the dominant modern methods are evaluated at scale (Section \ref{sec:ch4_scale}). The standard datasets, ground-truth definitions and evaluation paradigms on which both rest were summarised above.}
\fi

\section{A Practitioner's Decision Guide}\label{sec:practitioner}

The preceding sections have argued that the families of approximate nearest neighbour search differ chiefly in which of the three stages of Definition~\ref{def:lens} they act on, and that no family dominates along every axis. For a practitioner, this reframes method selection. The question is which resource, whether memory, query latency, indexing or update cost, or interpretability, is scarcest in the application at hand, rather than which algorithm is best in the abstract. We collect in Table~\ref{tab:decision} the guidance that follows from the survey. One caveat is important: the relative performance of these methods depends strongly on the data and the hardware. The figures reported by the standardised benchmarks~\cite{Aumuller20, Simhadri22} should therefore be consulted, and a short evaluation conducted on one's own data, before a design is committed to.

\begin{table}[!t]
\centering
\small
\caption[A practitioner's decision guide]{A decision guide for selecting an approximate nearest neighbour method, organised by the binding resource constraint; each recommendation acts on a particular axis of Definition~\ref{def:lens}, and the tools column lists representative implementations.}
\label{tab:decision}
\begin{tabular}{@{}p{2.7cm} p{2.7cm} p{2.9cm} p{2.0cm}@{}}
\toprule
\textbf{Binding constraint} & \textbf{Recommended approach} & \textbf{Representative tools} & \textbf{Axis invested} \\
\midrule
Small corpus, recall-critical, memory ample & exact or graph search & flat index; HNSW (FAISS, hnswlib) & organisation \\
In-memory, low latency, $10^{6}$--$10^{8}$ items & navigable proximity graph & HNSW, NSG (hnswlib, FAISS) & organisation \\
Billion-scale, memory-bound & graph or inverted file over PQ codes, on-disk re-rank & DiskANN; IVF-PQ (FAISS); ScaNN & compression $+$ organisation \\
Extreme memory; on-device or RAG serving & binary/scalar quantisation $+$ re-rank & binary embeddings; vector DBs (Milvus, Qdrant, pgvector) & compression \\
High update churn or streaming & incrementally updatable index & in-place IVF; incremental-insert graph & organisation $+$ compression \\
Interpretability or auditability required & binary learning-to-hash codes & supervised or unsupervised hashing & projection $+$ quantisation \\
\bottomrule
\end{tabular}
\end{table}

\longshort{Three regimes account for the majority of deployments. When the corpus is small enough to fit comfortably in memory and recall is paramount, a graph index such as HNSW~\cite{Malkov20} is usually the first choice. It acts on the organisation axis, searches essentially exact distances, and attains high recall at low latency. Its memory footprint is larger, and acceptable at this scale. As the corpus grows into the hundreds of millions and memory becomes the binding constraint, the compression axis can no longer be ignored. The preferred designs co-design compression with organisation: either a graph over product-quantised vectors with on-disk re-ranking, as in DiskANN~\cite{Subramanya19}, or an inverted file over product-quantised codes~\cite{Jegou11}, as provided by libraries such as FAISS~\cite{Johnson19} and ScaNN~\cite{Guo20}. At the extreme of memory pressure, characteristic of on-device retrieval and of serving retrieval-augmented generation over very large corpora, binary or low-bit quantisation of learned embeddings, followed by a re-ranking pass, recovers most of the accuracy at a fraction of the footprint~\cite{Yamada21}. This option is increasingly available as a first-class feature of vector databases~\cite{Wang21}.}{Three regimes account for the majority of deployments. When the corpus fits comfortably in memory and recall is paramount, a graph index such as HNSW~\cite{Malkov20} is the first choice; it acts on the organisation axis to attain high recall at low latency, at a footprint acceptable at that scale. As the corpus grows into the hundreds of millions and memory binds, the preferred designs co-design compression with organisation: a graph over product-quantised vectors with on-disk re-ranking (DiskANN~\cite{Subramanya19}) or an inverted file over product-quantised codes~\cite{Jegou11}, as provided by FAISS~\cite{Johnson19} and ScaNN~\cite{Guo20}. At the extreme of memory pressure, on-device retrieval and retrieval-augmented serving over very large corpora, binary or low-bit quantisation of learned embeddings followed by a re-ranking pass recovers most of the accuracy at a fraction of the footprint~\cite{Yamada21}. It is increasingly a first-class feature of vector databases~\cite{Wang21}.}

\iflong
\paragraph{Operating points.} Concrete starting values help to instantiate these recommendations, although they must be tuned to the data. For HNSW, a maximum out-degree $M$ of $16$--$64$ and a construction width $\mathit{efConstruction}$ of $100$--$500$ are typical; the query width $\mathit{efSearch}$ is raised until the target recall is met. For an inverted file, the number of lists is commonly set to between $\sqrt{N}$ and $4\sqrt{N}$ for $N$ database items, and the number probed per query, $\mathit{nprobe}$, trades recall against latency. For product quantisation, a typical compromise between accuracy and footprint is $M = D/8$ to $D/4$ subspaces at eight bits each ($256$ centroids per subspace), giving of the order of $M$ bytes per vector. Memory governs the crossover from an in-memory graph to a compressed or on-disk index. A graph stores the full vectors and its edge lists, so once their combined size approaches the available memory, an inverted-file-plus-PQ or DiskANN design becomes preferable. These figures are indicative only; the standardised benchmarks~\cite{Aumuller20, Simhadri22} catalogue current operating points on representative data.

Two further considerations cut across the scale axis. Where the corpus is subject to frequent insertion and deletion, compact codes over an inverted file have historically been easier to maintain than a graph, whose global edge structure is awkward to update incrementally. Recent in-place update schemes for graph and quantised indexes have narrowed this gap, so the choice now turns largely on the maturity of the available implementation. Where a code must be inspected or its behaviour explained, a learned binary code is more directly inspectable than a dense embedding traversed by a graph: each bit is a known half-space test, so a match can be traced to the bits on which it agrees. This advantage is modest and application-specific.
\fi
In every case the lens supplies the same discipline: identify the scarcest resource, then choose the axis that consumes least of it.\shortonly{ Where codes must be inspected or audited, a learned binary code is more directly inspectable than a dense embedding traversed by a graph: each bit is a known half-space test. This is the basis of the final row of Table~\ref{tab:decision}.}

\iflong
\section{Beyond Retrieval: The Quantisation Axis in Model Serving}\label{sec:llm}
The PQO lens was developed to organise similarity search, but its quantisation axis describes a second memory problem that has become more pressing still: the serving of large language models. We draw the connection here because the two literatures have developed separately, and each contains results the other could use.

\paragraph{The key-value cache as a retrieval index.} A transformer attends, at each step, over the keys and values it has cached for every preceding token, and the attention weights are a softmax over the inner products between the query and the cached keys. This is a soft nearest-neighbour retrieval. The key-value cache is a growing collection of stored vectors that is read on every forward pass, and at long context lengths it dominates the memory of inference. Compressing it therefore applies the quantisation axis of the lens to the vectors that attention retrieves over. The methods that have emerged read directly as settings of that axis. Per-token and per-channel scalar quantisation to two or four bits, with the dense region around zero handled by asymmetric or non-uniform thresholds (KIVI~\cite{Liu24kivi}, KVQuant~\cite{Hooper24}), is the embedding quantisation of Section~\ref{sec:resurgence} transplanted to a different store. The separate handling of a few outlier channels is the familiar approach of allocating bits where they preserve neighbourhoods rather than uniformly.

\input{supp/llm_weights}

\paragraph{Transferable and non-transferable results.} The targets differ in ways that matter. A database of vectors is queried by a hard top-$k$, a key cache by a soft attention average, and a weight matrix is multiplied rather than queried. The error that compression must respect is, respectively, rank preservation, attention fidelity, and end-to-end loss. The design choices, however, and much of their analysis, are shared, and at present they are pursued in disconnected communities with little citation between them. Key-value and weight quantisation are the quantisation axis exercised on stores other than a retrieval index. The survey's two decades of results on threshold placement, bit allocation, rotation, and codebook design are a body of prior art for a problem that the model-serving literature is in part rediscovering. The exchange should run in both directions: the outlier-channel phenomenon that dominates weight quantisation, for instance, restates the imbalanced-variance problem that the projection axis of Section~\ref{sec:ch2_projection} was built to address.
\fi

\section{Conclusion}

We have surveyed learning to hash, and more broadly compact-code approaches to approximate nearest neighbour search, through a single factorisation into projection, quantisation, and organisation stages. Under this factorisation, a literature spanning two decades and several apparently distinct communities resolves into a few recurring design decisions, pursued with progressively more data-driven machinery. We organise the field around its geometry rather than its objectives. This complements the objective-based account of Wang et al.~\cite{Wang18}, the standard reference for classical binary hashing, and extends it across the deep, product-quantisation, graph, and retrieval-augmented eras.

\longshort{Several broad lessons emerge:

\begin{itemize}
  \item \textbf{Data-aware binarisation improves on static rules.} Optimising one or more thresholds per projected dimension generally yields more discriminative codes than a single static threshold at zero.
  \item \textbf{Not all projections are equally informative.} Allocating thresholds non-uniformly, with finer quantisation for the more informative projections, improves effectiveness over uniform allocation.
  \item \textbf{Learned projections improve on random ones, and jointly learned projections improve further.} Supervising the placement of the hashing hypersurfaces typically improves on random hyperplanes. Deep methods that learn projection and quantiser together against a single objective improve in turn on the two-stage pipelines that preceded them.
  \item \textbf{The quantisation axis extends beyond binary codes.} The product-quantisation family pursues the same compression goal with vector codebooks and asymmetric distances, trading the speed of Hamming comparison for finer resolution.
  \item \textbf{The organisation of the codes is a third, largely independent degree of freedom.} Graph-based indexes, which act on this axis almost to the exclusion of the others, dominate in-memory benchmarks and mark where the projection--quantisation perspective reaches its limit.
  \item \textbf{Compact codes have returned to the centre of large-scale retrieval.} When memory rather than computation is the binding constraint, as at the scale of retrieval-augmented generation, the questions that originally defined learning to hash define large-scale retrieval once again.
\end{itemize}}{Several broad lessons emerge. Data-aware binarisation improves on static rules: one or more learned thresholds per projected dimension yield more discriminative codes than a single static threshold at zero. Not all projections are equally informative, so non-uniform threshold allocation improves on uniform. Learned projections improve on random ones, and deep methods that learn projection and quantiser together against a single objective improve in turn on the two-stage pipelines that preceded them. The quantisation axis extends beyond binary codes; the product-quantisation family trades the speed of Hamming comparison for the finer resolution of vector codebooks and asymmetric distances. The organisation of the codes is a third, largely independent degree of freedom, on which graph-based indexes dominate in-memory benchmarks and mark where the projection--quantisation perspective reaches its limit. Compact codes have returned to the centre of large-scale retrieval: when memory rather than computation binds, as at retrieval-augmented scale, the questions that originally defined learning to hash define large-scale retrieval once again.}

We tested the lens through a reproducible measurement (Section~\ref{sec:ch4_measured}), released as the openly available, extensible \textsc{BitBudget} benchmark. Three results stand out. First, quantisation is the more byte-efficient axis under a fixed memory budget: a one-bit code with re-ranking recovers full floating-point quality, with a scanned code a thirty-second of the float's size, on six of the seven embedders measured. Second, the orderings the lens anticipates persist as the embedding is enlarged, and the compression result persists to an $8.8$-million-passage corpus; the one boundary is organisational, a flat binary shortlist failing at ninety million vectors (Section~\ref{sec:ch4_scale_measured}). Third, given labels, a compact supervised code more than doubles the semantic-retrieval quality of a far larger task-agnostic floating-point vector. The one informative exception, an embedding whose off-centre geometry a zero threshold mishandles, confirmed the first lesson: threshold placement matters as much as the number of bits allocated.

\shortonly{We also observed recurring \emph{evaluation pitfalls}: inconsistent ground-truth definitions, split protocols that reuse one database for training and final evaluation, and under-specified metric choices. Section~S8 of the supplementary material sets out the remedies alongside the methodology.}
\iflong\input{supp/eval_pitfalls}\fi

\paragraph{Open problems.} Several questions remain open, and the lens helps to frame them\longshort{:
\begin{itemize}
  \item \textbf{Co-design of the three stages.} The three stages have largely been optimised in isolation, yet the systems that perform best at scale already couple them. A principled account of their joint design remains to be developed.
  \item \textbf{The geometry of the embedding space.} The lens takes the embedding as given, yet the difficulty of search has migrated upstream, into the distribution of the embeddings themselves. Hubness~\cite{Radovanovic10}, anisotropy, and high local intrinsic dimensionality~\cite{Amsaleg15} can degrade every subsequent stage. A single-vector embedding's capacity is bounded by its dimensionality, through the sign-rank of the relevance matrix, independently of training~\cite{Weller25}. How index design should adapt to this geometry, and whether representation and index are better co-designed, remains largely unexplored.
  \item \textbf{Memory at retrieval-augmented scale.} As corpora reach billions of learned embeddings, the question is how aggressively they may be compressed, and how compact codes are best combined with graph indexes as a compression layer.
  \item \textbf{Updatable organisation.} Graph indexes are costly to modify. In-place update schemes have made substantial recent progress~\cite{Xu23}, yet the co-design of freshness with aggressive compression remains open.
  \item \textbf{Auditable compact codes.} As retrieval increasingly grounds generative models, codes whose behaviour can be explained and verified grow in practical importance.
  \item \textbf{Generative retrieval and semantic identifiers.} Understanding learned discrete identifiers as quantisation codes, and characterising the trade-offs between generating identifiers and searching an explicit index, remains open.
  \item \textbf{Human alignment.} Whether gains in ranking metrics correspond to downstream usefulness, particularly within retrieval-augmented pipelines, remains insufficiently studied.
\end{itemize}}{. \textbf{Co-design of the three stages}: the three stages have largely been optimised in isolation, yet the systems that perform best at scale already couple them. A principled account of their joint design remains to be developed. \textbf{The geometry of the embedding space}: the lens takes the embedding as given, yet the difficulty of search has migrated upstream, into the distribution of the embeddings themselves. Hubness~\cite{Radovanovic10}, anisotropy, and high local intrinsic dimensionality~\cite{Amsaleg15} can degrade every subsequent stage. A single-vector embedding's capacity is bounded by its dimensionality, through the sign-rank of the relevance matrix, independently of training~\cite{Weller25}. How index design should adapt to this geometry, and whether representation and index are better co-designed, remains largely unexplored. \textbf{Memory at retrieval-augmented scale}: as corpora reach billions of learned embeddings, the question is how aggressively they may be compressed, and how compact codes are best combined with graph indexes as a compression layer. \textbf{Updatable organisation}: graph indexes are costly to modify. In-place update schemes have made substantial recent progress~\cite{Xu23}, yet the co-design of freshness with aggressive compression remains open. \textbf{Auditable compact codes}: as retrieval increasingly grounds generative models, codes whose behaviour can be explained and verified grow in practical importance. \textbf{Generative retrieval and semantic identifiers}: understanding learned discrete identifiers as quantisation codes, and characterising the trade-offs between generating identifiers and searching an explicit index, remains open. \textbf{Human alignment}: whether gains in ranking metrics correspond to downstream usefulness, particularly within retrieval-augmented pipelines, remains insufficiently studied.}

Taken together, these developments suggest that projection, quantisation, and organisation form a durable vocabulary for similarity search, one that has accommodated the transition from hand-crafted features to learned embeddings and from millions of items to billions. Because the literature continues to move quickly, we complement this static account with a continuously updated, community-curated index at \url{https://awesomepapers.io/similarity-search}~\cite{AwesomePapers}, which mirrors the organisation adopted here and tracks developments beyond this survey.

\bibliographystyle{ACM-Reference-Format}
\bibliography{sn-bibliography}

\end{document}

%% file: sn-preamble.tex
\usepackage{multirow}
\usepackage[title]{appendix}
\usepackage{manyfoot}
\usepackage{algorithmicx}
\usepackage{algpseudocode}
\usepackage{subfig}
\usepackage{rotating}
\usepackage{morefloats}
\usepackage{capt-of}
\usepackage{float}
\usepackage{array}
\usepackage{colortbl}
\usepackage{tikz}
\usetikzlibrary{calc,positioning}
\usepackage{adjustbox}
\usepackage{afterpage}

\usepackage[norelsize, linesnumbered, ruled, lined, boxed, commentsnumbered]{algorithm2e}

\theoremstyle{acmplain}

\theoremstyle{acmdefinition}
\newtheorem{definition}{Definition}

\newcommand{\RE}{\mathbb{R}}
\newcommand{\x}{\mathbf{x}}
\newcommand{\h}{\mathbf{h}}
\newcommand{\y}{\mathbf{y}}
\newcommand{\uu}{\mathbf{u}}
\newcommand{\bb}{\mathbf{b}}
\newcommand{\w}{\mathbf{w}}

\newcommand{\q}{\mathbf{q}}
\newcommand{\cc}{\mathbf{c}}

\newcommand{\T}{\intercal}

\newif\ifsupp
\suppfalse
\ifdefined\SUPP \supptrue \fi

\newif\iflong
\longtrue                 % default = long (arXiv); also holds for the supplement build
\ifdefined\CSUR \longfalse \fi   % the CSUR submission is always the condensed build
\newcommand{\longonly}[1]{\iflong #1\fi}
\newcommand{\shortonly}[1]{\iflong\else #1\fi}
\newcommand{\longshort}[2]{\iflong #1\else #2\fi}

%% file: supp/pipeline_figure.tex
%% Projection+quantisation pipeline schematic (fig:ch1_pipeline).
%% Relocated from the CSUR main paper; the pipeline is described in the text.
\begin{figure}[!t]
\centering
\hspace{-0.3in}\subfloat[\textbf{Projection}]{\includegraphics[width=110mm, height=75mm]{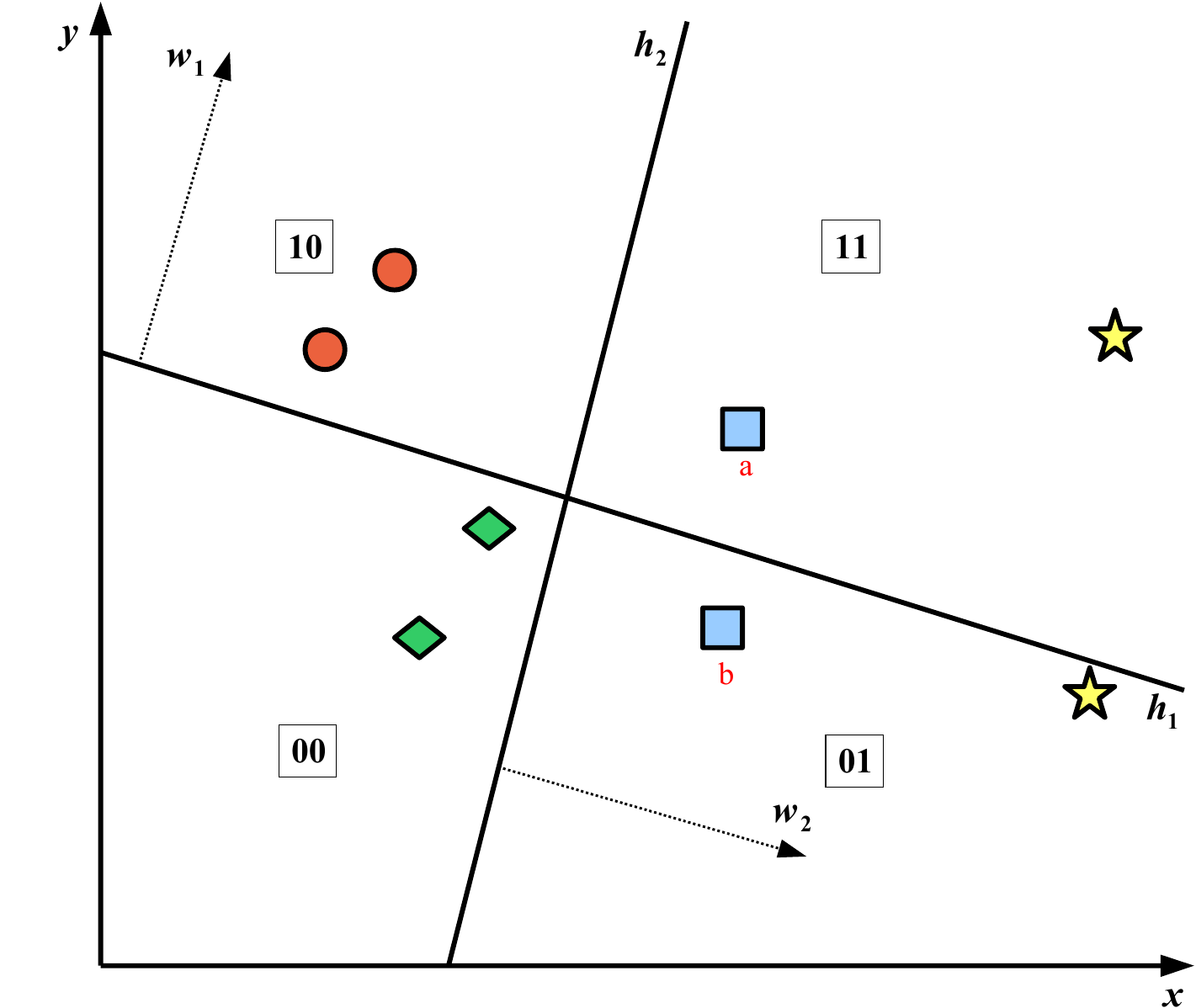}\label{fig:ch1_pipeline_1}}\\
\hspace{0.15in}
\subfloat[\textbf{Quantisation}]{\includegraphics[width=\textwidth, height=46mm, keepaspectratio]{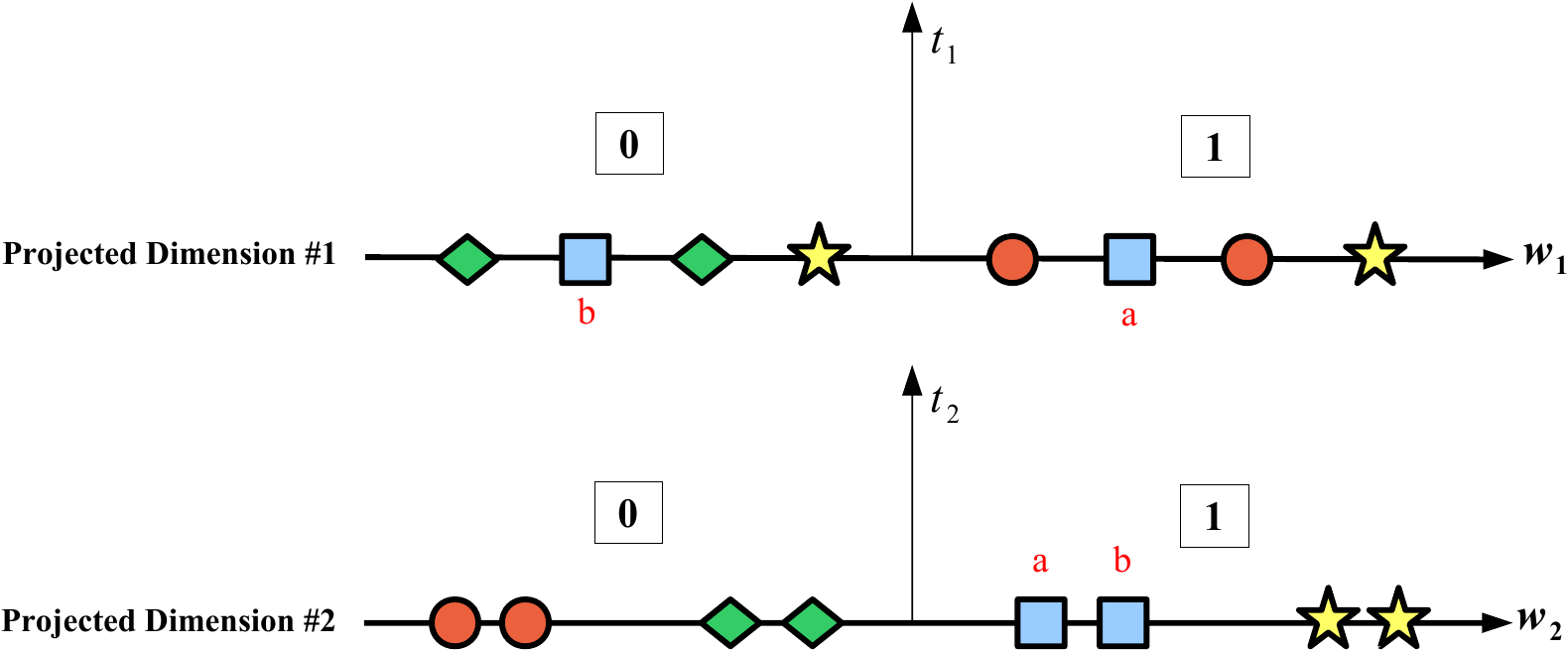}\label{fig:ch1_pipeline_2}}
\caption[The projection and quantisation steps of hashcode generation]{The projection and quantisation operations. In Figure \protect\subref{fig:ch1_pipeline_1}, two hyperplanes $\h_{1}$, $\h_{2}$ with normal vectors $\w_{1}, \w_{2}$ partition a 2D space into four buckets; data-points appear as coloured shapes, similar points sharing a colour and shape, and each data-point's hashcode is the dot-product of its feature representation onto the normal vectors. Thresholding at $t_{1}$, $t_{2}$ binarises the resulting projected dimensions (Figure \protect\subref{fig:ch1_pipeline_2}) and the bits are concatenated into a 2-bit hashcode (the unfilled squares): point $a$'s projections exceed both $t_{1}$ and $t_{2}$, so its hashcode is `11', the label of the top-right region in Figure \protect\subref{fig:ch1_pipeline_1}.}
\label{fig:ch1_pipeline}
\end{figure}

%% file: supp/surveys_table.tex
%% Prior-surveys comparison (tab:surveys) + its discussion.
%% Relocated from the CSUR main paper; positioning summarised there.
To make the differentiation from prior surveys concrete, Table~\ref{tab:surveys} contrasts the present work with the principal incumbents. The comparison covers their organising principle, the method families they cover, and their temporal range. Each incumbent is individually strong. We intend to extend them rather than to supplant them: we add the organisation axis and carry the resulting factorisation across the eras they predate. The taxonomy of~\cite{Chi17} and the supervision-level organisation of~\cite{Wang16} predate the deep era. The canonical objective-based organisation of~\cite{Wang18}, extended to deep methods by~\cite{Luo23}, treats quantisation as a loss formulation rather than a geometric stage. It reaches neither product quantisation, graph indexes, nor the vector-database era. The experimental comparison of~\cite{Li20} maps methods' relative performance but offers no unifying account of why methods differ. The family-specific surveys of product quantisation~\cite{Matsui18} and graph indexes~\cite{WangM21} are authoritative within their families and do not cross them. No incumbent spans the classical, deep, quantisation, graph, and RAG-era literatures under a single organising principle, and none provides the practitioner guidance of Section~\ref{sec:practitioner}.

\begin{table}[!t]
\centering
\small
\caption[The present survey contrasted with prior surveys]{The present survey contrasted with the principal prior surveys of learning to hash and ANN search. No incumbent organises the classical, deep, product-quantisation, graph, and RAG-era literatures under a single principle.}
\label{tab:surveys}
\resizebox{\textwidth}{!}{%
\begin{tabular}{@{}p{3.0cm} p{3.4cm} p{3.8cm} p{2.6cm}@{}}
\toprule
\textbf{Survey} & \textbf{Organising principle} & \textbf{Families / scope} & \textbf{Coverage} \\
\midrule
Chi \& Zhu~\cite{Chi17} (CSUR'17) & taxonomy by data type and mechanism & randomised and adaptive hashing & pre-deep \\
Wang et al.~\cite{Wang16} (Proc.\ IEEE'16) & level of supervision & unsupervised / semi / supervised hashing & pre-deep, brief deep coda \\
Wang et al.~\cite{Wang18} (TPAMI'18) & similarity-preserving objective & binary hashing; quantisation as a loss & to $\sim$2017; no graphs, PQ systems, or RAG \\
Luo et al.~\cite{Luo23} (TKDD'23) & deep-hashing objective & deep hashing only & deep era; no classical, PQ, graph, or RAG \\
Li et al.~\cite{Li20} (TKDE'20) & empirical benchmark & graph, LSH, PQ, tree indexes & systems; no unifying account \\
Matsui et al.~\cite{Matsui18} (ITE'18) & PQ family taxonomy & product quantisation only & no hashing, graphs, or RAG \\
Wang et al.~\cite{WangM21} (VLDB'21) & empirical comparison & graph indexes only & systems; no compact codes \\
\textbf{This survey} & projection--quantisation--organisation lens & LSH, binary hashing, deep, PQ, graph, binary-embedding / RAG & 1999--2025; with a practitioner decision guide \\
\bottomrule
\end{tabular}%
}
\end{table}

%% file: body/background.tex
This section reviews prior research on hashing-based approximate nearest neighbour (ANN) search. We begin with the standard nearest neighbour (NN) problem, its computational challenges, and the motivation for approximate variants. We then describe Locality-Sensitive Hashing (LSH), a seminal method that offers ANN search with sublinear query complexity. The data-oblivious design and simplistic assumptions of LSH in turn motivate the more recent data-driven approaches, which improve retrieval effectiveness.

We structure this background around the two components of the \emph{classical} hashcode-generation pipeline: (i) binary quantisation and (ii) projection function learning. This split mirrors the architecture of most classical hashing algorithms. The third axis of the lens, the organisation of the codes for search, is the modern addition introduced from Section~\ref{sec:lens} onward.

\subsection{Preliminaries and Notation}

\shortonly{We adopt standard conventions throughout: bold lowercase for vectors and bold uppercase for matrices, with a dataset $\mathbf{X} \in \mathbb{R}^{N \times D}$ of $N$ points in $D$ dimensions. We give full notation in Section~S1 of the supplementary material.}

\iflong
We adopt the following conventions. We write vectors in bold lowercase (e.g., $\mathbf{x}$) and matrices in bold uppercase (e.g., $\mathbf{X}$). The $(i,j)$-th entry of $\mathbf{X}$ is $X_{ij}$, and scalar-valued functions are regular lowercase, e.g., $d(\cdot,\cdot)$. A vector $\mathbf{x} = [x_1, x_2, \ldots, x_D]^{\mathsf{T}}$ is a column vector. A matrix $\mathbf{X} = [\mathbf{x}_1, \mathbf{x}_2, \ldots, \mathbf{x}_N]^{\mathsf{T}} \in \mathbb{R}^{N \times D}$ represents $N$ points in $D$-dimensional space. Its $c$-th column is $\x^c = \mathbf{X}_{\bullet c}$, and its $r$-th row is $\x_r = \mathbf{X}_{r \bullet}$.
\fi

Given a dataset $\mathbf{X} \in \mathbb{R}^{N \times D}$ of $N$ points $\x_i \in \mathbb{R}^{D}$, we aim to learn $K$ binary hash functions $\{ h_k : \mathbb{R}^{D} \rightarrow \{0,1\} \}_{k=1}^K$. Their outputs concatenate into a binary embedding $g : \mathbb{R}^{D} \rightarrow \{0,1\}^{K}$ with $g(\x_i) = [h_1(\x_i), h_2(\x_i), \ldots, h_K(\x_i)] = \bb_i$. We require that the hashcodes $\bb_i$ and $\bb_j$ are similar (e.g., under Hamming distance) whenever the inputs $\x_i$ and $\x_j$ are similar in the original space.

The remainder of this section reviews how such similarity-preserving hash functions have been constructed, focusing on quantisation and projection. The projection--quantisation perspective organises the survey as a whole. It spans the classical methods of Sections~\ref{sec:ch2_quantisation}--\ref{sec:ch2_projection} and the deep, product-quantisation, graph-based, and binary-embedding families taken up later. We introduce it in Section~\ref{sec:lens} and summarise it in Table~\ref{tab:lens}.

These models are evaluated primarily on image datasets, but they apply to any data representable as fixed-length vectors, including text and audio. Performance may vary by modality; eigen-decomposition methods, for example, often excel on low-dimensional visual features but scale poorly to high-dimensional sparse text.

\subsection{Approximate Nearest Neighbour (ANN) Search}
\label{sec:ch2_ann_search}

We now formalise the NN search problem and its widely adopted approximation. The classic 1-NN problem finds the point $NN(\q)$ closest to a query $\q \in \mathbb{R}^{D}$ in a database $\mathbf{X}$ under a distance function $d(\cdot,\cdot)$. The $k$-NN problem extends this by retrieving the $k$ closest neighbours.

\iflong
We quantify similarity with a distance function $d: \mathbb{R}^{D} \times \mathbb{R}^{D} \rightarrow \mathbb{R}_{\geq 0}$, and the 1-NN problem is:

\begin{equation}
NN(\mathbf{q}) = \arg\min_{\x_i \in \mathbf{X}} d(\x_i,\q)
\label{eqn:ch2_one_nn_search}
\end{equation}

Distance functions used in practice include the $l_p$-norm:

\begin{equation}
d_{pnorm}(\x_i,\x_j) = \left( \sum_{k=1}^{D} |x_{ik} - x_{jk}|^{\rho} \right)^{1/\rho}
\label{eqn:ch2_pnorm}
\end{equation}

\noindent where $\rho \in \mathbb{R}_{+}$. Common settings are $\rho = 1$ (Manhattan) and $\rho = 2$ (Euclidean). This Minkowski family is a metric for $\rho \geq 1$; for $\rho < 1$ it is a quasi-norm that violates the triangle inequality. Another frequently used metric is the cosine distance:

\begin{equation}
d_{cosine}(\x_i,\x_j) = 1 - \frac{\sum_{k=1}^{D} x_{ik}x_{jk}}{\sqrt{\sum_{k=1}^{D} x_{ik}^{2}} \sqrt{\sum_{k=1}^{D} x_{jk}^{2}}}
\label{eqn:ch2_cosine}
\end{equation}

\noindent For binary hashcodes, Hamming distance is the standard metric:

\begin{equation}
d_{hamming}(\bb_i,\bb_j) = \sum_{k=1}^{K} \mathbb{I}[b_{ik} \neq b_{jk}]
\label{eqn:ch2_hamming}
\end{equation}

\noindent where $\mathbb{I}[\cdot] = 1$ if the condition holds and $0$ otherwise. Intuitively, the Hamming distance counts the bits that differ between two binary vectors.
\fi

These generic functions are data-agnostic: they assume distances capture semantic similarity across all domains. This expectation is often unrealistic, particularly in high dimensions, and yields suboptimal retrieval. Metric learning adapts the distance function to the dataset, often improving nearest neighbour quality significantly~\cite{Kulis13}. This data-adaptive principle underpins many recent advances in hashing-based ANN search, particularly projection function design (see Section~\ref{sec:ch2_projection}).

Brute-force search compares the query against all $N$ points in $\mathcal{O}(ND)$ time, which is prohibitive at scale. Spatial indexes such as KD-trees~\cite{Bentley75}, quad-trees~\cite{Finkel74}, X-trees~\cite{Berchtold97}, and SR-trees~\cite{Katayama97} help in low dimensions. In high dimensions they degrade to linear time, a failure known as the \emph{curse of dimensionality}~\cite{Beyer99, Weber98}.

Approximate Nearest Neighbour (ANN) search retrieves neighbours sufficiently close to the query, trading accuracy for efficiency. The $(c,R)$-approximate nearest neighbour problem formalises this trade-off. The related $R$-near neighbour problem omits the approximation factor $c$.

\iflong
\begin{definition}[Randomised $c$-approximate $R$-near neighbour problem]
\label{def:ch2_rapprox_nn}
Given a set of $N$ data points in a $D$-dimensional space and a query $\mathbf{q}$, if some data point lies within distance $R$ of $\mathbf{q}$, return a data point within distance $cR$ of $\mathbf{q}$ with probability at least $1 - \delta$, where $\delta > 0$ and $c > 1$.
\end{definition}

\input{supp/auto_ch2_approx_nn}

The approximation factor $c$ controls how close a retrieved neighbour must be to the optimal one. Larger $c$ relaxes the requirement and can substantially improve query efficiency. A related problem, $R$-near neighbour search, omits the approximation factor:

\begin{definition}[Randomised $R$-near neighbour problem]
\label{def:ch2_rnn}
Given a set of $N$ data points in a $D$-dimensional space and a query $\mathbf{q}$, if some data point lies within distance $R$ of $\mathbf{q}$, return one such point (or report that none exists) with probability at least $1 - \delta$, where $\delta > 0$ and $R > 0$.
\end{definition}
\fi

We next introduce Locality-Sensitive Hashing (LSH), a widely studied framework that solves these ANN decision problems efficiently in high dimensions.

\subsection{Locality-Sensitive Hashing (LSH)}
\label{sec:ch2_lsh}

Hashing-based ANN search pre-processes the dataset $\mathbf{X} \in \mathbb{R}^{N \times D}$ so that query-time retrieval is far faster than brute-force search. Locality-Sensitive Hashing (LSH)~\cite{Indyk98} is among the most influential algorithms in this domain. It offered the first sub-linear time solution to the randomised $(c,R)$-approximate near neighbour problem and is widely applied across computer vision and retrieval.\longonly{ Examples include large-scale object recognition~\cite{Dean13}, pose estimation~\cite{Shakhnarovich03}, image indexing~\cite{Chum08}, and shape matching~\cite{Grauman04}.}

LSH hashes input vectors into buckets such that similar points are more likely to share a bucket. Nearest neighbour search then reduces to probing a small number of candidate buckets. The method rests on a family of hash functions with the locality-sensitive property: closer points collide with higher probability than distant ones.

\iflong
\begin{definition}[Locality-sensitive hash function family]
\label{def:ch2_lsh_family}
A hash function family $\mathcal{H}$ is $(R, cR, P_1, P_2)$-sensitive with respect to a distance function $d(\cdot, \cdot)$ if, for any two points $\mathbf{p}, \mathbf{q} \in \mathbb{R}^{D}$:

\[
\begin{aligned}
\text{if } d(\mathbf{p}, \mathbf{q}) \le R & \Rightarrow \Pr_{\mathcal{H}}[h(\mathbf{p}) = h(\mathbf{q})] \ge P_1 \\
\text{if } d(\mathbf{p}, \mathbf{q}) \ge cR & \Rightarrow \Pr_{\mathcal{H}}[h(\mathbf{p}) = h(\mathbf{q})] \le P_2
\end{aligned}
\]
\end{definition}

\noindent
Here $P_1 > P_2$ and $c > 1$ are required for the family to be useful. These conditions ensure that close pairs collide more often than distant pairs.\longonly{ Hash function families with this property have been developed for several distance measures, including $L_p$ norms~\cite{Datar04}, cosine similarity~\cite{Charikar02}, Jaccard similarity~\cite{Broder97}, and angular distances on hyperspheres~\cite{Terasawa07}.}

We amplify the gap between $P_1$ and $P_2$ by concatenating $K$ independently sampled hash functions $h_k \in \mathcal{H}$ into a composite function:

\begin{equation}
g(\mathbf{x}) = [h_1(\mathbf{x}), h_2(\mathbf{x}), \ldots, h_K(\mathbf{x})]
\label{eqn:ch2_lsh_embed_fn}
\end{equation}

\noindent
This maps each point to a $K$-bit hashcode and reduces the chance that distant points collide. The composite function $g$ is drawn from a higher-order family $\mathcal{G}$ and indexes data into $2^K$ buckets per hashtable.

To maintain recall while increasing precision, we use multiple such hashtables, indexed by $L$ independent $g_l$ functions.\longonly{ Figure~\ref{fig:ch2_lsh_table} shows the effect of $L$ on the collision probability.} The collision probability across at least one of the $L$ tables is $P^{''}_1 = 1 - (1 - P_1^K)^L$. We tune this for the desired trade-off between recall, precision, and query time.\longonly{ Figure~\ref{fig:ch2_lsh_key} illustrates the interaction between these parameters.}

\begin{algorithm}[!t]
\DontPrintSemicolon
\KwIn{Data $\mathbf{X} \in \mathbb{R}^{N \times D}$, $L$ embedding functions $g_l$, each composed of $K$ hash functions from family $\mathcal{H}$}
\KwOut{$L$ hashtables $H_1, \ldots, H_L$ with indexed data}
\For{$l \gets 1$ \textbf{to} $L$} {
    \For{$j \gets 1$ \textbf{to} $N$} {
        Insert $\x_j$ into bucket $H_l[g_l(\x_j)]$\;
    }
}
\Return{$H_1, \ldots, H_L$}
\caption{LSH Pre-processing}
\label{alg:ch2_lsh_preprocess}
\end{algorithm}

The pre-processing step (Algorithm~\ref{alg:ch2_lsh_preprocess}) indexes data points into multiple hashtables. During querying, we hash the query point with the same $g_l$ functions and draw candidate neighbours from matching buckets:

\begin{algorithm}[!t]
\DontPrintSemicolon
\KwIn{Query $\mathbf{q} \in \mathbb{R}^{D}$, hashtables $H_1, \ldots, H_L$, embedding functions $g_l$}
\KwOut{Set $S$ of approximate nearest neighbours}
\For{$l \gets 1$ \textbf{to} $L$} {
    \For{each $\x_j$ in $H_l[g_l(\mathbf{q})]$} {
        \If{$d(\x_j, \mathbf{q}) < R$}{
            Add $\x_j$ to $S$\;
        }
    }
}
\Return{$S$}
\caption{LSH Querying Step}
\label{alg:ch2_lsh_query}
\end{algorithm}

Different query strategies support the two decision problems. The $(c,R)$-approximate problem typically uses early stopping, e.g., after retrieving $3L$ candidates. The $R$-near problem examines all collisions exhaustively.
\fi

\shortonly{The locality-sensitive property is defined over a family $\mathcal{H}$ that is $(R, cR, P_1, P_2)$-sensitive with $P_1 > P_2$. We amplify the gap between $P_1$ and $P_2$ by concatenating $K$ hash functions into a $K$-bit code and replicating across $L$ hashtables; $K$ and $L$ jointly govern the precision/recall trade-off (Section~S1 of the supplementary material).}

LSH effectiveness depends heavily on the hash family and on $K$ and $L$, whose selection is well documented~\cite{Petrovic12, Osborne14, Rajaraman11}. LSH remains foundational and underpins the extensions discussed below.

\subsection{LSH with Sign Random Projections}
\label{sec:ch2_lsh_sign_random}

A particularly important instance of LSH uses random hyperplane projections to measure inner product or cosine similarity. This family, $\mathcal{H}_{cosine}$, is a frequent baseline in the learning-to-hash literature. Each hash function composes a projection function, a dot product against a random hyperplane normal, with a quantisation function, which thresholds the sign at zero. This gives the projection--quantisation decomposition that organises this survey in concrete form.\shortonly{ We sample the normals from a zero-mean unit-variance Gaussian, and each bit records the side of the hyperplane on which a point falls. The family's defining property is that the collision probability falls linearly with the angle between two vectors, $\Pr[h(\mathbf{p}) = h(\mathbf{q})] = 1 - \theta(\mathbf{p}, \mathbf{q})/\pi$~\cite{Goemans95, Charikar02}. The $K$/$L$ amplification of Section~\ref{sec:ch2_lsh} applies unchanged.}

\iflong
The inner product similarity is:

\begin{equation}
\begin{aligned}
s(\mathbf{p}, \mathbf{q}) & = \sum_{k=1}^{D} p_k q_k = \mathbf{p}^{\intercal} \mathbf{q}
\end{aligned}
\label{eqn:ch2_cosine2}
\end{equation}

When vectors are $L_2$-normalised to the unit sphere, this becomes cosine similarity, which measures the angle between them. As the angle increases, similarity and the collision probability under $\mathcal{H}_{cosine}$ both decrease.

We sample $K$ random hyperplanes from a zero-mean, unit-variance Gaussian. Each $h_k$ assigns a bit to a point $\mathbf{q}$ according to the side of the hyperplane on which it lies:

\begin{equation}
\begin{aligned}
h_k(\mathbf{q}) & = \frac{1}{2}(1 + \text{sgn}(\mathbf{w}_k^{\intercal} \mathbf{q}))
\end{aligned}
\label{eqn:ch2_cosine_hash_function}
\end{equation}

\noindent
Here $\mathbf{w}_k$ is the normal vector of the $k$-th hyperplane and \text{sgn} is the sign function (with $\text{sgn}(0) = -1$). This hash function composes a projection function $p_k : \mathbb{R}^D \rightarrow \mathbb{R}$, which computes the dot product, with a quantisation function $q_k : \mathbb{R} \rightarrow \{0,1\}$, which thresholds at zero.

The key property of $\mathcal{H}_{cosine}$ is that the collision probability (i.e., $h(\mathbf{p}) = h(\mathbf{q})$) decreases linearly with the angle $\theta$ between the vectors:

\begin{equation}
\Pr_{\mathcal{H}_{cosine}}(h(\mathbf{p}) = h(\mathbf{q})) = 1 - \frac{\theta(\mathbf{p}, \mathbf{q})}{\pi}
\label{eqn:ch2_cosine_angle}
\end{equation}

This angular sensitivity makes $\mathcal{H}_{cosine}$ effective for cosine-similarity ANN search. The amplification strategy of Section~\ref{sec:ch2_lsh} applies here too: we concatenate $K$ hash functions and use $L$ hashtables to increase discriminative power and collision selectivity.

\begin{figure}[!t]
\centering
\subfloat[Hashcode length $K$]{\includegraphics[width=0.47\textwidth]{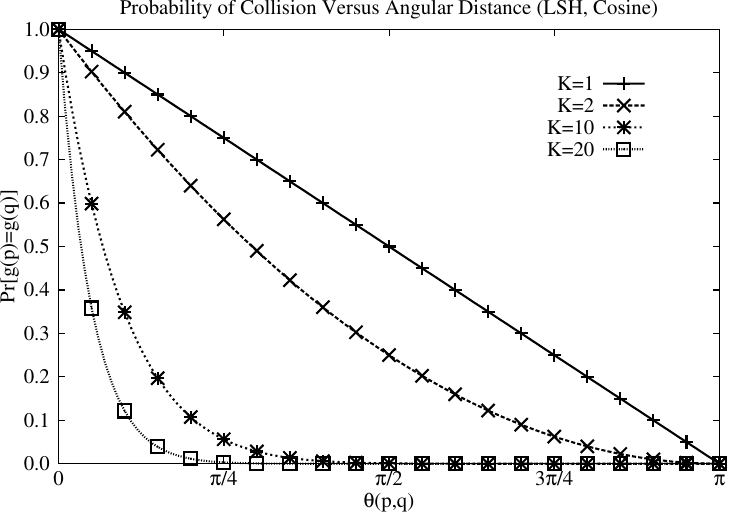}\label{fig:ch2_lsh_key}}\hfil
\subfloat[Number of hashtables $L$]{\includegraphics[width=0.47\textwidth]{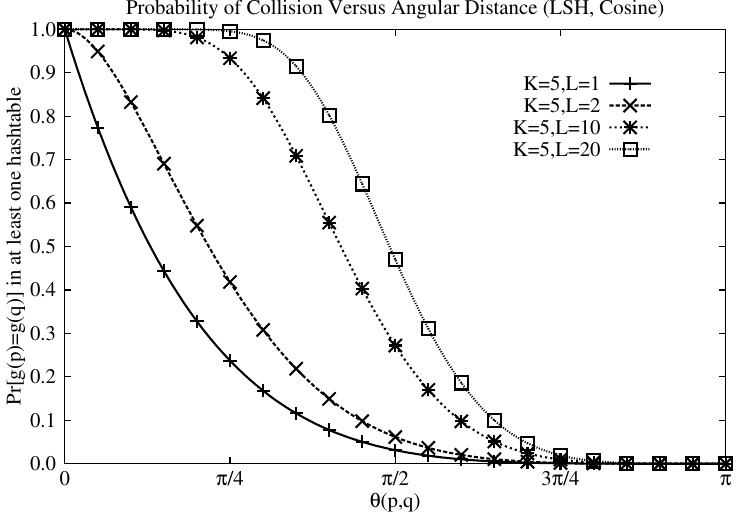}\label{fig:ch2_lsh_table}}
\caption[Collision probability under the cosine LSH family]{Probability of two hashcodes $g(\mathbf{p})$ and $g(\mathbf{q})$ matching under the $\mathcal{H}_{cosine}$ family. \protect\subref{fig:ch2_lsh_key}~As the hashcode length $K$ grows, the collision probability declines sharply, and more steeply for vectors with larger angular separation. \protect\subref{fig:ch2_lsh_table}~With $K=5$ fixed, increasing the number of hashtables $L$ raises the likelihood that nearby points collide in at least one bucket (adapted from~\cite{Petrovic12}).}
\label{fig:ch2_lsh_collision}
\end{figure}

An upper bound on the per-query cost with this hash family, before de-duplicating candidates across the $L$ tables, is:
\[
\mathcal{O}(KDL) + \mathcal{O}(L) + \mathcal{O}\left(\frac{NDL}{2^{\min(K,D)}}\right)
\]
\noindent
The terms are hashcode computation, bucket lookup, and distance comparisons with candidates, respectively. Increasing $K$ reduces candidates through more selective buckets but raises hashing cost. Increasing $L$ improves recall by spreading the query across more hashtables, at the expense of runtime.
\fi

This framework underpins many modern hashing methods. However, the random projection and thresholding design of LSH is data-independent and can be inefficient. It often requires many bits and multiple hashtables for adequate performance. Recent work addresses this by learning data-dependent projection and quantisation functions. We review these data-driven approaches in the following sections.

%% file: supp/auto_ch2_approx_nn.tex
%% Relocated from the CSUR main paper to the supplement (fig:ch2_approx_nn).
\begin{figure}[!t]
\centering
\includegraphics[width=\textwidth, height=70mm, keepaspectratio]{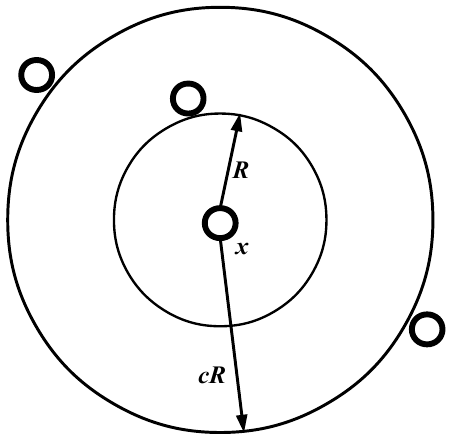}
\caption[The $(c,R)$-approximate NN problem]{The $(c,R)$-approximate NN problem. Many applications accept a retrieved data point (circle) within distance $cR$ of the query point $\mathbf{q}$, where $R$ is the distance to the exact NN.}
\label{fig:ch2_approx_nn}
\end{figure}

%% file: supp/lens_figure.tex
%% PQO-lens schematic (fig:lens). Relocated from the CSUR main paper;
%% the factorisation is given as Definition~\ref{def:lens} and Table~\ref{tab:lens} there.
\begin{figure}[!t]
\centering
\resizebox{\textwidth}{!}{%
\begin{tikzpicture}[
  font=\small, >=stealth,
  stage/.style={draw, rounded corners, minimum width=2.9cm, minimum height=1.15cm, align=center, fill=gray!12},
  ch/.style={align=center, font=\footnotesize, text=black!70}]
  \node (x) at (-0.4,0) {$\x \in \RE^{D}$};
  \node[stage] (p) at (3.0,0) {\textbf{Projection} $p$\\embed};
  \node[stage] (q) at (7.1,0) {\textbf{Quantisation} $q$\\compress};
  \node[stage] (i) at (11.2,0) {\textbf{Organisation} $\mathcal{I}$\\search};
  \node (out) at (13.9,0) {neighbours};
  \draw[->] (x) -- (p);
  \draw[->] (p) -- node[above, font=\footnotesize] {$\y \in \RE^{M}$} (q);
  \draw[->] (q) -- node[above, font=\footnotesize] {$\cc \in \mathcal{C}$} (i);
  \draw[->] (i) -- (out);
  \draw[gray, dashed] (p) -- (3.0,-1.1);
  \draw[gray, dashed] (q) -- (7.1,-1.1);
  \draw[gray, dashed] (i) -- (11.2,-1.1);
  \node[ch] at (3.0,-1.95) {random\\ unsupervised\\ supervised\\ deep network};
  \node[ch] at (7.1,-1.95) {single threshold\\ multi-threshold\\ vector codebook};
  \node[ch] at (11.2,-1.95) {hashtable\\ inverted list\\ graph};
  \node[ch, text=black!55] at (3.0,-3.3) {\emph{where to place}\\ \emph{the projections}};
  \node[ch, text=black!55] at (7.1,-3.3) {\emph{where to place}\\ \emph{the thresholds}};
  \node[ch, text=black!35] at (11.2,-3.3) {\emph{(how to organise}\\ \emph{the codes)}};
\end{tikzpicture}%
}
\caption{The PQO lens (Definition~\ref{def:lens}): a similarity-search system factorises into a projection $p$ that embeds a data-point, a quantisation $q$ that compresses the embedding into a code drawn from an alphabet $\mathcal{C}$, and an organisation structure $\mathcal{I}$ that searches the codes. Each stage admits a spectrum of increasingly data-driven choices (listed beneath it); a named method selects one choice per stage, as set out in Table~\ref{tab:lens}. The first two stages pose the design questions the learning-to-hash literature has overwhelmingly targeted; we take up the organisation of the codes (parenthesised) separately in Section~\ref{sec:lens_limits}.}
\label{fig:lens}
\end{figure}

%% file: body/quant.tex
\shortonly{\label{sec:ch2_hq_quant}\label{sec:ch2_mhq_quant}\label{sec:ch2_npq_quant}\label{sec:ch2_vbq_quant}}

A vector database obtains most of its memory savings on the quantisation axis. Production systems that serve binary, eight-bit, or product-quantised embeddings each select a configuration on this axis. The measurement of Section~\ref{sec:ch4_measured} finds it the axis on which a fixed memory budget yields more accuracy per byte. In the terms of the PQO lens (Section~\ref{sec:lens}), we examine the quantisation stage $q$. We take the projection as given and ask how the real-valued output of a projected dimension is best partitioned into discrete codes. Converting real-valued projections into bits is a key step in generating LSH codes (Section~\ref{sec:ch2_lsh_sign_random}). We focus on methods that reduce the information loss this discretisation introduces, going beyond the simple sign function\longshort{ of Equation~\ref{eqn:ch2_cosine_hash_function}}{ of LSH}; we clarify below what counts as an improvement. Quantisation, the mapping of an infinite-cardinality representation to a finite discrete set, has been studied extensively in information theory~\cite{Gray06}. We restrict attention to methods developed specifically for hashing-based ANN search. Before turning to them, we note that Figure~S3 of the supplementary material organises the survey's families, classical and modern alike, by the axis on which each principally acts.

\iflong\input{supp/auto_ch2_hierarchy}\fi

\iflong
\begin{figure}[!t]
\centering
\hspace{-0.05in}\subfloat[\textbf{Vector Quantisation}]{\includegraphics[width=0.42\textwidth, height=55mm, keepaspectratio]{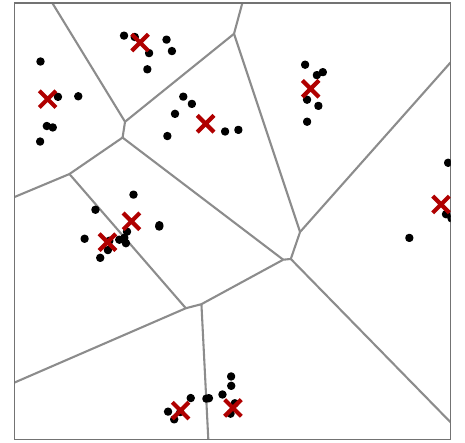}\label{fig:ch2_partitions_1}}
\hspace{0.55in}
\subfloat[\textbf{Scalar Quantisation}]{\includegraphics[width=0.42\textwidth, height=55mm, keepaspectratio]{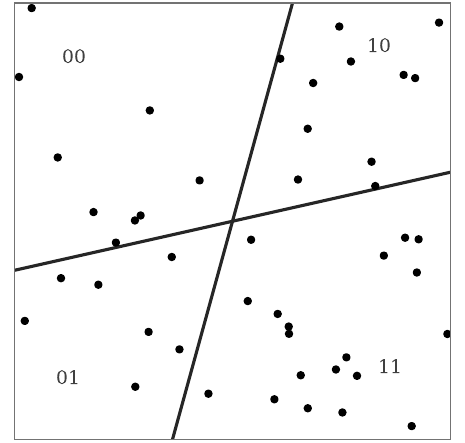}\label{fig:ch2_partitions_2}}
\caption[Vector and scalar quantisation]{Nearest neighbour search typically employs two variations on quantisation. Vector quantisation (Figure \protect\subref{fig:ch2_partitions_1}) partitions the feature space into Voronoi cells (\cite{Jegou11}); centroids are marked with a cross, data-points (illustrative) appear as black dots, and the distance between query and database points is computed by determining the distance to their closest centroids. Scalar quantisation (Figure \protect\subref{fig:ch2_partitions_2}) is frequently used to binarise a real-valued projection resulting from a dot product of a data-point with a hyperplane normal vector; the space is partitioned with multiple such hyperplanes, each usually contributing 1-bit to the final hashcode, which is effectively the index of the polytope-shaped region containing the associated data-point, the four regions here carrying the two-bit codes shown.}
\label{fig:ch2_partitions}
\end{figure}
\fi

Two primary categories of quantisation have been proposed for nearest neighbour search. Scalar quantisation thresholds the real-valued projection of a data-point onto a hyperplane normal to yield a binary code, and computes distances directly between such codes (Section~\ref{sec:ch2_sbq_quant}). Vector quantisation assigns each data-point to the nearest codebook centroid (Section~\ref{sec:pq}). \shortonly{We confine this section to scalar quantisation, the form that converts projected dimensions to bits and so meets the projection axis at the heart of the PQO lens. The two families trade reconstruction error against query-time decoding cost. Section~S2 of the supplementary material makes this trade precise, and we revisit it when we take up product quantisation in Section~\ref{sec:pq}.}

\iflong
Scalar and vector quantisation are distinguished by whether the quantiser's input and output are scalar or vector quantities. Scalar quantisation is commonly applied to the real-valued projections obtained by computing the dot product between a data-point’s feature vector and the normal vectors of a set of random hyperplanes partitioning the feature space\longonly{ (Figure~\ref{fig:ch2_partitions})}. As we discuss further in Section \ref{sec:ch2_sbq_quant}, each dot product produces a scalar value, which is then thresholded to yield a binary code ($0/1$).

In contrast, vector quantisation assigns each data-point to the nearest centroid from a codebook discovered, for example, via the k-means algorithm~\cite{Lloyd82}. Each input vector is thus represented by one of a much smaller set of centroids. K-means divides the space into Voronoi regions and so provides a more flexible, data-driven partitioning\longonly{ (see Figure~\ref{fig:ch2_partitions})}. We take vector quantisation up in depth in Section~\ref{sec:pq}; it has often been found more effective in computer vision tasks because of its lower reconstruction error~\cite{Jegou11}. Its principal drawback is the need to store a lookup table at query time to retrieve inter-cluster Euclidean distances from centroid indices. This decoding step introduces additional computational overhead: distance computation with vector quantisation has been reported to be 10–20 times slower than Hamming distance comparisons of binary hash codes on standard datasets~\cite{He13}.

Scalar quantisation avoids this overhead because distances can be computed directly from binary codes. This property has proven advantageous in applications such as mobile product search~\cite{Feng12}. Hyperplane-based scalar quantisation also partitions the space efficiently. $K$ hyperplanes in general position partition the space into $\sum_{j=0}^{D}\binom{K}{j}$ regions, which attains the full $2^{K}$ only when $K \le D$. In the high-bit regime $K \gg D$ that retrieval typically employs, the count grows only polynomially, as $\mathcal{O}(K^{D})$; yet a vector quantiser matching even this many cells would require an explicit codebook of comparable cardinality. Efficient methods, such as product quantisation, have been proposed to scale vector quantisation to this regime~\cite{Jegou11}. More recently, researchers have begun exploring approaches that unify the strengths of scalar and vector quantisation. We refer the reader to~\cite{He13} and references therein for a detailed survey of this line of work.
\fi

\iflong\input{supp/auto_ch2_sbq_quant1}\fi

In the context of hashing-based ANN search, a scalar quantiser
\[
q_{k} : \mathbb{R} \rightarrow \{0,1\}^{B}
\]
maps a real-valued projection $y_{i} \in \mathbb{R}$ to either a single-bit
(Section~\ref{sec:ch2_sbq_quant}) or multi-bit
(Section~\ref{sec:ch2_quant_summary}) binary codeword
\[
\mathbf{c}_{i} \in \mathcal{C}, \quad
\mathbf{c}_{i} \in \{0,1\}^{B}, \quad
i \in \{1,2,\ldots,T+1\},
\]
where $T$ denotes the number of quantisation thresholds,
$B$ the number of bits per projected dimension,
and $\mathcal{C}$ the binary codebook.

\iflong
Following~\cite{Kong12a}, we define a \emph{projected dimension}
$\mathbf{y}^{k}\in \mathbb{R}^{N_{\text{trd}}}$ as the set of real-valued projections
\[
\{y^{k}_{i} \in \mathbb{R}\}_{i=1}^{N_{\text{trd}}}
\]
obtained from all data-points
$[\mathbf{x}_{1}, \mathbf{x}_{2}, \ldots, \mathbf{x}_{N_{\text{trd}}}]$
for a given hyperplane $\mathbf{h}_{k}$, where each projection is computed as
\[
y^{k}_{i} = \mathbf{w}_{k}^{\intercal} \mathbf{x}_{i}.
\]
\fi

The quantisation function $q_{k}$ binarises each projected dimension
$\mathbf{y}^{k} \in \mathbb{R}^{N_{\text{trd}}}$ independently by placing one or more thresholds
along the dimension, where $N_{\text{trd}}$ denotes the number of training data-points. Projected values that fall into a given thresholded region are
assigned the corresponding codeword.

\iflong
Figure~\ref{fig:ch2_sbq_quant1} shows a simple example, with the projected
dimension drawn as a line and sample data-points as coloured shapes.
In this illustration, a single threshold partitions the dimension into two disjoint regions.
Projections below the threshold are assigned the codeword `0', and projections above
the threshold are assigned the codeword `1'. The corresponding codebook is
\[
\mathcal{C} = \{\mathbf{c}_{i} \in \{0,1\} \mid i \in \{1,2\}\}.
\]
\fi
\shortonly{In the simplest case a single threshold partitions the dimension into two regions. Projections below it take the codeword `0', projections above it take `1', and the codebook is $\mathcal{C} = \{0, 1\}$.}

\iflong
Formally, we denote the set of threshold positions along a projected dimension as
\[
\boldsymbol{\tau}_{k} = [t_{1}, t_{2}, \ldots, t_{T}], \quad t_{i} \in \mathbb{R}, \quad
t_{1} \leq t_{2} \leq \cdots \leq t_{T},
\]
with extremities defined as $t_{0} = -\infty$ and $t_{T+1} = +\infty$.
These thresholds partition the dimension into $T+1$ disjoint regions
\[
\mathcal{R}_{i} = \{ y_{j} \mid t_{i-1} < y_{j} \leq t_{i}, \, y_{j} \in \mathbf{y}^{k} \},
\quad i \in \{1,\ldots,T+1\}.
\]

Most scalar quantisation schemes adopt $T = 2^{B}-1$ thresholds for a budget of $B$ bits
per projected dimension. Each of the $T+1$ regions
$\{\mathcal{R}_{i} \subset \mathbf{y}^{k}\}_{i=1}^{T+1}$
is then associated with a unique codeword $\mathbf{c}_{i} \in \mathcal{C}$.
\fi

\iflong\input{supp/auto_ch2_quant_schemes}\fi

\iflong
The retrieval effectiveness of quantisation depends strongly on both the choice of codebook and the positioning of quantisation thresholds~\cite{Moran13a,Kong12a,Kong12b}. An effective encoding scheme must preserve the relative distances between data-points in the input space when they are mapped to binary hashcodes. Data-points that are far apart in the original feature space should be distant in Hamming space, and nearby data-points should be assigned similar codes. Ideally, the encoding of thresholded regions yields a smooth, monotonic increase in Hamming distance as the original-space distance grows. Threshold placement is equally critical. A poorly positioned threshold that bisects a region dense in true nearest neighbours scatters related data-points across different regions and inflates their Hamming distance. Suboptimal encoding or thresholding can therefore assign dissimilar codes to semantically related data-points, severely degrading retrieval accuracy. The state-of-the-art quantisation algorithms reviewed in this section address these issues by jointly proposing encoding schemes and threshold optimisation strategies that preserve relative neighbourhood structure.
\fi

\shortonly{The retrieval effectiveness of a quantiser depends on its codebook and on where its thresholds fall. A good encoding makes Hamming distance grow monotonically with original-space distance. A threshold that bisects a region dense in true neighbours scatters related points across codes and degrades accuracy.}

From the above discussion, three properties distinguish and categorise scalar quantisation methods\footnote{Throughout the remainder of this section, the term \emph{quantisation} refers specifically to scalar quantisation; the vector and product quantisers of Section~\ref{sec:pq} are signalled explicitly.}: (i) the encoding scheme used to assign symbols to thresholded regions; (ii) the strategy used to determine threshold positions, including whether a learning procedure is used for optimisation; and (iii) the number of thresholds allocated per dimension. Table~S1 of the supplementary material summarises the fixed-criterion quantisation methods along these three axes of variability. \longshort{In the following sections we provide detailed descriptions. Section~\ref{sec:ch2_sbq_quant} introduces the traditional Single-Bit Quantisation (SBQ) approach. Sections~\ref{sec:ch2_hq_quant}--\ref{sec:ch2_vbq_quant} present more recent multi-threshold methods: Hierarchical Quantisation (HQ), Double-Bit Quantisation (DBQ), and Manhattan Hashing Quantisation (MHQ), followed by the neighbourhood-preserving Neighbourhood Preserving Quantisation (NPQ) and the variable-allocation Variable Bit Quantisation (VBQ).}{In the following sections we describe Single-Bit Quantisation (SBQ; Section~\ref{sec:ch2_sbq_quant}) and the multi-threshold exemplar Double-Bit Quantisation (DBQ; Section~\ref{sec:ch2_dbq_quant}) in full. We summarise the remaining methods, Hierarchical Quantisation (HQ), Manhattan Hashing Quantisation (MHQ), Neighbourhood Preserving Quantisation (NPQ) and the variable-allocation Variable Bit Quantisation (VBQ), in Section~\ref{sec:ch2_quant_summary}.}

\subsection{Single Bit Quantisation (SBQ)}\label{sec:ch2_sbq_quant}

Single-Bit Quantisation (SBQ) is the binarisation strategy employed in the majority of existing hashing methods. A single threshold $t_{k}$ partitions a projected dimension $\mathbf{y}^{k}$ into two regions. Projected values at or below the threshold are assigned the bit `0', and values above it are assigned the bit `1' (under the convention $sgn(0)=-1$ adopted in Equation~\ref{eqn:ch2_sbq_quant} below). Formally, given a set of $K$ hyperplane normal vectors $[\mathbf{w}_{1}, \ldots, \mathbf{w}_{K}]$, SBQ generates the $k^{\text{th}}$ hashcode bit for a data-point $\mathbf{x}_{i}$ as shown in Equation~\ref{eqn:ch2_sbq_quant}.

\begin{equation}
h_{k}(\x_{i}) = \frac{1}{2}(1+sgn({\w^{\intercal}_k}\x_i - t_{k}))\\
\label{eqn:ch2_sbq_quant}
\end{equation}

In this quantisation scheme, each hyperplane contributes one bit to the hashcode of a data-point. For \emph{zero-centred} data the threshold reduces to the origin ($t_{k}=0$). No learning is involved, so both threshold computation and query encoding are $\mathcal{O}(1)$.\longonly{ More generally we threshold the projected dimension at the mean, $t_{k} = \tfrac{1}{N_{\text{trd}}} \sum_{i=1}^{N_{\text{trd}}} \mathbf{w}_{k}^{\intercal} \mathbf{x}_{i}$, or at the median, at $\mathcal{O}(N_{\text{trd}})$ cost. Figure~\ref{fig:ch2_sbq_quant1} provides a simple illustration of SBQ.}

\longshort{The multi-threshold quantisation algorithms described in Sections~\ref{sec:ch2_hq_quant}--\ref{sec:ch2_vbq_quant}}{The multi-threshold quantisation algorithms summarised in Section~\ref{sec:ch2_quant_summary}} address a fundamental limitation of SBQ that arises from its use of a single threshold for binarisation. SBQ may assign different bits to data-points that lie very close together along a projected dimension, while assigning identical bits to points that lie much farther apart~\cite{Kong12a,Kong12b,Moran13a,Moran13b}. This behaviour conflicts with the central objective of hashing, namely that nearby data-points in the original feature space should be assigned similar codes. This limitation of SBQ can therefore reduce retrieval effectiveness.

\longonly{Figure~\ref{fig:ch2_sbq_prob} illustrates the problem. Consider the data-points $a$ and $b$ (yellow stars), which lie close to one another in the projected space but fall on opposite sides of the threshold. Despite their proximity, SBQ assigns them different bits and so increases their Hamming distance. By projecting them nearby, the hash function has already indicated that $a$ and $b$ are likely to be close in the original feature space.\footnote{This assumes that the projection function preserves neighbourhood structure. Randomised LSH guarantees this in expectation, while data-dependent projection functions explicitly learn hyperplanes to encourage similar projections for related points; see Section~\ref{sec:ch2_projection}.} SBQ disrupts this structure by assigning opposing codes. Conversely, points $b$ (yellow star) and $c$ (red circle) lie far apart in the projected space, which implies they are distant in the original feature space. Both nevertheless receive the same bit, which brings their codes artificially close in Hamming space.}

\iflong\input{supp/auto_ch2_sbq_prob}\fi

This limitation of SBQ is particularly pronounced in practice because the vanilla scheme places the threshold at zero, where the density of projected values is often highest. This distributional pattern holds for many projection functions commonly used in practice\longonly{ (Figure~\ref{fig:ch2_proj_hist})}. A natural remedy is to partition each projected dimension into multiple regions and assign a multi-bit encoding, which reduces the likelihood of separating true neighbours across a single threshold. All multi-threshold quantisation schemes share the same core principle: they preserve neighbourhood structure through a combination of multi-bit codebooks and threshold optimisation. \longshort{We now turn to one of the earliest multi-threshold schemes, introduced in Section~\ref{sec:ch2_hq_quant}.}{We turn next to the first projection-agnostic remedy, Double-Bit Quantisation; we return to the broader multi-threshold family in the summary of Section~\ref{sec:ch2_quant_summary}.}

\iflong
\begin{figure}[!t]
\centering
\hspace{-0.3in}\subfloat[\textbf{Locality-Sensitive Hashing (LSH)}]{\includegraphics[width=0.42\textwidth, height=55mm, keepaspectratio]{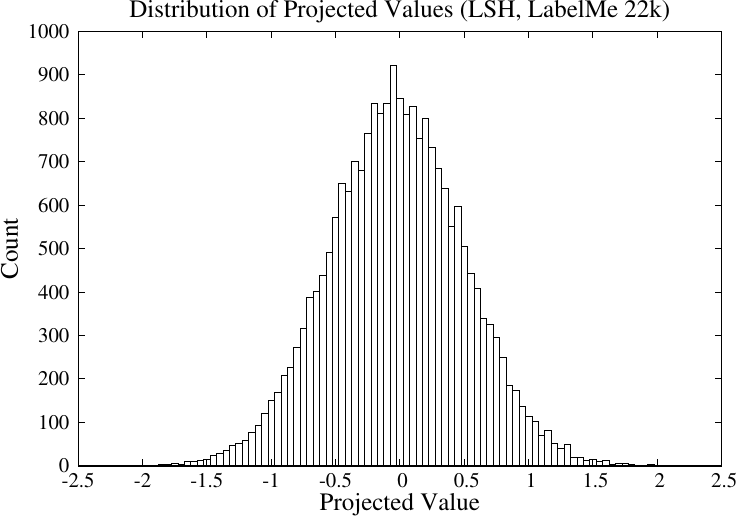}\label{fig:ch2_proj_hist_1}}
\subfloat[\textbf{Principal Component Analysis (PCA)}]{\includegraphics[width=0.42\textwidth, height=55mm, keepaspectratio]{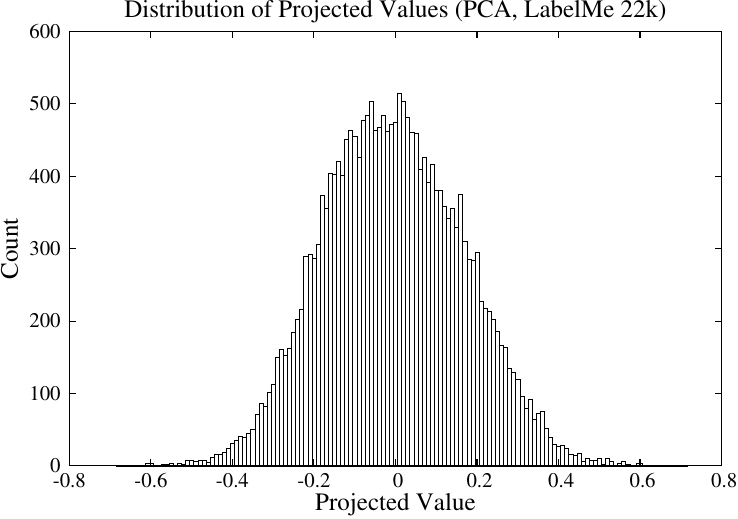}\label{fig:ch2_proj_hist_2}}
\caption[Distribution of projected values for LSH and PCA dimensions]{Distribution of projected values for two randomly chosen projected dimensions, (a) LSH and (b) PCA, evaluated on the LabelMe 22k image dataset~\cite{Torralba08a} with images represented as GIST features~\cite{Oliva01}. In both cases the highest density of projected values occurs near zero. The Double-Bit Quantisation (DBQ) algorithm~\cite{Kong12b}, described in Section~\ref{sec:ch2_dbq_quant}, explicitly avoids placing thresholds near zero, since doing so would separate many true nearest neighbours.}
\label{fig:ch2_proj_hist}
\end{figure}
\fi

\iflong\input{supp/quant_multibit}
\subsection{Variable Bit Quantisation (VBQ)}\label{sec:ch2_vbq_quant}

The quantisers discussed so far allocate the same number of bits to every projected dimension. This is the uniform-allocation assumption $A_{2}$ of Section~\ref{sec:contributions}. Variable Bit Quantisation (VBQ)~\cite{Moran13b, Moran16thesis} relaxes it by distributing a fixed bit budget non-uniformly across the dimensions. VBQ measures each dimension's neighbourhood-preservation quality directly, rather than allocating by variance in the manner the spectral projections of Section~\ref{sec:ch2_projection} suggest, and allocates bits where they yield the greatest gain.

An obvious alternative is to allocate by variance, in the manner of PCAH and Spectral Hashing. These methods assign more, higher-frequency bits to the higher-variance directions and so taper allocation smoothly down the eigenvalue spectrum. VBQ does not allocate by variance because variance is only a proxy for discriminative quality. A high-variance dimension whose spread arises from within-class scatter or a few outliers may separate true neighbours poorly, whereas a lower-variance dimension may separate them cleanly. VBQ therefore measures each dimension's neighbourhood-preservation quality directly, by the same $F_{\beta}$-measure that NPQ employs, computed as a function of the number of thresholds the dimension is allocated. It then solves a budget-constrained assignment that allocates bits where they yield the greatest gain in that score. The resulting allocation is consequently a combinatorial function of measured neighbourhood preservation, rather than a smooth, monotone function of variance. A low-variance dimension that discriminates well may therefore receive more bits than a high-variance one that does not, and a dimension that preserves no neighbourhood structure may be dropped from the code entirely.
\fi
\iflong\else
\subsection{Double-Bit Quantisation (DBQ)}\label{sec:ch2_dbq_quant}

Double-Bit Quantisation (DBQ)~\cite{Kong12b} is the first projection-agnostic multi-threshold remedy and among the most widely cited. It places two thresholds per projected dimension, yielding three regions encoded so that adjacent regions differ by a single bit. It positions the thresholds by a squared-error criterion: the projected values are clustered to minimise within-region variance, with the central region's mean pinned to zero so that no threshold falls in the dense region around the origin that SBQ bisects. An outward sweep finds the optimal pair exactly at $\mathcal{O}(N_{\text{trd}}\log N_{\text{trd}})$ cost. Under the lens, DBQ operates on the quantisation axis only: the projection is taken as given, and the information loss of binarisation is reduced purely by where, and how many, thresholds are placed. Its squared-error surrogate is, however, only a proxy for neighbourhood preservation, a gap the semi-supervised $F_{\beta}$ criteria of NPQ and VBQ address (Section~\ref{sec:ch2_quant_summary}).
\fi

\iflong
\subsection{A Link to the Discretisation of Continuous Attributes}\label{sec:ch2_discretisation}

We briefly consider how this research area relates to the well-studied problem of \emph{discretisation of continuous attributes} in machine learning~\cite{Dougherty95,Garcia13}. Many machine learning models, such as Na\"{\i}ve Bayes~\cite{Bishop06,Yang09}, transform continuous attributes into nominal attributes via discretisation prior to training. The mechanism underlying this transformation is closely related to the quantisation process in multi-threshold hashing. In both cases, a set of cut-points (thresholds) partitions the attributes (or projected dimensions), yielding a non-overlapping division of the continuous domain. Real-valued numbers falling within each region are then assigned a discrete symbol corresponding to that region. The quantisation algorithms studied in this section restrict these discrete symbols to binary codewords, whereas discretisation methods in machine learning allow a broader range of categorical values.

The discretisation literature is extensive. It proposes a wide variety of strategies for determining cut-points, ranging from unsupervised approaches (e.g., equal-interval partitioning) to supervised methods~\cite{Fayyad93} and multivariate extensions, in which attributes are discretised jointly~\cite{Mehta05,Kerber92}. Given the maturity of this field, we believe established ideas in discretisation hold significant potential to inform the design of future scalar quantisation algorithms for hashing.
\fi

\subsection{A Brief Summary}\label{sec:ch2_quant_summary}

\longshort{In this section we reviewed six scalar quantisation algorithms proposed for hashing-based ANN search.}{We have now met six scalar quantisation algorithms: SBQ and DBQ in full above, and the remaining four (HQ, MHQ, NPQ, VBQ) in Section~S2 of the supplementary material.} Each method transforms real-valued projections into binary codewords, which are concatenated to form the hashcodes of data-points. All of the algorithms follow the same basic principle: one or more thresholds partition the projected dimension, and each region is assigned a codeword (either a single bit or multiple bits).

Single-Bit Quantisation (SBQ; Section~\ref{sec:ch2_sbq_quant}) is the standard approach adopted in most hashing models. It places a single threshold, typically at zero, and is valued for its simplicity and computational efficiency. However, as discussed, SBQ can incur high quantisation error by assigning different codes to related data-points.

\longshort{The five further algorithms, Hierarchical Quantisation (HQ)~\cite{Liu11}, Double-Bit Quantisation (DBQ)~\cite{Kong12b}, Manhattan Hashing Quantisation (MHQ)~\cite{Kong12a}, Neighbourhood Preserving Quantisation (NPQ)~\cite{Moran13a}, and the variable-allocation Variable Bit Quantisation (VBQ; Section~\ref{sec:ch2_vbq_quant}), seek}{Beyond the DBQ of Section~\ref{sec:ch2_dbq_quant}, the multi-threshold family comprises Hierarchical Quantisation (HQ)~\cite{Liu11}, Manhattan Hashing Quantisation (MHQ)~\cite{Kong12a}, Neighbourhood Preserving Quantisation (NPQ)~\cite{Moran13a}, and the variable-allocation Variable Bit Quantisation (VBQ; Section~\ref{sec:ch2_vbq_quant}), all seeking} to overcome this limitation by introducing novel encoding schemes and threshold optimisation strategies. Their optimisation criteria differ. HQ employs a spectral graph partitioning objective, while DBQ and MHQ minimise objectives related to squared error and variance. NPQ and VBQ instead position thresholds, and in VBQ's case allocate the bit budget itself, by a direct $F_{\beta}$-measure of neighbourhood preservation computed from must-link and cannot-link pairs. VBQ's non-uniform allocation relaxes assumption $A_{2}$ of Section~\ref{sec:contributions} by allocating bits where they discriminate rather than where variance is high. Their encoding schemes also vary, yet all are designed with the shared goal of maximising the preservation of relative distances between data-points in the resulting binary space.

We now turn to a complementary family of methods in Section~\ref{sec:ch2_projection}, which generate the projections these algorithms subsequently quantise.

%% file: supp/auto_ch2_hierarchy.tex
%% Relocated from the CSUR main paper to the supplement (fig:ch2_hierarchy).
\begin{figure}[!t]
\centering
\resizebox{\textwidth}{!}{%
\begin{tikzpicture}[
  font=\small,
  axis/.style={draw, very thick, rounded corners, fill=gray!20, text width=3.2cm, align=center, minimum height=1cm, inner sep=3pt},
  fam/.style={draw, rounded corners, fill=white, text width=3.2cm, align=center, font=\scriptsize, inner sep=3pt, minimum height=8mm},
  coup/.style={draw, dashed, thick, rounded corners, fill=gray!8, text width=12.5cm, align=center, font=\scriptsize, inner sep=4pt},
  node distance=3.5mm and 7mm,
]
\node[axis] (P) {\textbf{Projection}~$p$\\\textit{where to embed}};
\node[axis, right=of P] (Q) {\textbf{Quantisation}~$q$\\\textit{where to threshold}};
\node[axis, right=of Q] (I) {\textbf{Organisation}~$\mathcal{I}$\\\textit{how to search}};
\node[fam, below=of P] (P1) {data-independent:\\ LSH, SKLSH};
\node[fam, below=of P1] (P2) {unsupervised:\\ PCAH, SH, ITQ, AGH};
\node[fam, below=of P2] (P3) {supervised:\\ KSH, BRE, STH, GRH};
\node[fam, below=of P3] (P4) {cross-modal:\\ CVH, CMSSH, CRH, PDH, IMH, RCMH};
\node[fam, below=of P4] (P5) {adaptive-dimension:\\ Matryoshka};
\node[fam, below=of Q] (Q1) {single-bit:\\ SBQ};
\node[fam, below=of Q1] (Q2) {multi-threshold:\\ HQ, DBQ, MHQ, NPQ, VBQ};
\node[fam, below=of Q2] (Q3) {vector / product:\\ PQ, OPQ, ScaNN, RaBitQ};
\node[fam, below=of Q3] (Q4) {binary \& scalar:\\ sign (1-bit), int8};
\node[fam, below=of I] (I1) {hashtables:\\ LSH buckets};
\node[fam, below=of I1] (I2) {inverted lists:\\ IVF, IVFADC, SPANN};
\node[fam, below=of I2] (I3) {proximity graphs:\\ HNSW, NSG, DiskANN, CAGRA};
\node[fam, below=of I3] (I4) {generative:\\ DSI, semantic IDs};
\draw[thick] (P) -- (P1);  \draw[thick] (Q) -- (Q1);  \draw[thick] (I) -- (I1);
\node[coup, anchor=north] (C) at ([yshift=-7mm]P5.south -| Q) {\textbf{Methods that deliberately couple the axes:}~ deep hashing (DPSH, HashNet) couples $p\!+\!q$;~ DiskANN and ScaNN couple $q\!+\!\mathcal{I}$;~ binary-plus-re-ranking and generative semantic identifiers couple $q\!+\!\mathcal{I}$.};
\end{tikzpicture}%
}
\caption[The families organised by the lens]{We organise the families of compact-code nearest-neighbour search by the projection--quantisation--organisation lens of Section~\ref{sec:lens}: each family sits under the axis on which it principally acts, and the methods that deliberately couple two axes, namely deep hashing and the modern indexed and generative designs, sit apart in the band below. A method that appears under an axis and again in the band (ScaNN, DiskANN) acts principally on the named axis but deliberately engages a second; LSH appears under both projection and organisation because its random hyperplanes and its multi-table buckets are distinct choices on different axes. This taxonomy sorts methods by the stage they advance rather than by mechanism or supervision level.}
\label{fig:ch2_hierarchy}
\end{figure}

%% file: supp/auto_ch2_sbq_quant1.tex
%% SBQ illustration kept in the main paper (factored into supp/ for tidiness) (fig:ch2_sbq_quant1).
\begin{figure}[!t]
\centering
\includegraphics[width=\textwidth, height=25mm, keepaspectratio]{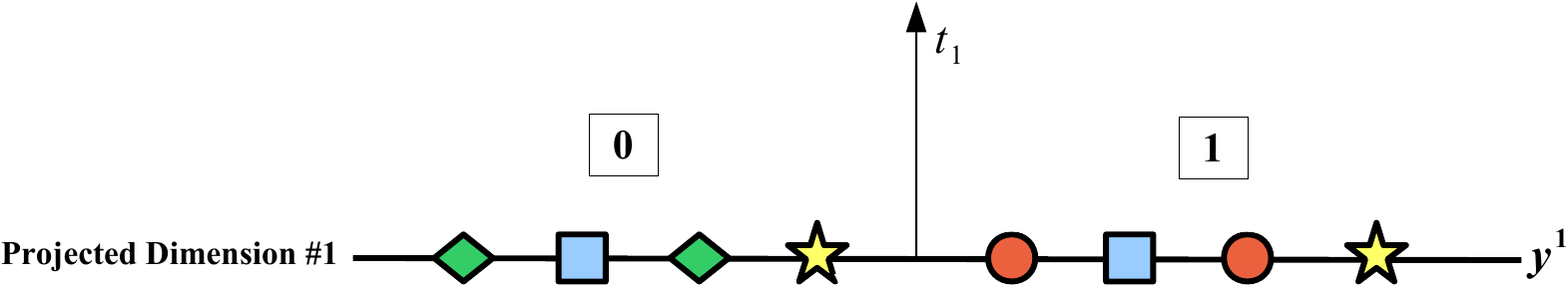}
\caption[Single-bit quantisation (SBQ)]{Illustration of single-bit quantisation (SBQ). A threshold $t_{1}\in \mathbb{R}$ is applied to a projected dimension. Values below the threshold (left, coloured shapes) receive a bit `0', and values above the threshold (right) receive a bit `1'.}
\label{fig:ch2_sbq_quant1}
\end{figure}

%% file: supp/auto_ch2_quant_schemes.tex
%% Relocated from the CSUR main paper to the supplement (tab:ch2_quant_schemes).
\begin{table}
\centering
\begin{tabular}{@{}l l c c l c@{}}
\toprule
\small{\textbf{Method}} & \small{\textbf{Encoding} } & \small{\textbf{Optimisation}} & \small{\textbf{Thresholds (T)}} & \small{\textbf{Complexity}} & \small{\textbf{Section}} \\
\midrule
\small{SBQ} & \small{0/1} & \small{Mean thresholding} & \small{1} & \small{$\mathcal{O}(1)$} & \small{\ref{sec:ch2_sbq_quant}}\\
\small{HQ} & \small{00/01/10/11} & \small{Spectral partitioning} & \small{1 and 2} & \small{$\mathcal{O}(CN_{trd}^{+})$} & \small{\ref{sec:ch2_hq_quant}}\\
\small{DBQ} & \small{00/10/11} & \small{Squared error} & \small{2} & \small{$\mathcal{O}(N_{trd}\log N_{trd})$} & \small{\ref{sec:ch2_dbq_quant}} \\
\small{MHQ} & \small{NBC} & \small{1D K-means} & \small{\hspace{0.1in}3+} & \small{$\mathcal{O}(2^{B}N_{trd})$} & \small{\ref{sec:ch2_mhq_quant}} \\
\bottomrule
\end{tabular}
\caption[Categorisation of existing quantisation models]{The fixed-criterion scalar quantisation schemes, single-bit SBQ and multi-threshold HQ, DBQ and MHQ, categorised along the three main dimensions of variability; the neighbourhood-preserving schemes NPQ and VBQ (Sections~\ref{sec:ch2_npq_quant} and~\ref{sec:ch2_vbq_quant}) position thresholds and allocate bits by a direct $F_{\beta}$ measure, and we describe them in the text. NBC stands for Natural Binary Encoding and is explained in Section \ref{sec:ch2_mhq_quant}; time complexity is for positioning thresholds along a single projected dimension. $C$ is the number of anchor points, $N_{trd}$ is the number of training data-points, $N_{trd}^{+}$ is the number of training data-points with positive projected value for the given projected dimension, and $B$ is the number of bits per projected dimension.}
\label{tab:ch2_quant_schemes}
\end{table}

%% file: supp/auto_ch2_sbq_prob.tex
%% SBQ illustration kept in the main paper (factored into supp/ for tidiness) (fig:ch2_sbq_prob).
\begin{figure}[!t]
\centering
\includegraphics[width=\textwidth, height=25mm, keepaspectratio]{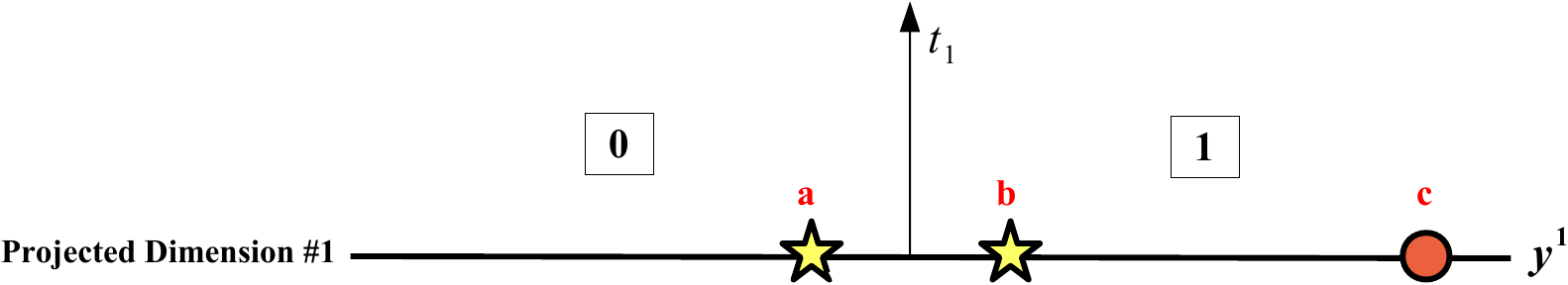}
\caption[Limitation of Single-Bit Quantisation (SBQ)]{Illustration of the limitation of Single-Bit Quantisation (SBQ). True nearest neighbours such as points $a$ and $b$ lie close together along the projected dimension, yet SBQ assigns them different bits. Conversely, points $b$ and $c$ are not nearest neighbours and are more widely separated in the projected space, yet both lie above the threshold and receive the same bit (`1').}
\label{fig:ch2_sbq_prob}
\end{figure}

%% file: supp/quant_multibit.tex
%% Extended multi-bit scalar quantisation methods (HQ, DBQ, MHQ, NPQ).
%% Inlined by the arXiv-long build of sn-article.tex (inside \iflong) and by the
%% supplement (Appendix A); dropped from the CSUR main build, where the summary in
%% Section~\ref{sec:ch2_quant_summary} and the re-homed labels stand in for it.
\subsection{Hierarchical Quantisation (HQ)}\label{sec:ch2_hq_quant}

Hierarchical Quantisation (HQ)~\cite{Liu11}, introduced for Anchor Graph Hashing, was the first scheme to assign more than one bit to a projected dimension. The scheme first thresholds the projection at zero in the manner of SBQ, producing one bit. A further pair of thresholds $t_{2}, t_{3}$ then subdivides each of the two resulting half-lines, producing a second bit. HQ places the two thresholds by minimising a graph-Laplacian objective, $\mathbf{f}^{\T}\mathbf{L}\mathbf{f} = \tfrac{1}{2}\sum_{ij}\hat{S}_{ij}(f_{i}-f_{j})^{2}$, subject to a balance constraint, and the optimal thresholds then follow in closed form. Intuitively, the objective makes it likely that true neighbours separated by the first threshold are reunited by the second. Allocating two bits to each of $\lfloor K/2 \rfloor$ low-order eigenvectors in this way was shown to outperform a single bit on all $K$. This result is consistent with the observation that the lowest-order Laplacian eigenvectors carry most of the neighbourhood structure. HQ is limited in its generality: the construction is tied to graph-Laplacian projections and does not transfer to arbitrary hyperplanes. This limitation motivates the projection-agnostic methods that follow.
\subsection{Double Bit Quantisation (DBQ)}\label{sec:ch2_dbq_quant}

Double-Bit Quantisation (DBQ)~\cite{Kong12b} removes that dependence on the projection. It places two thresholds per dimension, yielding three regions encoded so that adjacent regions differ by a single bit. It positions the thresholds by a squared-error criterion: the projected values are clustered to minimise their within-region variance. The central region's mean is pinned to zero, so no threshold falls in the dense region around the origin. A simple sweep finds the optimal thresholds exactly by moving the two of them outward from zero and examining every projected value, at a cost of $\mathcal{O}(N_{\text{trd}}\log N_{\text{trd}})$. DBQ rests on the assumption that true neighbours have small squared distance along each projected dimension. This assumption is often reasonable, but it has been shown to be far from optimal. A semi-supervised objective that does not rely solely on the quality of the projection can obtain higher effectiveness~\cite{Moran13a}.
\subsection{Manhattan Hashing Quantisation (MHQ)}\label{sec:ch2_mhq_quant}

Manhattan Hashing Quantisation (MHQ)~\cite{Kong12a} lifts the restriction to two bits. It allocates $B$ bits per dimension by placing $2^{B}-1$ thresholds at the midpoints between $2^{B}$ one-dimensional $k$-means centroids. Its distinctive contribution is on the encoding side. MHQ numbers the regions left to right and labels them by natural binary codes, which do not preserve adjacency under Hamming distance. MHQ therefore compares codes by the \emph{Manhattan} distance between their region indices. This distance increases smoothly with separation along the dimension, so it preserves the relative ordering of separations within it. Note that Manhattan distance forgoes the hardware-level efficiency of the Hamming \texttt{XOR}-and-\texttt{POPCOUNT} comparison~\cite{Wang15}. More general quantisers later resolve this tension by supporting both a binary--Hamming and an index--Manhattan reading of the same codes~\cite{Moran13b}.
\subsection{Neighbourhood Preserving Quantisation (NPQ)}\label{sec:ch2_npq_quant}

The schemes above position their thresholds by criteria that are, at best, indirect proxies for retrieval quality. HQ uses a graph-Laplacian objective tied to its projection, DBQ uses squared error, and MHQ uses $k$-means. Neighbourhood Preserving Quantisation (NPQ)~\cite{Moran13a, Moran16thesis} instead positions the thresholds to preserve neighbourhood structure directly, and it does so semi-supervisedly. Given a set of must-link and cannot-link pairs, NPQ scores a candidate partition of a projected dimension by an $F_{\beta}$-measure. The measure rewards placing must-link pairs in the same region and cannot-link pairs in different regions. NPQ positions the thresholds to maximise this score. The objective is the quantity that retrieval actually depends upon, rather than a variance- or reconstruction-based surrogate. NPQ therefore tends to keep its thresholds out of the dense, neighbour-rich region around the origin that a static SBQ threshold bisects. Its semi-supervised criterion lets it draw on whatever labelling is available without requiring a fully supervised projection.

%% file: body/projection.tex
%% CSUR-short: re-home labels of long-only subsections onto this section so cross-references resolve.
\shortonly{\label{sec:ch2_crossmodal_projection}\label{sec:ch2_cvh_projection}\label{sec:ch2_crh_projection}\label{sec:ch2_cmssh_projection}\label{sec:ch2_pdh_projection}\label{sec:ch2_imh_projection}\label{sec:ch2_rcmh_projection}\label{sec:ch2_independent_projection}\label{sec:ch2_sklsh_projection}\label{sec:ch2_pcah_projection}\label{sec:ch2_sh_projection}\label{sec:ch2_agh_projection}\label{sec:ch2_itq_cca_projection}\label{sec:ch2_bre_projection}\label{sec:ch2_grh_projection}\label{sec:ch2_sth_projection}}

The projection axis determines the content of the hashcodes. A learned embedding model occupies this stage of the pipeline. The gap between a random hyperplane and a learned, semantically rich projection separates the first generation of hashing from the learned embeddings used in current retrieval systems. In Section~\ref{sec:ch2_lsh_sign_random}, we identified two key steps, projection and quantisation, which together generate similarity-preserving hashcodes in Locality-Sensitive Hashing (LSH). Applied in sequence, these steps determine on which side of each hyperplane a data-point lies. They append a `1' to the hashcode if the point falls on one side and a `0' otherwise. In Section~\ref{sec:ch2_quantisation}, we reviewed prior work on improving the quantisation step. These algorithms aim to preserve neighbourhood structure during binarisation more faithfully than the simple zero-thresholded sign rule of Equation~\longshort{\ref{eqn:ch2_projection_learning}}{\ref{eqn:ch2_sbq_quant}}.

We now turn to research that improves the \emph{projection} step itself, seeking to learn more effective hyperplanes for similarity-preserving hashing. In the terms of the PQO lens of Section~\ref{sec:lens}, this section examines the projection stage $p$. We ask how the hyperplanes onto which the data are projected may best be placed.

\iflong
\begin{equation}
h_{k}(\mathbf{x}_{i}) = \frac{1}{2}(1 + sgn({\w^{\intercal}_k}\x_i - t_{k})) \\
\label{eqn:ch2_projection_learning}
\end{equation}

\noindent
Here, $\mathbf{w}_{k} \in \mathbb{R}^{D}$ denotes the hyperplane normal vector and $t_{k} \in \mathbb{R}$ the associated quantisation threshold. Equation~\ref{eqn:ch2_projection_learning} defines the linear hash function adopted in most hashing research. As discussed in Section~\ref{sec:ch2_quantisation}, most quantisation models operate independently of the projection stage; Anchor Graph Hashing~\cite{Liu11} is the exception. They assume that some existing projection method has already generated the projections to be binarised.
\fi
\shortonly{The linear hash function of Equation~\ref{eqn:ch2_sbq_quant} composes a hyperplane normal $\w_{k}$ with a threshold $t_{k}$, and most hashing research adopts this form. As discussed in Section~\ref{sec:ch2_quantisation}, quantisation models operate independently of the projection stage, with Anchor Graph Hashing~\cite{Liu11} the exception. They assume the projections to be binarised are already given.}

We focus on algorithms that aim to generate projections which preserve relative distances between data-points along the resulting projected dimensions. For a linear hash function, this corresponds to positioning a set of $K$ hyperplanes in the input feature space so that similar data-points are likely to fall within the same polytope-shaped region. These regions form the hashtable buckets used for indexing and retrieval.

\iflong
Formally, we generate a projected dimension $\mathbf{y}^{k} \in \mathbb{R}^{N_{\text{trd}}}$ from a hyperplane $\mathbf{h}_{k} \in \mathbb{R}^{D}$ as follows. We project the data-points $\{\mathbf{x}_{i} \in \mathbb{R}^{D}\}_{i=1}^{N_{\text{trd}}}$ onto the normal vector $\mathbf{w}_{k} \in \mathbb{R}^{D}$ via the dot product operation $\mathbf{w}_{k}^{\intercal}\mathbf{x}_{i}$. This places each point on the one-dimensional axis defined by the normal, and so induces the projected dimension that the quantiser then thresholds\longonly{ (cf.\ Figure~\ref{fig:ch1_pipeline})}.
\fi

\iflong\input{supp/projection_schemes_table}\fi

In Section~\ref{sec:ch2_lsh}, we introduced Locality-Sensitive Hashing (LSH), a seminal early method for solving the ANN search decision problems\longshort{ defined in Definitions~\ref{def:ch2_rapprox_nn}--\ref{def:ch2_rnn}}{ of approximate and near-neighbour retrieval}. \iflong As discussed in Section~\ref{sec:ch2_lsh_sign_random}, LSH for inner-product similarity samples hyperplanes uniformly from the unit sphere. The asymptotic guarantee is that, as the number of hyperplanes increases, the Hamming distance between hashcodes approximates the cosine similarity between data-points.\footnote{\cite{Charikar02} applied the random-hyperplane rounding of~\cite{Goemans95}. They showed that the expected Hamming distance between two bit vectors formed using hash functions sampled from $\mathcal{H}_{\text{cosine}}$ approximates the angle between the corresponding vectors in the input feature space.} \fi Nevertheless, as noted in Section~\ref{sec:ch2_lsh}, randomly sampled hyperplanes often lack discrimination. They risk partitioning regions of the input space dense with related data-points. In practice, many hyperplanes (bits) and multiple hash tables are required for acceptable retrieval effectiveness. Longer hashcodes and more tables, however, increase the memory footprint of an LSH system.

Recent work has therefore focused on learning hyperplanes adapted to the data distribution, which yields more compact and discriminative hashcodes~\cite{Liu11,Liu12,Weiss08,Gong11,Raginsky09,Kulis09b,Zhang10}. These methods form the focus of this part of the review. Non-hyperplane partitions also exist, most prominently the hypersphere boundaries of Spherical Hashing~\cite{Heo12}; we confine the discussion to the hyperplane and kernel families that dominate the literature. We divide projection learning methods for hashing-based ANN into three categories based on the extent to which the data distribution informs the construction of hashing hyperplanes: \emph{data-independent}, \emph{data-dependent but unsupervised}, and \emph{data-dependent and supervised}. A fourth, cross-modal, setting is treated in Section~S3 of the supplementary material; Table~S2 there summarises these approaches, with their hash functions and training complexities. Under the PQO lens, the categories form a graded relaxation of where the hyperplanes are placed. Data-independent methods fix the hyperplanes without consulting the data. Unsupervised methods adapt them to the data distribution, supervised methods to a user-defined notion of similarity, and cross-modal methods to correspondences across feature spaces.

\iflong
In the following sections, we review representative work under each category. The field spans a wide variety of techniques for generating hash functions, including random projections, kernel functions, spectral methods, and boosting. The literature is too extensive for exhaustive coverage. We therefore focus on influential models that have become widely adopted, particularly those accompanied by publicly available codebases. This grounds our discussion in results collected on comparable datasets and experimental protocols, and ensures that comparisons are made against competitive baselines. For broader surveys, we refer the reader to~\cite{Wang14a} and~\cite{Grauman13}.

Finally, note that all of the hashing models reviewed here assume search over a single hash table ($L=1$), as is standard in the literature. Methods that explicitly \emph{learn} multiple tables in a data-dependent manner form a promising subfield but lie beyond the scope of this review. See~\cite{Hao11} and~\cite{Liu13} for representative research in this direction.
\fi
\shortonly{Under each category we review influential, publicly available baselines, so that comparisons rest on comparable datasets and protocols. Broader surveys, and the data-dependent learning of multiple hash tables, which we do not treat here, are noted in Section~S3 of the supplementary material. All models reviewed assume search over a single hash table ($L=1$).}

\subsection{The Four Properties of an Effective Hashcode}\label{sec:ch2_effective_hashcodes}

Before discussing individual models for projection, we consider several properties that contribute to the effectiveness of a hashcode for nearest neighbour search. These properties may be read as desiderata on the projection stage $p$ and the quantiser $q$ of Section~\ref{sec:lens}. The projection methods reviewed below pursue them by differing means. The seminal work on Spectral Hashing (SH) by~\cite{Weiss08} codified four such properties ($E_{1}$--$E_{4}$):

\begin{itemize}
\item{$E_{1}$: Hashcodes for similar data-points should have \emph{low Hamming distance}}.
\item{$E_{2}$: Hashcodes should be \emph{efficiently computable} for novel query points}.
\item{$E_{3}$: Each bit should take the values 0 and 1 with \emph{equal probability}}.
\item{$E_{4}$: Different bits should be \emph{pairwise independent}}.
\end{itemize}

\shortonly{In a nutshell, $E_{1}$ is the neighbourhood-preservation criterion already encountered with LSH and binary quantisation. $E_{2}$ requires cheap \emph{out-of-sample} encoding for novel queries. $E_{3}$ asks each hyperplane to split the data evenly, which by the maximum-entropy principle maximises the information each bit carries. $E_{4}$ seeks compactness by removing redundant, correlated bits. The majority of data-dependent methods developed since Spectral Hashing aim to satisfy as many of these properties as possible. We return to how effectively each method does so in the sections that follow.}

\iflong
We have already highlighted the importance of the first property ($E_{1}$) in the context of LSH (Section~\ref{sec:ch2_lsh}) and binary quantisation (Section~\ref{sec:ch2_quantisation}). The remaining criteria ($E_{2}$--$E_{4}$) are equally significant. Property $E_{2}$ is crucial for practical deployment. Given a novel data-point, its hashcode must be computed rapidly so that query time remains low. The learning-to-hash literature calls this problem \emph{out-of-sample extension}. LSH provides a straightforward solution: multiply the query vector by a matrix whose columns are hyperplane normals, then apply sign thresholding (Section~\ref{sec:ch2_lsh}).

Properties $E_{3}$ and $E_{4}$ address the \emph{efficiency} and \emph{compactness} of hashcodes, respectively. Property $E_{3}$ requires each hyperplane to partition the dataset evenly. Written in the $\pm 1$ convention of the sign-valued hash functions, this is the balance condition $\sum_{i=1}^{N_{\text{trd}}} \operatorname{sgn}(\mathbf{w}_{k}^{\intercal}\mathbf{x}_{i}-t_{k})=0$; equivalently, the corresponding $0/1$ bit is set on half of the data. By the principle of maximum entropy, this maximises the information carried by each bit~\cite{Baluja08} and ensures balanced bucket occupancy in the hashtable.\footnote{\cite{Wang12} showed that the NP-hard constraint $E_{3}$ can be relaxed by demonstrating its equivalence to variance maximisation for the $k^{\text{th}}$ bit. Strict enforcement of $E_{3}$ may, however, be sub-optimal if it partitions clusters of related data-points across different buckets. Multiple independent hashtables can often mitigate this issue.} Balanced partitions also prevent the degenerate case where a query requires examining an excessively large number of candidates within a single bucket.

Property $E_{4}$ seeks compactness by eliminating redundant bits that encode overlapping information about the input space. An ideal hashing scheme minimises the number of bits required to represent the data, conserving both storage and computation. The majority of data-dependent projection methods developed since~\cite{Weiss08} aim to learn hyperplanes that produce hashcodes satisfying as many of these four properties as possible. In the sections that follow, we examine how effectively the data-independent, unsupervised, and supervised approaches jointly preserve these properties during optimisation.
\fi

\iflong
\subsection{Data-Independent Projection Methods}\label{sec:ch2_independent_projection}

Aside from Locality-Sensitive Hashing (LSH), reviewed in Section~\ref{sec:ch2_lsh}, we consider one additional data-independent method. Locality-Sensitive Hashing from Shift-Invariant Kernels (SKLSH)~\cite{Raginsky09} extends LSH to preserve kernel-based similarity (Section~\ref{sec:ch2_sklsh_projection}).

\subsubsection{Locality-Sensitive Hashing from Shift Invariant Kernels (SKLSH)}\label{sec:ch2_sklsh_projection}

Locality-Sensitive Hashing from Shift-Invariant Kernels (SKLSH) extends LSH to preserve similarity as defined by a kernel function $\kappa : \mathbb{R}^{D} \times \mathbb{R}^{D} \rightarrow \mathbb{R}$. Common examples include the Gaussian kernel,
\[
\kappa(\mathbf{x}_{i}, \mathbf{x}_{j}) = \exp\!\left(-\gamma \|\mathbf{x}_{i}-\mathbf{x}_{j}\|^{2}/2\right),
\]
and the Laplacian kernel,
\[
\kappa(\mathbf{x}_{i}, \mathbf{x}_{j}) = \exp\!\left(-\gamma \|\mathbf{x}_{i}-\mathbf{x}_{j}\|_{1}/2\right),
\]
where $\gamma \in \mathbb{R}$ is the kernel bandwidth parameter.

Conceptually, SKLSH resembles LSH but employs a different hash function family $\mathcal{H}$ in order to preserve kernel-based similarity. The objective is to construct an embedding $g : \mathbb{R}^{D} \rightarrow \{0,1\}^{K}$ with the following property. If two data-points are highly similar under the kernel (i.e., $\kappa(\mathbf{x}_{i}, \mathbf{x}_{j}) \approx 1$), their hashcodes overlap strongly ($d_{\text{Hamming}}(g(\mathbf{x}_{i}), g(\mathbf{x}_{j})) \approx 0$); the converse holds when $\kappa(\mathbf{x}_{i}, \mathbf{x}_{j}) \approx 0$.

To achieve this,~\cite{Raginsky09} define a low-dimensional projection $\Psi^{K}: \mathbb{R}^{D} \rightarrow \mathbb{R}^{K}$ using the random Fourier features of~\cite{Rahimi07}. These features guarantee that the inner product between transformed data-points approximates the value of a \emph{shift-invariant} kernel,\footnote{A shift-invariant kernel is defined as $\kappa(\mathbf{x}_{i}, \mathbf{x}_{j}) = \hat{\kappa}(\mathbf{x}_{i} - \mathbf{x}_{j})$.}
\[
\Psi_{k}(\mathbf{x}_{i}) \cdot \Psi_{k}(\mathbf{x}_{j}) \approx \hat{\kappa}(\mathbf{x}_{i}-\mathbf{x}_{j}).
\]
The explicit random Fourier features mapping is given in Equation~\ref{eqn:ch2_random_fourier_features}.

\begin{equation}
\Psi_{k}(\x_{i}) = \sqrt{2}cos(\w^{\intercal}_{k}\x_{i}+t_{k})
\label{eqn:ch2_random_fourier_features}
\end{equation}

\noindent
For the Gaussian kernel, $\mathbf{w}_{k} \sim \mathcal{N}(0,\gamma \mathbf{I}_{D \times D})$ and $t_{k} \sim \text{Unif}[0,2\pi]$. The key contribution of~\cite{Raginsky09} is to employ this embedding as the basis of a novel hash function (Equation~\ref{eqn:ch2_sklsh_hash_function}).

\begin{equation}
h_{k}(\x_{i}) = \frac{1}{2}[1+sgn(cos(\w^{\intercal}_{k}\x_{i}+t_{k})+t_{k^{'}})]
\label{eqn:ch2_sklsh_hash_function}
\end{equation}

\noindent
Here, $\operatorname{sgn}$ denotes the sign function, adjusted so that $\operatorname{sgn}(0) = -1$, and $t_{k'} \sim \text{Unif}[-1,1]$.~\cite{Raginsky09} prove that hashing data-points with $K$ randomly sampled hash functions yields a binary embedding whose Hamming distance approximates the desired shift-invariant kernel similarity. The hyperplanes are sampled randomly, so the training time complexity of the algorithm is only $\mathcal{O}(DK)$. SKLSH therefore satisfies property $E_{2}$ of an effective hashcode, namely efficient computation of hashcodes.

\fi
\subsection{Data-Dependent (Unsupervised) Projection Methods}\label{sec:ch2_dependent_projection}

In this section, we provide a critical appraisal of data-dependent hashing methods that learn hyperplanes in an unsupervised manner, i.e., without supervisory information such as pairwise similarity constraints or class labels. All of the unsupervised approaches reviewed here learn hashing hyperplanes by formulating a \emph{trace minimisation/maximisation} problem. They solve this problem either in closed form as an eigenvalue problem or via singular value decomposition (SVD). These methods build directly on well-established techniques for linear and non-linear dimensionality reduction, most notably Principal Components Analysis (PCA) and Laplacian Eigenmaps (LapEig).\longonly{ Matrix factorisation is central to the learning-to-hash literature, including many methods not covered in detail here. We therefore begin with a brief introduction to this solution strategy before reviewing individual algorithms in Sections~\ref{sec:ch2_pcah_projection}--\ref{sec:ch2_agh_projection}.\footnote{For further details on trace optimisation and eigenproblems for dimensionality reduction, see the survey of~\cite{Kokiopoulou11}.}}

\longonly{There are two main strategies for dimensionality reduction of a dataset $\mathbf{X} \in \mathbb{R}^{N \times D}$ into a lower-dimensional representation $\mathbf{Y} \in \mathbb{R}^{N \times K}$, where $K \ll D$. The first approach seeks an explicit linear transformation characterised by a projection matrix $\mathbf{W} \in \mathbb{R}^{D \times K}$, such that $\mathbf{Y} = \mathbf{XW}$. PCA is a canonical example of this \emph{projective} strategy. The second approach computes a non-linear embedding $\mathbf{Y} \in \mathbb{R}^{N \times K}$ directly, without an explicit mapping function. Methods of this type, such as LapEig, typically impose neighbourhood constraints so that nearby data-points in the original space remain close in the reduced space.}
\iflong
Despite their differences, both categories can be expressed within a unified framework using a standard trace maximisation objective (Equation~\ref{eqn:ch2_unsupervised_template}).

\begin{equation}
\begin{aligned}
\text{argmax}_{\mathbf{V}\in \RE^{N \times K}} \quad & tr(\mathbf{V}^{\intercal}\mathbf{A}\mathbf{V})\\
\text{subject to } & \mathbf{V}^{\intercal}\mathbf{1}=0\\
& \mathbf{V}^{\intercal}\mathbf{B}\mathbf{V}=\mathbf{I}^{K\times K}
\end{aligned}
\label{eqn:ch2_unsupervised_template}
\end{equation}
\vspace{0.1in}

\noindent
Here, $\mathbf{A}$ is a symmetric matrix, $\mathbf{B}$ is a positive-definite matrix, $\mathbf{V}$ is an orthonormal\footnote{An orthonormal matrix $\mathbf{V}$ is a square matrix whose columns and rows are orthogonal unit vectors, satisfying $\mathbf{V}^{\intercal}\mathbf{V}=\mathbf{V}\mathbf{V}^{\intercal}=\mathbf{I}$.} matrix, and $\operatorname{tr}(\mathbf{A})=\sum_{i} A_{ii}$. The exact specification of these matrices depends on the projection function under consideration. We define $\mathbf{A}$, $\mathbf{B}$, and $\mathbf{V}$, including their dimensionalities, in Sections~\ref{sec:ch2_pcah_projection}--\ref{sec:ch2_agh_projection}.

\longonly{Intuitively, for Principal Component Analysis (PCA) we have $\mathbf{A} = \mathbf{X}^{\intercal}\mathbf{X}$, $\mathbf{V} = \mathbf{W} \in \mathbb{R}^{D \times K}$, and $\mathbf{B} = \mathbf{I} \in \mathbb{R}^{D \times D}$. Maximising the trace (Equation~\ref{eqn:ch2_unsupervised_template}) in this case is equivalent to finding the principal directions that capture maximum variance in the input space.}

The trace maximisation in Equation~\ref{eqn:ch2_unsupervised_template} is solved as a generalised eigenvalue problem, $\mathbf{A}\mathbf{v}_{i} = \lambda_{i}\mathbf{B}\mathbf{v}_{i}$, whose $K$ leading eigenvectors $\mathbf{v}_{i}$ (eigenvalues $\lambda_{i}$) become the hashing hyperplanes~\cite{Saad11a,Kokiopoulou11}.\longonly{ Much of the unsupervised learning-to-hash literature reduces to this template. We shape the optimisation problem into the trace maximisation form of Equation~\ref{eqn:ch2_unsupervised_template}, and then solve for the $K$ eigenvectors of the resulting eigenvalue problem. Standard solvers such as {\tt eigs} or {\tt svd} in Matlab address this optimisation efficiently.}

The design of an unsupervised data-dependent hashing function typically follows four steps:

\begin{enumerate}
\item Reformulate the problem as a matrix trace minimisation/maximisation (Equation~\ref{eqn:ch2_unsupervised_template}).
\item Solve the optimisation as an eigenvalue problem or via SVD; the $K$ eigenvectors (or right-singular vectors) serve as the normal vectors of the hashing hyperplanes.
\item Address the imbalanced variance resulting from matrix factorisation.
\item Construct an out-of-sample extension in the case of a non-projective mapping.
\end{enumerate}
\vspace{0.1in}

We observe these four design principles across the unsupervised methods reviewed in this section: PCA Hashing (PCAH, Section~\ref{sec:ch2_pcah_projection}), Spectral Hashing (SH, Section~\ref{sec:ch2_sh_projection}), Iterative Quantisation (ITQ, Section~\ref{sec:ch2_itq_projection}), and Anchor Graph Hashing (AGH, Section~\ref{sec:ch2_agh_projection}). In three of these methods (PCAH, SH, ITQ), PCA first extracts the directions of maximum variance, which then serve as hashing hyperplanes. The main contributions of these approaches lie in Step~3. Each proposes a different strategy to mitigate the impact of imbalanced variance across hyperplanes, which otherwise reduces the quality of hashcodes derived from lower principal components. The final method, AGH, adopts a different approach. It computes an eigenfunction extension of graph Laplacian eigenvectors and bases hash function learning on the Laplacian Eigenmap algorithm.

\fi
\shortonly{The leading eigenvectors of the resulting eigenproblem become the hyperplanes. The chief design concern of these methods is the imbalanced variance that matrix factorisation induces across those directions.}
Ultimately, all these methods aim to learn $K$ hash functions $\{h_{k} : \mathbb{R}^{D} \rightarrow \{0,1\}\}_{k=1}^{K}$ that can be concatenated to generate binary hashcodes for unseen data-points.\longonly{ Figure~\ref{fig:ch2_data_depen_diag} summarises one interpretation of the relationships among these models.}

\iflong
\subsubsection{Principal Components Analysis Hashing (PCAH)}\label{sec:ch2_pcah_projection}

Principal Components Analysis hashing (PCAH) provides the most basic data-dependent projection~\cite{Wang10a, Hotelling33}. The hashing hyperplanes are the leading principal directions of the mean-centred data, obtained as the top-$K$ singular vectors of $\mathbf{X}$, and each projection is thresholded at zero. PCAH is the natural data-dependent baseline against which the unsupervised methods below define themselves. Its main weakness is the source of much that follows, the \emph{imbalanced variance problem}. The orthogonality of the principal directions forces successive bits onto directions of diminishing variance, so the later, low-variance directions yield unreliable bits. The methods that follow each address this defect. Spectral and Anchor Graph Hashing reallocate bits to the more informative directions; Iterative Quantisation (Section~\ref{sec:ch2_itq_projection}) rotates the projected space so that variance is spread evenly across the bits.
\subsubsection{Spectral Hashing (SH)}\label{sec:ch2_sh_projection}

Spectral Hashing (SH)~\cite{Weiss08} was among the earliest data-dependent schemes and is widely regarded as the work that drew the computer-vision community to learning to hash. It also supplied the four properties of an effective hashcode used above. SH casts binarisation as graph partitioning. It seeks codes that minimise the similarity-weighted Hamming distance $\sum_{ij} S_{ij}\lVert \y_{i}-\y_{j}\rVert^{2} = \operatorname{tr}(\mathbf{Y}^{\T}(\mathbf{D}-\mathbf{S})\mathbf{Y})$ subject to bit-balance ($\mathbf{Y}^{\T}\mathbf{1}=\mathbf{0}$) and bit-independence ($\mathbf{Y}^{\T}\mathbf{Y}=N_{\text{trd}}\mathbf{I}$) constraints, where $\mathbf{S}$ is a Gaussian affinity matrix and $\mathbf{D}$ its degree matrix. The integer problem is NP-hard, so SH relaxes it spectrally~\cite{Shi00}, taking the codes to be the smallest-eigenvalue eigenvectors of the graph Laplacian $\mathbf{D}-\mathbf{S}$. For out-of-sample extension, SH assumes a uniform data distribution and evaluates analytic one-dimensional Laplacian eigenfunctions along the PCA directions. The eigenvalue of an eigenfunction decreases with the variance of its direction. SH therefore allocates more bits, of progressively higher frequency, to the most informative hyperplanes. This relaxes LSH's uniform-allocation assumption $A_{2}$ on the projection side, and this relaxation explains the improvement of SH over PCAH~\cite{Liu11, Moran13b}. Its principal weakness, the unrealistic uniform-distribution assumption, motivates Anchor Graph Hashing.
\fi
\shortonly{Three unsupervised methods besides ITQ address the imbalanced-variance problem of PCA-derived codes by different means. PCA Hashing (PCAH)~\cite{Wang10a, Hotelling33} is the plainest: it takes the leading principal directions as hyperplanes and thresholds at zero, so its later, low-variance bits tend to be unreliable. Spectral Hashing (SH)~\cite{Weiss08}, the work that drew the vision community to learning to hash, casts binarisation as graph partitioning and, through analytic Laplacian eigenfunctions, allocates more high-frequency bits to the high-variance directions. Anchor Graph Hashing (AGH)~\cite{Liu11} keeps that data-adaptive allocation while replacing SH's full affinity, and its uniform-distribution assumption, with a sparse anchor graph that also supplies an efficient out-of-sample map. Iterative Quantisation, the worked exemplar below, takes the complementary route: it rotates the projected space so that the variance is spread evenly across the bits.}

\subsubsection{Iterative Quantisation (ITQ)}\label{sec:ch2_itq_projection}

\shortonly{Iterative Quantisation (ITQ)~\cite{Gong11} takes the complementary route to Spectral Hashing. Rather than reallocating bits across directions, it learns an orthogonal rotation $\mathbf{R}$ of the PCA-projected space that minimises the quantisation error $\lVert\mathbf{B}-\mathbf{Y}\mathbf{R}\rVert_{F}^{2}$ between the projections $\mathbf{Y}=\mathbf{X}\mathbf{W}$ and their signs $\mathbf{B}=\operatorname{sgn}(\mathbf{Y}\mathbf{R})$. A $k$-means-like alternation finds the two, and the rotation step has a closed-form orthogonal-Procrustes solution~\cite{Schonemann66}. By spreading the variance evenly across the bits, ITQ approximately satisfies bit balance ($E_{3}$) and independence ($E_{4}$), both inherited from the PCA projection~\cite{Wang10a}. We class ITQ as a projection rather than a quantisation method because it does not itself threshold; it rotates the hyperplanes so that the subsequent single-bit quantiser incurs less error. The full derivation and algorithm are given in Section~S3 of the supplementary material.}\iflong
Spectral Hashing (SH) implicitly allocates more bits to hyperplanes that capture a greater proportion of the variance in the input space, which partially mitigates the imbalanced variance problem. Iterative Quantisation (ITQ) takes a different approach. It explicitly balances the variance across the PCA hyperplanes by learning a rotation of the feature space. Specifically, ITQ introduces an iterative scheme, reminiscent of the $k$-means algorithm, to learn a rotation matrix $\mathbf{R} \in \mathbb{R}^{K \times K}$ such that the projections onto the principal directions $\mathbf{W} \in \mathbb{R}^{D \times K}$ minimise the quantisation error. Equation~\ref{eqn:ch2_itq_projection_obj} expresses this objective in matricial form.

\begin{equation}
\begin{aligned}
\text{argmin}_{\mathbf{B}\in \RE^{N_{trd}\times K},\mathbf{R}\in \RE^{K \times K}} \quad & \|\mathbf{B}-\mathbf{Y}\mathbf{R}\|^{2}_{F}\\
\text{where } & \mathbf{B} \in \left\{-1,1\right\}^{N_{trd} \times K} \\
\text{subject to } & \mathbf{R}^{\intercal}\mathbf{R}=\mathbf{I}_{K\times K}
\end{aligned}
\label{eqn:ch2_itq_projection_obj}
\end{equation}
\vspace{0.1in}

\noindent
\longonly{Equation~\ref{eqn:ch2_itq_projection_obj} is an instance of the orthogonal Procrustes problem~\cite{Schonemann66}, in which one matrix is transformed into another under an orthogonal constraint so as to minimise the sum of squared residuals.}

Here $\mathbf{R}$ minimises the squared Euclidean distance between the projections $\mathbf{Y} = \mathbf{X}\mathbf{W}$ and their binarised counterparts $\mathbf{B} = \operatorname{sgn}(\mathbf{Y}\mathbf{R})$, where the PCA hyperplanes are stacked as the columns of $\mathbf{W}$. Estimating $\mathbf{R}$ requires $\mathbf{B}$, and estimating $\mathbf{B}$ requires $\mathbf{R}$, so the two are found by alternation.

Beginning from a random rotation, and in the manner of $k$-means, the scheme alternates between solving for $\mathbf{B}$ with $\mathbf{R}$ fixed and for $\mathbf{R}$ with $\mathbf{B}$ fixed~\cite{Gong11}.\longonly{ Algorithm~\ref{alg:ch2_itq_algorithm} summarises the procedure.}

\iflong
\begin{algorithm}[!t]
\DontPrintSemicolon
\KwIn{Data-points $\mathbf{X}\in \RE^{N_{\text{trd}}\times D}$, PCA hyperplanes $\mathbf{W}\in \RE^{D\times K}$, number of iterations $T$, randomly initialised rotation matrix $\mathbf{R}\in \RE^{K \times K}$}
\KwOut{Optimised rotation matrix $\mathbf{R}\in \RE^{K \times K}$}
$\mathbf{Y} \gets \mathbf{XW}$\; \tcp*{Project data onto PCA hyperplanes}
\For{$m \gets 1$ \textbf{to} $T$}{
  $\mathbf{B} \gets \operatorname{sgn}(\mathbf{Y}\mathbf{R})$\; \label{alg:ch2_itq_algorithm_sbq} \tcp*{Rotate data using $\mathbf{R}$ and quantise}
  $\mathbf{U}\boldsymbol{\Sigma}\mathbf{V}^{\intercal} \gets \mathrm{SVD}(\mathbf{B}^{\intercal}\mathbf{Y})$\; \label{alg:ch2_itq_algorithm_svd} \tcp*{Perform SVD on $\mathbf{B}^{\intercal}\mathbf{Y}$}
  $\mathbf{R} \gets \mathbf{V}\mathbf{U}^{\intercal}$\; \tcp*{Update $\mathbf{R}$ to minimise Eq.~\ref{eqn:ch2_itq_projection_obj} for fixed $\mathbf{B}$}
}
\KwRet{$\mathbf{R}$}
\caption[Iterative Quantisation (ITQ)]{{\sc Iterative Quantisation (ITQ)}~\cite{Gong11}}
\label{alg:ch2_itq_algorithm}
\end{algorithm}
\fi

\longonly{With $\mathbf{B}$ fixed, the optimal rotation follows in closed form from the SVD of $\mathbf{B}^{\intercal}\mathbf{Y}$~\cite{Hanson81, Arun87} (Line~\ref{alg:ch2_itq_algorithm_svd}); with $\mathbf{R}$ fixed, the optimal $\mathbf{B}$ is given by single-bit quantisation (Line~\ref{alg:ch2_itq_algorithm_sbq})~\cite{Gong11}.}

Beyond the neighbourhood-preservation and out-of-sample properties $E_{1}$ and $E_{2}$, ITQ approximately satisfies bit balance ($E_{3}$) and bit independence ($E_{4}$). Both are inherited from the variance-maximising, orthogonal PCA projection~\cite{Wang10a}.

\longonly{The SVD of the $K \times K$ matrix in Line~\ref{alg:ch2_itq_algorithm_svd} costs only $\mathcal{O}(K^{3})$. The per-iteration cost of ITQ is instead dominated by forming the product $\mathbf{B}^{\intercal}\mathbf{Y}$ that the SVD decomposes, at $\mathcal{O}(N_{\text{trd}}K^{2})$ over the $N_{\text{trd}}$ training points. Since $N_{\text{trd}} \gg K$, this term dominates the iteration, while the one-off PCA initialisation that precedes it remains the largest single cost. }Once learned, the rotation matrix $\mathbf{R}$ can be applied to construct ITQ hashcodes for unseen queries $\mathbf{q} \in \mathbb{R}^{D}$ as defined in Equation~\ref{eqn:ch2_itq_hash_function}.

\begin{equation}
g_{k}(\q)=\frac{1}{2}(1+sgn(\mathbf{R}\mathbf{W}^{\intercal}\q))
\label{eqn:ch2_itq_hash_function}
\end{equation}

\iflong\input{supp/auto_ch2_itq_projection}\fi

\noindent
Here we assume the data has been mean-centred, so that the quantisation threshold is fixed at $t_{k}=0$.

We close with two observations. First, ITQ exemplifies a two-step relaxation: learn a continuous embedding, then binarise it so as to minimise the error that binarisation introduces. This pattern recurs throughout the field as a tractable response to the NP-hardness of learning binary codes directly. BRE (Section~\ref{sec:ch2_bre_projection}) and later work~\cite{Liu14} take up the harder relaxation-free problem. Second, we class ITQ as a projection rather than a quantisation method because it does not itself threshold. It rotates the PCA hyperplanes so that the subsequent single-bit quantiser incurs less error, and its contribution is therefore to the placement of the projections.
\fi

\iflong
\subsubsection{Anchor Graph Hashing (AGH)}\label{sec:ch2_agh_projection}

Anchor Graph Hashing (AGH)~\cite{Liu11} optimises the same spectral objective as SH but removes its two computational bottlenecks, the $\mathcal{O}(N_{\text{trd}}^{2})$ affinity matrix and the expensive out-of-sample extension. It replaces the full affinity with a sparse, low-rank \emph{anchor graph}. Each data-point is described by its similarities to a small set of $C \ll N_{\text{trd}}$ $k$-means centroids, and the affinity is approximated as $\hat{\mathbf{S}} = \mathbf{Z}\Sigma^{-1}\mathbf{Z}^{\T}$. This reduces the eigenproblem to a $C\times C$ matrix and supplies a natural out-of-sample map: a query is first embedded non-linearly by its anchor similarities and then projected linearly, a kernelised feature map followed by thresholding. AGH thereby keeps the data-adaptive bit allocation of SH while dispensing with its uniform-distribution assumption. Its most accurate variant applies the hierarchical quantiser of Section~\ref{sec:ch2_hq_quant} to assign two bits per projected dimension.
\fi
\iflong
\subsubsection{A Brief Summary}

In our first exploration of data-dependent hashing algorithms, we surveyed several of the most widely studied \emph{unsupervised} approaches, which position the hashing hyperplanes according to the data distribution. Specifically, we reviewed Principal Components Analysis Hashing (PCAH) (Section~\ref{sec:ch2_pcah_projection}), Spectral Hashing (SH) (Section~\ref{sec:ch2_sh_projection}), Iterative Quantisation (ITQ) (Section~\ref{sec:ch2_itq_projection}), and Anchor Graph Hashing (AGH) (Section~\ref{sec:ch2_agh_projection}). All four models share a common foundation in dimensionality reduction. They derive their hyperplanes from either Principal Components Analysis (PCA) or Laplacian Eigenmaps (LapEig).

Three of the four unsupervised methods (PCAH, SH, ITQ) rely directly on PCA. They set the hashing hyperplanes to the right singular vectors obtained from a singular value decomposition (SVD) of the data matrix. Two of these methods (SH and ITQ) explicitly address \emph{variance imbalance}, where hyperplanes capturing less variance prove unreliable for hashing. PCAH consequently suffers degraded retrieval effectiveness as hashcode length increases, since additional bits are drawn from lower-quality hyperplanes. SH mitigates this effect by allocating more bits to high-variance directions, while ITQ applies a learned rotation to redistribute the variance evenly across hyperplanes. Both strategies improve retrieval performance relative to vanilla PCAH, which assigns one bit per principal component without adjustment.

By contrast, AGH adopts a LapEig-inspired dimensionality reduction. A neighbourhood graph is first constructed from the input data, and the resulting graph Laplacian eigenvectors define the hashing hyperplanes. LapEig is inherently non-projective, so the learned eigenvectors yield hashcodes only for the training points used in constructing the graph. A key contribution of AGH is its efficient out-of-sample extension, achieved via a Nyström approximation~\cite{Williams01}. This enables encoding of unseen query points at practical computational cost.
\fi

\iflong
With the exception of AGH, which makes a deliberate attempt to reduce training complexity, the primary drawback of these methods is the heavy computational overhead of matrix factorisation. Solving the SVD or eigenvalue problem typically requires $\mathcal{O}(\min(N^{2}D, ND^{2}))$ operations, which renders these approaches intractable for large-scale, high-dimensional datasets. As we observe in the following sections, this reliance on matrix factorisation recurs across both supervised and unsupervised data-dependent hashing models.
\fi
\shortonly{With the exception of AGH, the chief drawback of these methods is the heavy cost of matrix factorisation. This reliance on the SVD or eigenproblem recurs across the supervised data-dependent models below.}

\subsection{Data-Dependent (Supervised) Projection Methods}\label{sec:ch2_supervised_projection}

We have so far reviewed representative data-independent and unsupervised data-dependent hashing models. Data-independent approaches preserve fixed similarity measures, such as cosine or kernel similarity. These measures are not data-adaptive and therefore often fail to align with user-defined similarity notions across diverse tasks. Unsupervised data-dependent models instead assume that discriminative hashcodes can be derived from projections that capture maximal variance in the input space. This assumption is problematic in practice: variance does not always distinguish between semantically different data-points.\longonly{ The problem is acute in real-world image datasets collected ``in the wild,'' where quality, resolution, and topic vary widely. The well-known \emph{semantic gap} in computer vision names this mismatch between low-level image features (e.g., GIST, SIFT) and the high-level semantic concepts they are intended to represent~\cite{Smeulders00}. Bridging this gap remains a central challenge in object recognition and image annotation~\cite{Moran15a}, as well as in the broader task of image retrieval.}

\iflong
\begin{figure}[!t]
\centering
\hspace{-0.3in}\subfloat[\textbf{Unsupervised}]{\includegraphics[width=0.42\textwidth, height=50mm, keepaspectratio]{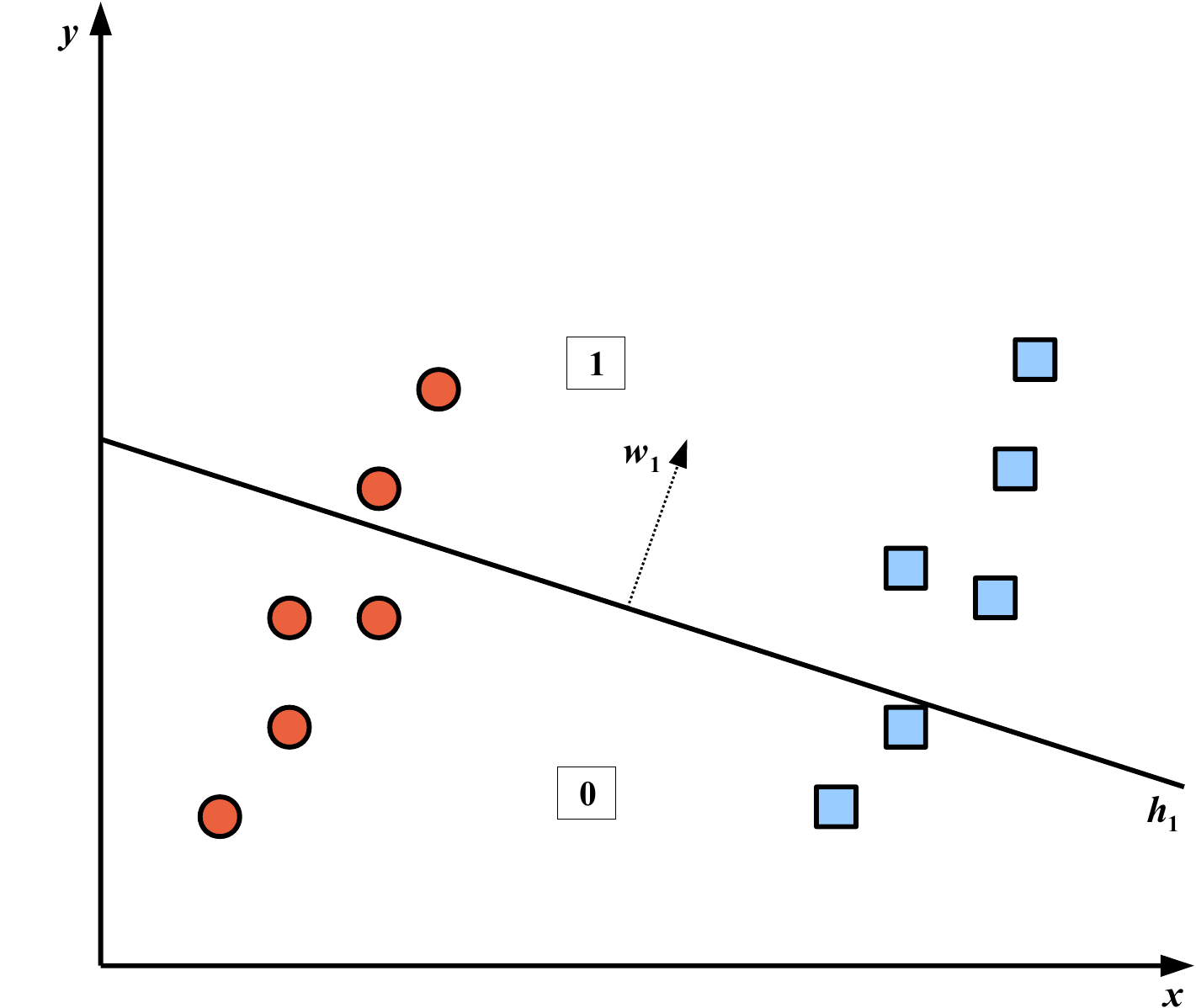}\label{fig:ch2_pca_vs_lda_projection_1}}
\subfloat[\textbf{Supervised}]{\includegraphics[width=0.42\textwidth, height=50mm, keepaspectratio]{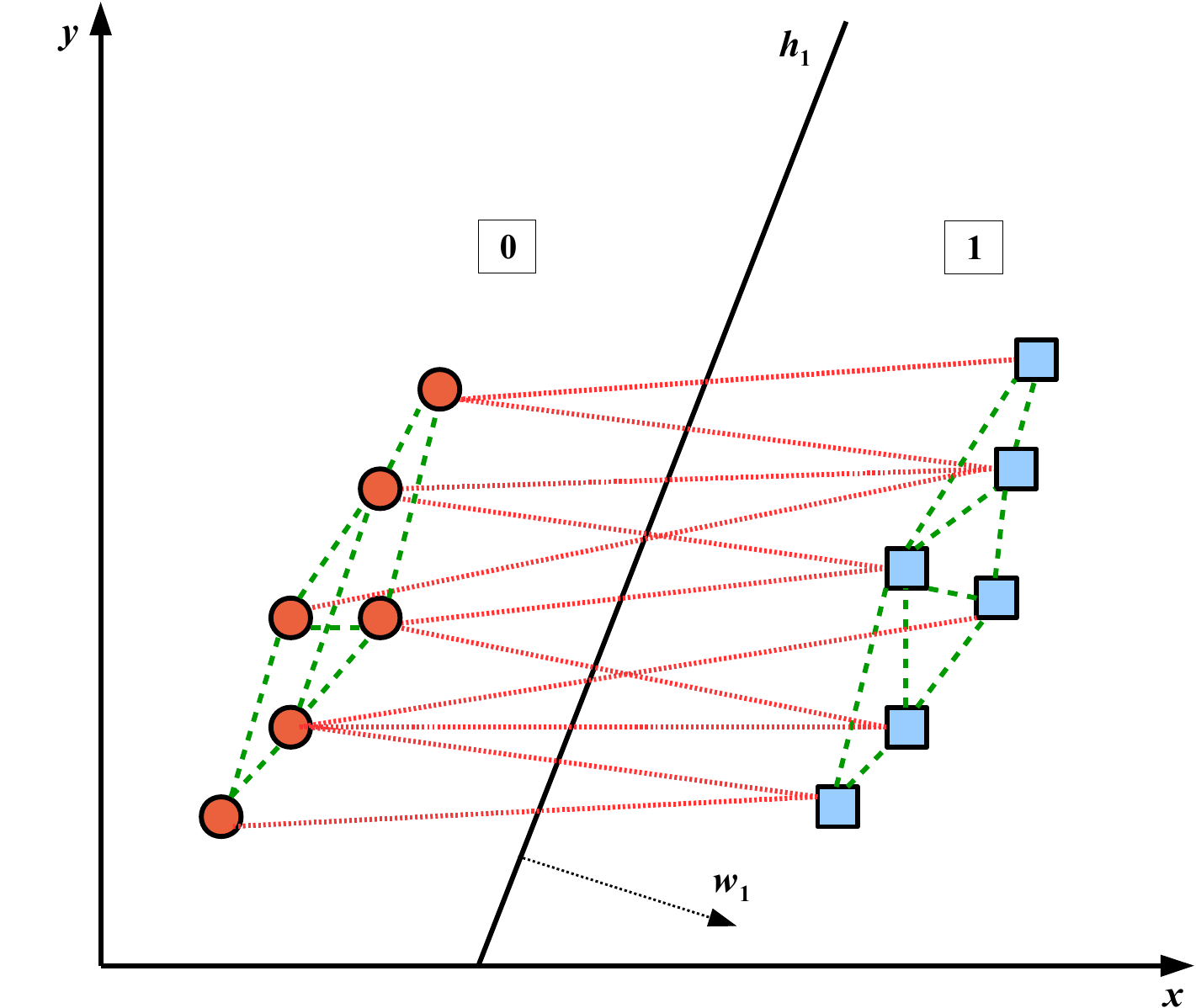}\label{fig:ch2_pca_vs_lda_projection_2}}
\caption[Supervised versus unsupervised projection function learning]{Supervised versus unsupervised projection function learning, showing how pairwise user-provided constraints can yield a more effective bucketing of the space than variance-based partitioning; shapes and colours denote 1-nearest neighbours. In Figure~\subref{fig:ch2_pca_vs_lda_projection_1}, the hyperplane $\h_{1}\in \RE^{D}$ is learnt via PCA, its normal vector $\w_{1}\in \RE^{D}$ pointing in the direction of maximum variance, and projecting onto $\w_{1}$ separates related data-points (same shapes) into different buckets. By contrast, Figure~\subref{fig:ch2_pca_vs_lda_projection_2} incorporates must-link (dotted lines) and cannot-link (solid lines) constraints, grouping related data-points in the same bucket.}
\label{fig:ch2_pca_vs_lda_projection}
\end{figure}
\fi

To address the semantic gap and better capture the complex relationships between data-points (e.g., whether two images depict the same semantic category such as a cat or a person), it is generally advantageous to incorporate even limited supervision. The supervision takes the form of class labels, or pairwise constraints indicating whether two data-points should or should not share the same hashcodes.\longonly{ In the visual search domain,~\cite{Grauman13} identify a range of supervisory sources, from explicit label annotation, to known correspondences between images, to user feedback on search results.}

\longonly{Semi-Supervised Hashing~\cite{Wang12}, among the most cited methods of the era, regularises a supervised objective with an entropy term over unlabelled data and sits between the two regimes. }We define a supervised hashing model as one that exploits, directly within hash function learning, the same type of information used to establish ground-truth similarity during evaluation (e.g., class labels, metric distances).\longonly{ Figure~\ref{fig:ch2_pca_vs_lda_projection} illustrates how supervision can guide hyperplane placement, producing more effective bucketings than purely variance-driven methods.}

As in the review of unsupervised methods, we focus on well-known supervised baselines whose implementations are publicly available and which the literature has widely adopted for reproducible and fair comparison. Specifically, we review ITQ with Canonical Correlation Analysis (ITQ+CCA)~\cite{Gong11}, Supervised Hashing with Kernels (KSH)~\cite{Liu12}, Binary Reconstructive Embedding (BRE)~\cite{Kulis09b}, and Self-Taught Hashing (STH)~\cite{Zhang10}. \longonly{These methods differ primarily in how they exploit supervisory information to construct an error signal that guides the positioning of the hashing hyperplanes. BRE and KSH, for example, minimise discrepancies between label information and hashcode distances. STH enforces similarity-preserving projections for points with the same label using a LapEig-inspired objective, while ITQ+CCA maximises correlations between data projections and label embeddings.}\longonly{ Figure~\ref{fig:ch2_supervised_diag} summarises the relationships between these models, together with the graph-regularised method (GRH) detailed below.}

\iflong
\begin{figure}[!t]
\centering
\subfloat[Unsupervised data-dependent methods, as transformations of PCA]{%
\begin{tikzpicture}[
  font=\footnotesize, >=stealth,
  hub/.style={draw, very thick, rounded corners, fill=gray!18, text width=2.7cm, align=center, inner sep=4pt},
  meth/.style={draw, rounded corners, fill=white, text width=3.2cm, align=center, font=\scriptsize, inner sep=3pt, minimum height=11mm},
  lbl/.style={font=\scriptsize, text=black!55, fill=white, inner sep=1.2pt}]
  \node[hub] (P) {\textbf{PCA}\\ leading variance directions};
  \node[meth, above left=8mm and 14mm of P]  (PCAH) {\textbf{PCAH}\\ threshold at zero; imbalanced variance};
  \node[meth, above right=8mm and 14mm of P] (ITQu) {\textbf{ITQ}\\ rotate to spread variance evenly across the bits};
  \node[meth, below left=8mm and 14mm of P] (SH) {\textbf{SH}\\ allocate bits by Laplacian eigenfunctions};
  \node[meth, below right=8mm and 14mm of P] (AGH) {\textbf{AGH}\\ anchor-graph Laplacian eigenmap};
  \draw[->] (P) -- (PCAH) node[lbl,pos=0.5,sloped]{threshold};
  \draw[->] (P) -- (ITQu) node[lbl,pos=0.5,sloped]{rotate};
  \draw[->] (P) -- (SH) node[lbl,pos=0.5,sloped]{eigenfunctions};
  \draw[->] (SH) -- (AGH) node[lbl,pos=0.5]{anchor graph};
\end{tikzpicture}%
\label{fig:ch2_data_depen_diag}}\\[3mm]
\subfloat[Supervised methods, as choices of supervision signal and optimisation]{%
\begin{tikzpicture}[
  font=\footnotesize, >=stealth,
  hub/.style={draw, very thick, rounded corners, fill=gray!18, text width=3.0cm, align=center, inner sep=4pt},
  meth/.style={draw, rounded corners, fill=white, text width=3.2cm, align=center, font=\scriptsize, inner sep=3pt, minimum height=12mm},
  disc/.style={meth, fill=gray!12},
  lbl/.style={font=\scriptsize, text=black!55, fill=white, inner sep=1.2pt}]
  \node[hub] (H) {\textbf{Supervised hyperplane learning}\\[1pt] must-link / cannot-link constraints};
  \node[meth, above left=8mm and 15mm of H]  (ITQs) {\textbf{ITQ$+$CCA}\\ correlate projections with the label embedding};
  \node[meth, above right=8mm and 15mm of H] (KSH) {\textbf{KSH}\\ match labels to projection inner products};
  \node[meth, below left=10mm and 15mm of H] (STH) {\textbf{STH}\\ label-graph Laplacian codes, SVM out-of-sample};
  \node[disc, below right=10mm and 15mm of H] (BRE) {\textbf{BRE}\\ match labels to Hamming distances};
  \node[disc, below=13mm of H] (GRH) {\textbf{GRH}\\ regularise the codes on a signed label graph, $\mathbf{b}\!\leftarrow\! sgn(\mathbf{S}\mathbf{b})$};
  \draw[->] (H) -- (ITQs) node[lbl,pos=0.5,sloped]{correlation};
  \draw[->] (H) -- (KSH) node[lbl,pos=0.5,sloped]{inner product};
  \draw[->] (H) -- (STH) node[lbl,pos=0.5,sloped]{graph Laplacian};
  \draw[->] (H) -- (BRE) node[lbl,pos=0.5,sloped]{Hamming distance};
  \draw[->] (H) -- (GRH) node[lbl,pos=0.5]{graph on the codes};
\end{tikzpicture}%
\label{fig:ch2_supervised_diag}}
\caption[Relationship maps for the data-dependent projection methods]{Relationship maps for the data-dependent projection methods, each drawn as transformations of a shared starting point. \protect\subref{fig:ch2_data_depen_diag}~The unsupervised methods all begin from the PCA variance directions and differ in how they counter the imbalanced-variance problem: PCAH simply thresholds them, ITQ rotates the projected space so that variance is spread evenly across the bits, SH reallocates bits by Laplacian eigenfunctions, and AGH replaces SH's uniform-distribution assumption with an anchor graph. \protect\subref{fig:ch2_supervised_diag}~The supervised methods all start from must-link/cannot-link supervision and differ in how that supervision becomes an error signal (edge labels) and how the codes are optimised: the shaded methods are relaxation-free, with BRE retaining the sign function under coordinate descent and GRH (with its cross-modal extension RCMH) refining the discrete codes directly on a signed graph without a continuous relaxation, whereas the unshaded methods relax the problem to a continuous or spectral one.}
\label{fig:ch2_relationship_maps}
\end{figure}
\fi

\iflong
\subsubsection{ITQ $+$ Canonical Correlation Analysis (CCA)}\label{sec:ch2_itq_cca_projection}

Iterative Quantisation extends naturally to the supervised setting, since its rotation step (Section~\ref{sec:ch2_itq_projection}) is indifferent to how the projection directions are obtained. ITQ$+$CCA~\cite{Gong11} therefore replaces the unsupervised PCA projection with Canonical Correlation Analysis~\cite{Hardoon03}. CCA learns directions in the feature space that are maximally correlated with a second view formed from the labels. The usual ITQ rotation is then applied to minimise the quantisation error in the resulting supervised subspace. The effect is to bias the projection towards label-correlated directions while retaining the variance-balancing rotation of ITQ. This improves semantic neighbourhood preservation over unsupervised ITQ at essentially the same cost.
\subsubsection{Binary Reconstructive Embedding (BRE)}\label{sec:ch2_bre_projection}

Binary Reconstructive Embedding (BRE)~\cite{Kulis09b} is notable for avoiding the spectral relaxation altogether. Rather than solving for a real-valued projection and binarising afterwards, it retains the sign function and optimises the binary codes directly. The objective minimises the squared gap between the supervised similarity of a pair and the normalised Hamming distance of their kernel-based codes. The non-differentiable objective is handled by coordinate descent over the hyperplane entries, with a closed-form update per coordinate. BRE preserves neighbourhoods ($E_{1}$, $E_{2}$), but the bits are learned sequentially, so it does not enforce balance or independence. Its direct treatment of the discrete problem was nonetheless influential on later relaxation-free methods~\cite{Liu14, Shen15}.
\subsubsection{Supervised Hashing with Kernels (KSH)}\label{sec:ch2_ksh_projection}

Supervised Hashing with Kernels (KSH)~\cite{Liu12} adopts a kernelised hash function similar in spirit to AGH (Section~\ref{sec:ch2_agh_projection}) and BRE (Section~\ref{sec:ch2_bre_projection}), but couples it with a distinct, spectrally relaxed optimisation strategy. In practice, KSH often attains state-of-the-art retrieval effectiveness among supervised hashing baselines. It frequently serves as a de facto comparison point on standard image retrieval benchmarks. The kernelised hash function is
\begin{equation}
h_{k}(\mathbf{q}) \;=\; \operatorname{sgn}\!\Big(\sum_{j=1}^{C} W_{jk}\,\kappa(\mathbf{x}_{j},\mathbf{q}) \;+\; t_{k}\Big),
\label{eqn:ch2_ksh_hash_function}
\end{equation}
where $\kappa:\mathbb{R}^{D}\times\mathbb{R}^{D}\to\mathbb{R}$ is a kernel function, $t_{k}\in\mathbb{R}$ is a scalar threshold, and $\mathbf{W}\in\mathbb{R}^{C\times K}$ collects $K$ hyperplane normals in the anchor space. As in AGH and BRE, a small set of $C$ representative data-points ($C\ll N$) is sampled to serve as kernel anchors. A set of $N_{\text{trd}}$ data-points ($C<N_{\text{trd}}\ll N$) forms a supervised adjacency matrix $\mathbf{S}\in\{-1,1\}^{N_{\text{trd}}\times N_{\text{trd}}}$ for training.

KSH minimises the discrepancy between supervised similarity labels and inner products of (relaxed) hashcodes. Compared to BRE’s objective, KSH removes the sign function during optimisation and measures agreement via inner products rather than Euclidean distances:
\begin{equation}
\begin{aligned}
\underset{\mathbf{W}\in\mathbb{R}^{C\times K}}{\text{argmin}}\quad
& \sum_{ij}\Big\{\,S_{ij}-\tfrac{1}{K}\,g(\mathbf{x}_{i})^{\!\top}g(\mathbf{x}_{j})\,\Big\}^{2} \\
\text{subject to}\quad
& g(\mathbf{x}_{i})=\big[h_{1}(\mathbf{x}_{i}),\ldots,h_{K}(\mathbf{x}_{i})\big]^{\!\top}, \\
& h_{k}(\mathbf{x}_{i})=\operatorname{sgn}\!\Big(\sum_{j=1}^{C} W_{jk}\,\kappa(\mathbf{x}_{j},\mathbf{x}_{i})\Big).
\end{aligned}
\label{eqn:ch2_ksh_supervised_objective1}
\end{equation}

For an efficient solution, KSH relaxes the sign during optimisation and learns the $K$ bits sequentially. Each bit is initialised by an eigenproblem and refined by gradient descent on a smooth relaxation of the sign.
\iflong
As in BRE, inter-bit dependency is introduced through a residue $\mathbf{R}^{\,k-1} = K\mathbf{S} - \sum_{\ell=1}^{k-1}\mathbf{y}^{\ell}(\mathbf{y}^{\ell})^{\!\top}$, which up-weights the pairs misfit by the bits already learned; $\mathbf{y}^{\ell}$ denotes the $\ell$-th relaxed projection. With this residue, the per-bit problem reduces to maximising the Rayleigh quotient $(\mathbf{K}\mathbf{w}_{k})^{\!\top}\mathbf{R}^{\,k-1}(\mathbf{K}\mathbf{w}_{k})$ subject to $(\mathbf{K}\mathbf{w}_{k})^{\!\top}(\mathbf{K}\mathbf{w}_{k}) = N_{\text{trd}}$. The initial solution is the leading generalised eigenvector of
\[
\mathbf{K}^{\!\top}\mathbf{R}^{\,k-1}\mathbf{K}\,\mathbf{w}_{k} \;=\; \lambda\,\mathbf{K}^{\!\top}\mathbf{K}\,\mathbf{w}_{k},
\]
with $\mathbf{K}\in\mathbb{R}^{N_{\text{trd}}\times C}$ the kernel matrix between training points and anchors. The residue is a signed, indefinite matrix, so the relevant solution is the eigenvector of largest (most positive) generalised eigenvalue, with the norm constraint fixing the scale.
\fi

Despite being a non-linear model, KSH remains computationally tractable by (i) restricting $N_{\text{trd}}$ and $C$ to modest sizes\footnote{Typical choices are $N_{\text{trd}}\approx 1000$ and $C\approx 300$.}, and (ii) operating in the continuous (real-valued) domain during optimisation.
\iflong
The overall time complexity is
\[
\mathcal{O}\!\big(N_{\text{trd}}C K \;+\; N_{\text{trd}}^{2} C K \;+\; N_{\text{trd}} C^{2} K \;+\; C^{3} K\big),
\]
accounting for kernel evaluations, residue updates, matrix multiplications, and per-bit eigenproblems.
\fi
KSH does not explicitly enforce $E_{3}$ (bit balance) or $E_{4}$ (bit independence). It satisfies $E_{1}$ (neighbourhood preservation) and $E_{2}$ (efficient out-of-sample evaluation), producing highly discriminative hashcodes with fast query-time hash computation.

\subsubsection{Self-Taught Hashing (STH)}\label{sec:ch2_sth_projection}

Self-Taught Hashing (STH)~\cite{Zhang10} separates the two concerns explicitly. It first learns codes for the training points by a Laplacian Eigenmap on a label-derived neighbourhood graph, a normalised-cut variant of the SH objective. It then obtains the out-of-sample projection by training $K$ linear support vector machines to predict, at maximum margin, each bit of the binarised embedding. The novelty is the use of supervised binary classifiers for the out-of-sample extension, in place of the uniform-distribution eigenfunctions of SH or the anchor map of AGH. This keeps query encoding cheap, $\mathcal{O}(DK)$, while the supervision enters through the graph construction.
\fi
\shortonly{ITQ$+$CCA~\cite{Gong11} replaces PCA with Canonical Correlation Analysis, biasing the projection towards label-correlated directions before the usual ITQ rotation. Binary Reconstructive Embedding (BRE)~\cite{Kulis09b} avoids the spectral relaxation and optimises the binary codes directly by coordinate descent to match supervised similarities. Minimal loss hashing~\cite{Norouzi11mlh} pursues this discrete line with a quantisation-aware hinge loss, and Supervised Discrete Hashing~\cite{Shen15} carries it to scale. Graph Regularised Hashing (GRH)~\cite{Moran15c} likewise dispenses with the relaxation. It alternates a regularisation step that diffuses the binary codes over a signed supervised graph, $\mathbf{b}_{i}\leftarrow\operatorname{sgn}(\sum_{j}S_{ij}\mathbf{b}_{j})$, with a linear learning step for out-of-sample encoding. GRH is an early lightweight instance of the projection--quantiser coupling that the deep methods of Section~\ref{sec:deep} generalise (Section~S3 of the supplementary material). Self-Taught Hashing (STH)~\cite{Zhang10} separates the two concerns. It learns codes by a Laplacian eigenmap on a label graph and then fits per-bit support vector machines for the out-of-sample map.}

\iflong\else
\subsubsection{Supervised Hashing with Kernels (KSH)}\label{sec:ch2_ksh_projection}

Supervised Hashing with Kernels (KSH)~\cite{Liu12} is the de facto supervised comparison point on the standard benchmarks, and serves here as the worked supervised exemplar. It learns a kernelised hash function over a small set of $C$ anchor points, $h_{k}(\q) = \operatorname{sgn}\big(\sum_{j=1}^{C} W_{jk}\,\kappa(\x_{j},\q) + t_{k}\big)$. Training minimises the discrepancy between the supervised labels and the inner products of the relaxed codes, $\sum_{ij}\big\{S_{ij} - \tfrac{1}{K}\,g(\x_{i})^{\T} g(\x_{j})\big\}^{2}$. The sign is relaxed during optimisation, and the $K$ bits are learned sequentially, each against a residue that up-weights the pairs its predecessors misfit. This decorrelates the bits ($E_{4}$) without sacrificing the kernel's non-linearity. In the lens's terms, KSH represents the most effective use of the supervised projection axis before deep learning. Supervision shapes where the kernelised hypersurfaces fall, while the quantiser remains the simple sign. KSH thus follows the relax-then-binarise pattern, whose residual error the coupled deep objectives of Section~\ref{sec:deep} later remove; its optimisation and complexity are given in full in Section~S3 of the supplementary material.
\fi
\iflong
\subsubsection{Graph Regularised Hashing (GRH)}\label{sec:ch2_grh_projection}

Graph Regularised Hashing (GRH)~\cite{Moran15c, Moran16thesis} dispenses with the continuous relaxation altogether and operates directly on the binary codes. It refines them by an alternating procedure reminiscent of the expectation--maximisation algorithm. The inputs are an initial set of hashcodes, supplied by any base method, and a supervised affinity graph $\mathbf{S}$ in which must-link pairs carry positive weight and cannot-link pairs negative. GRH alternates two steps. In the \emph{regularisation} step, each code is replaced by the sign of the weighted sum of its graph neighbours' codes, $\mathbf{b}_{i} \leftarrow \operatorname{sgn}\!\big(\sum_{j} S_{ij}\mathbf{b}_{j}\big)$. The supervisory information thereby propagates through the code space in the manner of label propagation. In the \emph{learning} step, a set of linear hyperplanes is fitted to predict the regularised codes from the features; this supplies the out-of-sample extension and ties the codes back to the data. The two steps are iterated. The supervision enters only through $\mathbf{S}$, so the method accommodates whatever form the available labelling takes. Operating directly on the sign function, GRH avoids the continuous-to-binary error that the relaxation-based methods of this section incur.

We examine the regularisation step more closely. For each bit, it is a discrete diffusion of the form $\mathbf{b} \leftarrow \operatorname{sgn}(\mathbf{S}\mathbf{b})$, the synchronous linear-threshold dynamics familiar from the analysis of associative memories. Two features of GRH's design prevent that diffusion from degenerating. The first is the signed graph. If $\mathbf{S}$ were purely non-negative, the Perron--Frobenius theorem implies the dynamics would converge towards the constant code in which every point agrees and the bit carries no information. The cannot-link edges, which push dissimilar points apart, remove that degenerate consensus as an attractor and so preserve the balance of the bits. The second is the bounded iteration interleaved with the linear fit. This anchors the codes to the feature geometry and arrests the diffusion once it has injected the neighbourhood structure but before it can over-smooth. GRH is an early instance of coupling. The regularisation step is a data-dependent adjustment of where the codes fall, and the learning step is a projection fitted to honour it. The deep methods of Section~\ref{sec:deep} would later generalise this coupling by optimising the projection and the quantiser jointly against a single differentiable objective.
\fi
\iflong
\subsubsection{A Brief Summary}

We have reviewed four of the most widely studied approaches for incorporating supervision into the learning of hashing hyperplanes for unimodal ANN search. The models considered are ITQ$+$CCA (Section \ref{sec:ch2_itq_cca_projection}), Binary Reconstructive Embedding (BRE) (Section \ref{sec:ch2_bre_projection}), Supervised Hashing with Kernels (KSH) (Section \ref{sec:ch2_ksh_projection}), and Self-Taught Hashing (STH) (Section \ref{sec:ch2_sth_projection}). We also considered the graph-regularised method (GRH) of Section~\ref{sec:ch2_grh_projection}, which refines the binary codes directly rather than through a continuous relaxation. Each of these models learns a set of $K$ hyperplanes guided by must-link and cannot-link constraints. Must-link pairs should map to the same hashtable bucket, while cannot-link pairs should be separated. For example, two images of cats (must-link) should be collocated, whereas an image of a dog (cannot-link) should be placed in a different bucket.

The key differences between the models lie in how the available supervision is related to the projections or hashcodes to generate an error signal for hyperplane adjustment. BRE and KSH, for instance, both minimise the discrepancy between labels and hashcode similarities, but their formulations differ. BRE minimises the gap with respect to Hamming distances, while KSH focuses on inner products of the real-valued projections. Their optimisation strategies also diverge significantly. BRE retains the discrete sign function and tackles the NP-hard optimisation directly with a coordinate descent algorithm. KSH instead relaxes the objective into a continuous domain and applies gradient-based optimisation. ITQ$+$CCA and STH take yet different routes, using CCA embeddings or SVM-based out-of-sample extensions respectively. Together, these approaches illustrate the breadth of strategies available for supervised hyperplane learning.
\fi

\iflong
\subsection{Cross-Modality Projection Methods}\label{sec:ch2_crossmodal_projection}

All models discussed so far, including LSH and SKLSH (Sections \ref{sec:ch2_lsh}, \ref{sec:ch2_sklsh_projection}) and the data-dependent models of Sections \ref{sec:ch2_dependent_projection}--\ref{sec:ch2_supervised_projection}, address only \emph{unimodal} retrieval, where queries and database entries share the same feature representation. In this case, the learned hyperplanes partition a single modality. This is restrictive, since much real-world data is inherently multi-modal\footnote{We use the terms `modality' and `feature space' interchangeably.}. For example, an image on Flickr\footnote{\url{http://www.flickr.com}} may carry user-assigned tags and geolocation metadata in addition to its raw pixel values. Ideally, a retrieval system would allow queries across modalities. For instance, we may query with an image to retrieve relevant text tags (Figure \ref{fig:ch2_cross_modal_hash_table}), or query with GPS coordinates to retrieve corresponding images.

\begin{figure}[!ht]
\centering
\includegraphics[width=0.8\textwidth, height=100mm, keepaspectratio]{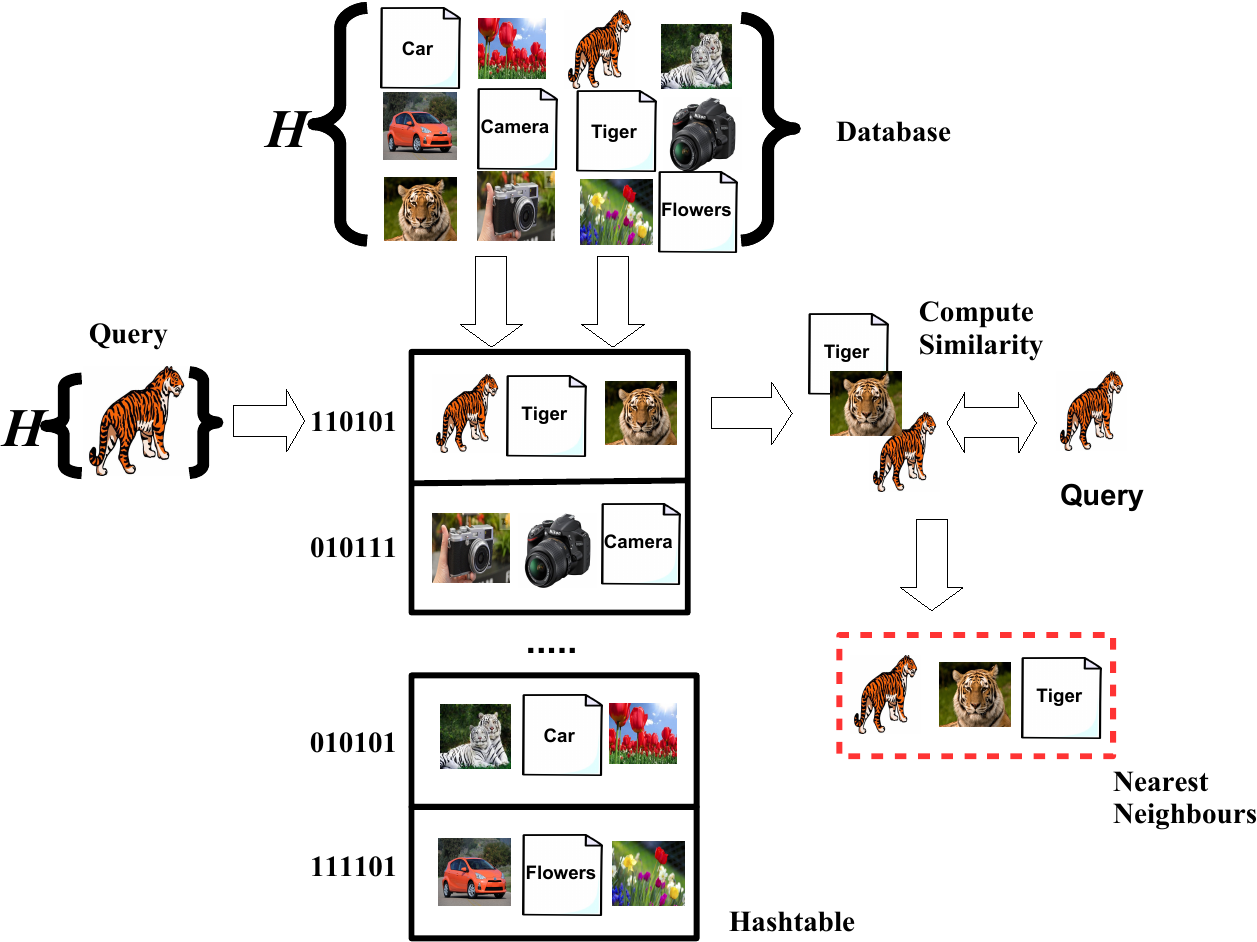}
\caption[Hashing related data-points in two different modalities into the same hashtable buckets]{Cross-modal hashing-based ANN search. In cross-modal hashing, we wish to partition the input space so that semantically similar data-points across modalities fall into the same hashtable buckets. Here, cross-modal hash functions $\mathcal{H}$ assign similar hashcodes to images and documents, enabling fast retrieval of related data across modalities.}
\label{fig:ch2_cross_modal_hash_table}
\end{figure}

Cross-modal hashing models extend unimodal methods by learning two sets of $K$ hyperplanes, one per modality: $\mathbf{W}\in \RE^{D_{x}\times K}$ for $\mathcal{X}$ and $\mathbf{U}\in \RE^{D_{z}\times K}$ for $\mathcal{Z}$. Let $\mathbf{X}\in \RE^{N_{trd}\times D_{x}}$ and $\mathbf{Z}\in \RE^{N_{trd}\times D_{z}}$ denote the feature matrices, where $D_{x}\neq D_{z}$ in general, and $N_{xz}$ the number of paired data-points across modalities ($N_{xz}\le N_{trd}$). An adjacency matrix $\mathbf{S}\in \{0,1\}^{N_{xz}\times N_{xz}}$ encodes semantic relationships across modalities ($S_{ij}=1$ if $(\x_{i},\mathbf{z}_{j})$ are related). The goal is to ensure that $d_{H}(g_{\mathcal{X}}(\x_{i}),g_{\mathcal{Z}}(\mathbf{z}_{j}))\approx 0$ when $S_{ij}=1$, and large otherwise. This extends the unimodal case by enforcing \emph{consistency} across two modalities. The methods reviewed below differ chiefly in how they impose it (Figure \ref{fig:ch2_cross_modal}).

\begin{figure}[!t]
\centering
\resizebox{\textwidth}{!}{%
\begin{tikzpicture}[
  font=\footnotesize, >=stealth,
  hub/.style={draw, very thick, rounded corners, fill=gray!18, text width=3.1cm, align=center, inner sep=4pt},
  meth/.style={draw, rounded corners, fill=white, text width=3.4cm, align=center, font=\scriptsize, inner sep=3pt, minimum height=11mm},
  disc/.style={meth, fill=gray!12}]
  \node[hub] (H) {\textbf{Cross-modal consistency}\\[1pt] two hyperplane sets $\mathbf{W},\mathbf{U}$; paired items $\to$ consistent codes};
  \node[meth, left=16mm of H] (CRH) {\textbf{CRH}\\ per-bit max-margin; intra- and cross-modal terms, boosted};
  \node[meth, above=4mm of CRH] (CVH) {\textbf{CVH}\\ CCA: maximise cross-modal correlation};
  \node[meth, below=4mm of CRH] (IMH) {\textbf{IMH}\\ intra-modal manifold $+$ pairs; semi-supervised};
  \node[meth, right=16mm of H] (PDH) {\textbf{PDH}\\ each modality's code labels the other; spectral independence};
  \node[meth, above=4mm of PDH] (CMSSH) {\textbf{CMSSH}\\ cross-covariance SVD; pairs only};
  \node[disc, below=4mm of PDH] (RCMH) {\textbf{RCMH}\\ regularise codes on a cross-modal graph (extends GRH)};
  \draw[->] (H) -- (CVH); \draw[->] (H) -- (CRH); \draw[->] (H) -- (IMH);
  \draw[->] (H) -- (CMSSH); \draw[->] (H) -- (PDH); \draw[->] (H) -- (RCMH);
  \draw[->, dashed, gray] (IMH) to[out=160,in=200] node[font=\scriptsize, text=black!60, left=1pt] {pairs only} (CVH);
\end{tikzpicture}%
}
\caption[Relationship between cross-modal projection methods for hashing-based ANN search]{Relationship between the six cross-modal projection methods. All learn two hyperplane sets, one per modality, that drive paired items to consistent codes; they differ in how that consistency is imposed. CVH maximises cross-modal correlation by CCA; CMSSH keeps only the cross-modal term, solved by a cross-covariance SVD; CRH adds intra-modal max-margin terms, coupled across bits by boosting; PDH predicts each modality's code from the other and refreshes bit independence spectrally; IMH adds an intra-modal manifold term and so exploits unpaired data, reducing to CVH when only the pairs are used (dashed); and the shaded RCMH regularises the binary codes directly on a cross-modal graph, the relaxation-free extension of GRH.}
\label{fig:ch2_cross_modal}
\end{figure}

Following the same principles as in previous sections, we restrict our review to well-established baselines with widely available implementations. We consider five established cross-modal baselines, and review a sixth (the author's own RCMH) separately below: Cross-View Hashing (CVH) (\cite{kumar11}, Section \ref{sec:ch2_cvh_projection}), Co-Regularised Hashing (CRH) (\cite{Zhen12}, Section \ref{sec:ch2_crh_projection}), Predictable Dual-View Hashing (PDH) (\cite{Rastegari13}, Section \ref{sec:ch2_pdh_projection}), Inter-Media Hashing (IMH) (\cite{Song13}, Section \ref{sec:ch2_imh_projection}), and Cross-Modal Semi-Supervised Hashing (CMSSH) (\cite{Bronstein10}, Section \ref{sec:ch2_cmssh_projection}).

\subsubsection{Cross-View Hashing (CVH)}\label{sec:ch2_cvh_projection}

Cross-View Hashing (CVH)\footnote{As is standard in the literature, we adopt the special case of CVH with only cross-modality supervision and one-to-one paired samples across modalities (Section~3.2 in~\cite{kumar11}).}~\cite{kumar11} mirrors ITQ$+$CCA (Section~\ref{sec:ch2_itq_cca_projection}) in using Canonical Correlation Analysis (CCA) to learn two sets of hyperplanes that maximise cross-modal correlation. There are two main differences relative to ITQ$+$CCA. First, CVH retains \emph{both} sets of hyperplane normals, $\mathbf{W}\in\mathbb{R}^{D_{x}\times K}$ and $\mathbf{U}\in\mathbb{R}^{D_{z}\times K}$, rather than only those of the visual modality. Second, CVH does not apply the post-hoc ITQ rotation to balance variance across hyperplanes. The hash functions for the two modalities are
\begin{equation}
\begin{aligned}
h^{\mathcal{X}}_{k}(\mathbf{x}_{i}) \;=\; \tfrac{1}{2}\!\big(1+\operatorname{sgn}(\mathbf{w}_{k}^{\!\top}\mathbf{x}_{i})\big), \\
h^{\mathcal{Z}}_{k}(\mathbf{z}_{i}) \;=\; \tfrac{1}{2}\!\big(1+\operatorname{sgn}(\mathbf{u}_{k}^{\!\top}\mathbf{z}_{i})\big).
\end{aligned}
\label{eqn:ch2_cvh_hash_function}
\end{equation}

\noindent As with ITQ$+$CCA, the asymptotic training complexity is $\mathcal{O}(N_{trd}D^{2}+D^{3})$ with $D=\max(D_{x},D_{z})$. CVH is one of the earliest cross-modal hashing methods and is widely used as a baseline. The models reviewed next (Sections~\ref{sec:ch2_crh_projection}--\ref{sec:ch2_imh_projection}) introduce alternative learning schemes for both hyperplane sets. These schemes typically achieve higher accuracy on standard image–text benchmarks.

\subsubsection{Co-Regularised Hashing (CRH)}\label{sec:ch2_crh_projection}

Co-Regularised Hashing (CRH)~\cite{Zhen12} learns the two sets of hyperplanes one bit at a time. Each bit is a max-margin problem combining an intra-modal margin term in each modality, a cross-modal consistency term that penalises disagreement between the projections of paired points, and $\ell_{2}$ regularisation. Successive bits are coupled by boosting, which reweights the pairs that earlier bits failed to satisfy. This induces the inter-bit dependence familiar from the unimodal supervised methods. The non-convex per-bit problem is solved by alternating between the two modalities with the concave--convex procedure.
\subsubsection{Cross-Modal Similarity-Sensitive Hashing (CMSSH)}\label{sec:ch2_cmssh_projection}

Cross-Modal Similarity-Sensitive Hashing (CMSSH)~\cite{Bronstein10} is the simplest of the family. It retains only the cross-modal consistency term and discards the intra-modal ones. Each bit maximises the agreement of the signed projections of related cross-modal pairs. Dropping the signs gives a bilinear objective $\mathbf{w}_{k}^{\T}\mathbf{C}\mathbf{u}_{k}$, whose solution is read off from the singular value decomposition of a weighted cross-covariance matrix $\mathbf{C}$; boosting again couples the bits across iterations. As the earliest cross-modal hashing method, CMSSH remains a standard baseline. Its reliance on cross-modal pairs alone, however, limits its accuracy relative to the methods that add intra-modal structure.
\subsubsection{Predictable Dual-View Hashing (PDH)}\label{sec:ch2_pdh_projection}

Predictable Dual-View Hashing (PDH)~\cite{Rastegari13} also learns the hyperplanes bit by bit, with two distinguishing features. It enforces pairwise bit independence by driving the off-diagonal code correlations towards zero, and it couples the modalities by using the current code of one modality as the training label for a max-margin classifier in the other. The method alternates between training the $2K$ classifiers, relabelling the codes with the resulting hyperplanes, and refreshing the codes through a spectral step that re-imposes independence. The spectral step works in the manner of the unsupervised methods of Section~\ref{sec:ch2_dependent_projection}.
\subsubsection{Inter-Media Hashing (IMH)}\label{sec:ch2_imh_projection}

Inter-Media Hashing (IMH)~\cite{Song13} is best read as the semi-supervised member of the family. In addition to the cross-modal pairs, it exploits unsupervised structure within each modality through a $k$-nearest-neighbour graph built over all training points, including those absent from the cross-modal supervision. Its objective combines an intra-modal Laplacian Eigenmap term per modality, a cross-modal consistency term, and a regression term that supplies the out-of-sample projection. In the limiting case where only the cross-modal pairs are used, IMH reduces to the CCA-based CVH. IMH preserves neighbourhoods in both modalities, and its use of abundant unlabelled within-modality structure typically improves accuracy where cross-modal pairs are scarce.

\subsubsection{Regularised Cross-Modal Hashing (RCMH)}\label{sec:ch2_rcmh_projection}

Regularised Cross-Modal Hashing (RCMH)~\cite{Moran15b} carries the graph-regularisation idea of GRH (Section~\ref{sec:ch2_grh_projection}) into the cross-modal setting. A single graph is built across the two modalities, with edges joining paired or similarly-labelled items. The binary codes are refined by the same alternating procedure: a regularisation step propagates codes along the cross-modal graph so that paired items converge upon consistent codes, and a learning step fits a linear hash function per modality for out-of-sample encoding. Operating directly on the codes, RCMH avoids the continuous relaxation of the spectral cross-modal methods. The supervision is carried entirely by the graph, so the method extends naturally to whatever cross-modal correspondences are available, whether explicit pairs or shared labels.
\subsubsection{A Brief Summary}

We reviewed five influential algorithms for hashing across heterogeneous feature spaces: Cross-View Hashing (CVH), Co-Regularised Hashing (CRH), Cross-Modal Similarity Sensitive Hashing (CMSSH), Predictable Dual-View Hashing (PDH), and Inter-Media Hashing (IMH). Each learns two sets of $K$ hyperplanes, one per modality, such that cross-modal neighbours receive consistent hashcodes. The unifying mechanism is an objective function with a cross-modal consistency term that penalises discrepancies between paired projections. More recent approaches (CRH, PDH, IMH) strengthen this formulation with intra-modal terms, regularising the embeddings by enforcing similarity within each modality. Despite these advances, the primary limitations remain the reliance on computationally expensive matrix factorisations or non-convex optimisation procedures.
\fi

%% file: supp/projection_schemes_table.tex
%% Full categorisation of projection-learning algorithms (data-independent,
%% unsupervised, supervised, cross-modal). Full-page summary table; abbreviated
%% out of the CSUR main paper and retained here and in the arXiv-long build.
\begin{table}[p]
\centering
\vspace*{\fill}
\rotatebox{90}{%
\begin{minipage}{\textheight}
\centering
\tiny   % shrink further than \small
\begin{adjustbox}{max width=\textheight,max totalheight=0.70\textwidth,center}
\begin{tabular}{l | l | l | l | l | l | l}
\textbf{Method} & \textbf{Dependency} & \textbf{Learning Paradigm} & \textbf{Hash Function} & \textbf{Training Complexity} & \textbf{Properties} & \textbf{Section} \\
\hline \hline 
LSH & Independent & Unsupervised & $sgn(\mathbf{w}^{\intercal}_{k}\mathbf{x} - t_{k})$ & $\mathcal{O}(KD)$ & $E_{2}$ & \ref{sec:ch2_lsh_sign_random}\\ 
SKLSH & Independent & Unsupervised & $sgn(\cos(\mathbf{w}^{\intercal}_{k}\mathbf{x}+t_{k})+t_{k^{'}})$ & $\mathcal{O}(KD)$ & $E_{2}$ & \ref{sec:ch2_sklsh_projection}\\
\hline
PCAH & Dependent & Unsupervised & $sgn(\mathbf{w}^{\intercal}_{k}\mathbf{x} - t_{k})$ & $\mathcal{O}(\min(N_{trd}^{2}D, N_{trd}D^{2}))$ & $E_{2}, E_{3}, E_{4}$ & \ref{sec:ch2_pcah_projection} \\
AGH & Dependent & Unsupervised & $sgn(\mathbf{w}^{\intercal}_{k}\mathbf{z} - t_{k})$ & $\mathcal{O}(N_{trd}CD + C^{3} + N_{trd}CK)$ & $E_{1}, E_{2}, E_{3}, E_{4}$ & \ref{sec:ch2_agh_projection} \\
ITQ & Dependent & Unsupervised & $sgn(\mathbf{R}\mathbf{W}^{\intercal}\mathbf{x})$ & $\mathcal{O}(\min(N_{trd}^{2}D, N_{trd}D^{2}) + T(N_{trd}K^{2}+K^{3}))$ & $E_{1}, E_{2}, E_{3}, E_{4}$ & \ref{sec:ch2_itq_projection} \\
SH & Dependent & Unsupervised & $sgn\!\left(\sin\!\big(\tfrac{\pi}{2}+\tfrac{j\pi}{b-a}(\mathbf{w}_{k}^{\intercal}\mathbf{x})\big)\right)$ & $\mathcal{O}(\min(N_{trd}^{2}D, N_{trd}D^{2}))$ & $E_{1}, E_{2}, E_{3}, E_{4}$ & \ref{sec:ch2_sh_projection} \\
\hline
STH & Dependent & Supervised & $sgn(\mathbf{w}^{\intercal}_{k}\mathbf{x} - t_{k})$ & $\mathcal{O}(N_{trd}DK + TN_{trd}^{2}K)$ & $E_{1}, E_{2}, E_{3}, E_{4}$ & \ref{sec:ch2_sth_projection} \\
KSH & Dependent & Supervised & $sgn(\mathbf{w}^{\intercal}_{k}\kappa(\mathbf{x}) - t_{k})$ & $\mathcal{O}(N_{trd}CK + N_{trd}^{2}CK + N_{trd}C^{2}K + C^{3}K)$ & $E_{1}, E_{2}$ & \ref{sec:ch2_ksh_projection} \\
BRE & Dependent & Supervised & $sgn(\mathbf{w}^{\intercal}_{k}\kappa(\mathbf{x}) - t_{k})$ & $\mathcal{O}(KN_{trd}^{2} + KN_{trd}\log N_{trd})$ & $E_{1}, E_{2}$ & \ref{sec:ch2_bre_projection} \\
GRH & Dependent & Supervised & $sgn(\mathbf{w}^{\intercal}_{k}\mathbf{x} - t_{k})$ & $\mathcal{O}(N_{trd}DK + TN_{trd}^{2}K)$ & $E_{1}, E_{2}, E_{3}$ & \ref{sec:ch2_grh_projection} \\
\hline
CVH & Dependent & Supervised & $sgn(\mathbf{w}^{\intercal}_{k}\mathbf{x} + t^{x}_{k}),\; sgn(\mathbf{u}^{\intercal}_{k}\mathbf{z} + t^{z}_{k})$ & $\mathcal{O}(N_{trd}D^{2} + D^{3}),\; D=\max(D_{x},D_{z})$ & $E_{1}, E_{2}, E_{3}, E_{4}$ & \ref{sec:ch2_cvh_projection} \\
CMSSH & Dependent & Supervised & $sgn(\mathbf{w}^{\intercal}_{k}\mathbf{x} + t^{x}_{k}),\; sgn(\mathbf{u}^{\intercal}_{k}\mathbf{z} + t^{z}_{k})$ & $\mathcal{O}(KTN_{trd}D),\; D=\max(D_{x},D_{z})$ & $E_{1}, E_{2}$ & \ref{sec:ch2_cmssh_projection} \\
CRH & Dependent & Supervised & $sgn(\mathbf{w}^{\intercal}_{k}\mathbf{x} + t^{x}_{k}),\; sgn(\mathbf{u}^{\intercal}_{k}\mathbf{z} + t^{z}_{k})$ & $\mathcal{O}(KTN_{trd}(D_{x}+D_{z}))$ & $E_{1}, E_{2}$ & \ref{sec:ch2_crh_projection} \\
PDH & Dependent & Supervised & $sgn(\mathbf{w}^{\intercal}_{k}\mathbf{x} + t^{x}_{k}),\; sgn(\mathbf{u}^{\intercal}_{k}\mathbf{z} + t^{z}_{k})$ & $\mathcal{O}(N_{trd}DK + TN_{trd}^{2}K)$ & $E_{1}, E_{2}, E_{4}$ & \ref{sec:ch2_pdh_projection} \\
IMH & Dependent & Supervised & $sgn(\mathbf{w}^{\intercal}_{k}\mathbf{x} + t^{x}_{k}),\; sgn(\mathbf{u}^{\intercal}_{k}\mathbf{z} + t^{z}_{k})$ & $\mathcal{O}(N_{trd}^{3})$ & $E_{1}, E_{2}, E_{3}, E_{4}$ & \ref{sec:ch2_imh_projection} \\
RCMH & Dependent & Supervised & $sgn(\mathbf{w}^{\intercal}_{k}\mathbf{x} + t^{x}_{k}),\; sgn(\mathbf{u}^{\intercal}_{k}\mathbf{z} + t^{z}_{k})$ & $\mathcal{O}(N_{trd}(D_{x}+D_{z})K + TN_{trd}^{2}K)$ & $E_{1}, E_{2}, E_{3}$ & \ref{sec:ch2_rcmh_projection} \\
\hline \hline
\end{tabular}%
\end{adjustbox}
\par\vspace{4pt}
\captionof{table}{Categorisation of existing projection learning algorithms, adapted in part from~\cite{Wang10b}; see \ifsupp Section~\ref{supp:projection}\else Section~\ref{sec:ch2_projection}\fi\ for details on each hash function, and note that the final six algorithms listed are cross-modal. The graph-regularised methods GRH and RCMH refine the discrete codes directly rather than through a continuous relaxation, and balance them through a signed (must-link/cannot-link) graph; they do not pursue bit independence ($E_{4}$), and the measurement of Section~\ref{sec:ch4_measured} examines the effect of the resulting redundancy. Notation: $N$ denotes the total number of data-points, $K$ the hashcode length, $D$ the data dimensionality (with $D_{x}, D_{z}$ the per-modality dimensionalities for the cross-modal methods), $C$ a set of anchor data-points ($C \ll N_{\text{trd}}$), $N_{\text{trd}}$ the number of training data-points ($C < N_{\text{trd}} \ll N$), $T$ the number of iterations, $[a,b]$ the range spanned by a projected dimension, and $j$ the integer frequency index of the Spectral Hashing eigenfunctions; typical values are $N_{\text{trd}}=1000$--$2000$, $K=32$--$128$, and $C=300$.}
\label{tab:ch2_projection_schemes}
\end{minipage}%
}
\vspace*{\fill}
\end{table}

%% file: supp/auto_ch2_itq_projection.tex
%% Relocated from the CSUR main paper to the supplement (fig:ch2_itq_projection).
\begin{figure}[!t]
\centering
\includegraphics[width=82mm, height=61mm]{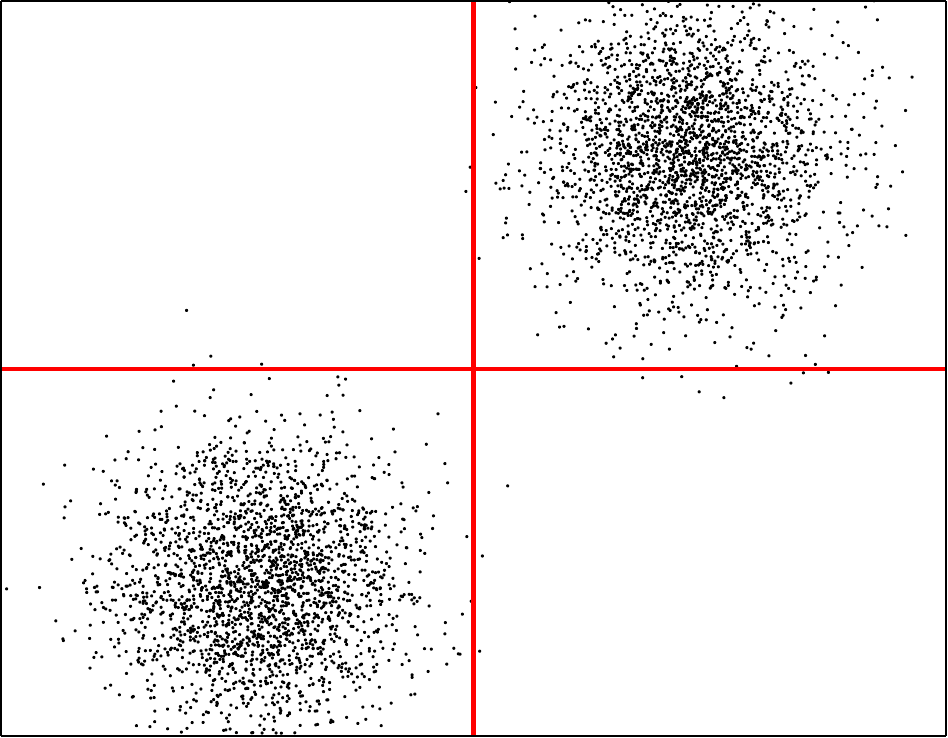}
\caption[The effect of an ITQ rotation on the feature space]{Illustration of the effect of ITQ on the feature space: a toy two-dimensional dataset after applying the learned rotation $\mathbf{R} \in \mathbb{R}^{K \times K}$ (100 iterations). The rotation distributes the variance more evenly across the two hyperplanes (shown as perpendicular lines), and the hyperplanes no longer intersect cluster centres. This reduces the quantisation error, the objective ITQ optimises~\cite{Gong11}.}
\label{fig:ch2_itq_projection}
\end{figure}

%% file: body/deep.tex
\shortonly{\label{sec:deep_unsupervised}\label{sec:deep_crossmodal}}

The methods surveyed so far overwhelmingly treat projection and quantisation as separable. They fix $p$ first and fit $q$ to the geometry it produces, even when both are learned. The exceptions are few; we note two. Minimal loss hashing~\cite{Norouzi11mlh} trains its projections against a quantisation-aware hinge loss. The shallow scheme of~\cite{Moran16} learns the projection hypersurfaces and their binarisation thresholds jointly within a single objective. Both anticipate the coupling that defines the deep era. When applied at high capacity, that coupling removes in full the deficit of Section~\ref{sec:classical_summary}: the classical field otherwise almost never couples the two. Under the PQO lens of Section~\ref{sec:lens}, deep hashing escalates two existing axes simultaneously rather than introducing a new paradigm. The projection $p$ becomes a high-capacity learned $f_{\theta}$, first a convolutional network and later a vision transformer~\cite{Chen22}. The quantiser $q$ is trained jointly against the same objective. The embedding is therefore shaped in anticipation of binarisation, and vice versa.

\iflong
The idea predates the convolutional wave. Semantic Hashing~\cite{Salakhutdinov09} trains a deep autoencoder and binarises its narrowest layer, so that Hamming neighbours are semantic neighbours. This is the first instantiation of ``$p$ as a learned network, $q$ as a thresholding of its output.'' The first convolutional method, CNNH~\cite{Xia14}, remained two-stage and so escalated only the projection. It factorises a pairwise-similarity matrix into approximate target codes and then regresses them with a network. DPSH~\cite{Li16} first learned the network and the codes together, end to end: it learns network and codes together, end to end, from pairwise labels.
\fi
\shortonly{The genealogy runs from Semantic Hashing's autoencoder~\cite{Salakhutdinov09}, through the still two-stage CNNH~\cite{Xia14}, to the end-to-end coupling of DPSH~\cite{Li16}.} We organise the rest of this section by the axes the lens names: differentiable binarisation (the quantisation axis, Section~\ref{sec:deep_binarisation}); the signal that shapes the projection (the supervision axis, Section~\ref{sec:deep_supervision}); operation without labels and across modalities (both folded into Section~\ref{sec:deep_supervision}); and the resolved and open questions of the deep era (Section~\ref{sec:deep_synthesis}).

\subsection{Differentiable binarisation: the quantisation axis in a network}\label{sec:deep_binarisation}

The central obstacle of deep hashing is the non-differentiability of its quantiser. The natural quantiser is the sign function $\cc = \mathrm{sgn}(\mathbf{y})$ with $\cc \in \{-1,+1\}^{K}$. It has zero gradient almost everywhere and is undefined at the origin. We therefore cannot place it na\"ively inside a back-propagated network. The literature offers four broad strategies, summarised in Table~S3 of the supplementary material. Each gives a distinct answer to the question of how to threshold differentiably.

\shortonly{\emph{Relax and penalise} replaces $\mathrm{sgn}$ with a smooth surrogate and adds a quantisation penalty (DPSH~\cite{Li16}, the Deep Hashing Network~\cite{Zhu16}, Deep Supervised Hashing~\cite{Liu16}). \emph{Continuation} anneals $\tanh(\beta\mathbf{y})$ towards the true quantiser (HashNet~\cite{Cao17}). \emph{Straight-through estimation} keeps the exact $\mathrm{sgn}$ forward and passes the gradient through unchanged (GreedyHash~\cite{Su18}). \emph{Discrete optimisation} never relaxes and solves the codes by alternating minimisation (Deep Supervised Discrete Hashing~\cite{Li17}).}

\iflong
\emph{Relax and penalise.} The earliest and commonest approach drops the binary constraint during training. It replaces $\mathrm{sgn}$ with a smooth surrogate, the identity or $\tanh$, and adds a quantisation penalty that pulls activations towards $\pm 1$. DPSH~\cite{Li16}, the Deep Hashing Network~\cite{Zhu16} and Deep Supervised Hashing~\cite{Liu16} all take this route. They differ mainly in the penalty and the pairwise loss. The codes are binarised only at indexing time, and the penalty controls the resulting quantisation error.

We make this concrete with the formulation of DPSH~\cite{Li16}, which has served as the reference instance of the pairwise approach. Let $\uu_i = f_{\theta}(\x_i) \in \RE^{K}$ denote the real-valued output of the network for the $i$-th image. Let $\mathcal{S} = \{s_{ij}\}$ be a set of pairwise labels, in which $s_{ij} = 1$ when $\x_i$ and $\x_j$ are similar and $s_{ij} = 0$ otherwise. We write $\Omega_{ij} = \tfrac{1}{2}\uu_i^{\T}\uu_j$ for one half of the inner product of the two outputs. A logistic likelihood then models the probability of the observed similarity:
\begin{equation}
p(s_{ij} \mid \uu_i, \uu_j) =
\begin{cases}
\sigma(\Omega_{ij}) & s_{ij}=1,\\
1 - \sigma(\Omega_{ij}) & s_{ij}=0,
\end{cases}
\qquad \sigma(x) = \frac{1}{1+e^{-x}},
\label{eqn:deep_dpsh_lik}
\end{equation}
which encourages a large positive inner product between the outputs of similar pairs and a large negative one between the outputs of dissimilar pairs. Maximising this likelihood over $\mathcal{S}$ is equivalent to minimising the negative log-likelihood. To this we add a quantisation penalty that draws each continuous output towards its binarised counterpart $\bb_i = \mathrm{sgn}(\uu_i)$:
\begin{equation}
\mathcal{L} = -\!\!\sum_{s_{ij}\in\mathcal{S}}\!\big(s_{ij}\Omega_{ij} - \log(1+e^{\Omega_{ij}})\big) \;+\; \eta\sum_{i=1}^{N}\lVert \bb_i - \uu_i \rVert_2^2 .
\label{eqn:deep_dpsh_loss}
\end{equation}
The first term shapes the projection: the geometry of the network output preserves the supervised neighbourhood structure. The second term, governed by the penalty weight $\eta$, controls the error incurred when that output is finally thresholded. In the terms of the lens, the two terms act on the two axes at once. The likelihood places the projection and the penalty places the thresholds. Their joint optimisation distinguishes the method from the two-stage pipelines of the preceding section.

\emph{Continuation.} HashNet~\cite{Cao17} closes the gap between a surrogate and its sign by construction, instead of penalising it. It replaces $\mathrm{sgn}$ with $\tanh(\beta \mathbf{y})$ and anneals $\beta \to \infty$. The projection is therefore optimised against a sequence of ever-sharper quantisers that converge on the true one, rather than against a fixed soft proxy.

We can state the continuation principle precisely. HashNet replaces the sign function with a scaled hyperbolic tangent, $\h_i = \tanh(\beta\, f_{\theta}(\x_i))$. It optimises a weighted maximum-likelihood objective of the same logistic form as the likelihood term of Equation~\ref{eqn:deep_dpsh_loss},
\begin{equation}
\mathcal{L}(\beta) = -\!\!\sum_{s_{ij}\in\mathcal{S}}\! w_{ij}\,\big(s_{ij}\Omega_{ij} - \log(1+e^{\Omega_{ij}})\big), \qquad \Omega_{ij} = \alpha\,\h_i^{\T}\h_j,
\label{eqn:deep_hashnet}
\end{equation}
in which the weights $w_{ij}$ compensate for the heavy imbalance between similar and dissimilar pairs that is typical of retrieval data. The scale $\alpha$ widens the range over which the logistic responds; HashNet sets $\alpha$ below one, whereas Equation~\ref{eqn:deep_dpsh_loss} fixed it at $\tfrac12$. The bandwidth $\beta$ increases over a sequence of training stages, $\beta_1 < \beta_2 < \cdots$. Because $\tanh(\beta z) \to \mathrm{sgn}(z)$ as $\beta \to \infty$, the network is optimised against a succession of progressively sharper quantisers that converge on the exact binariser. In contrast to the relax-and-penalise approach, the projection never trains against a fixed soft proxy. It trains against a quantiser that hardens as learning proceeds. This narrows the discrepancy between the codes used at training time and those used at retrieval.

\emph{Straight-through estimation.} GreedyHash~\cite{Su18} keeps the exact $\mathrm{sgn}$ in the forward pass, so the training codes are genuinely binary. It passes the gradient through unchanged on the backward pass; this is the straight-through estimator. The approach removes the relaxation gap at the cost of a biased gradient.

\emph{Discrete optimisation.} Deep Supervised Discrete Hashing~\cite{Li17} does not relax the binary constraint at any stage. It retains a hard binary constraint and solves for the codes by alternating minimisation interleaved with network updates. The quantiser stays exact throughout.
\fi

\iflong\input{supp/auto_deep_binarisation}\fi

\subsection{Supervision signals: the projection axis in a network}\label{sec:deep_supervision}

Binarisation concerns the quantiser; the training signal concerns the projection. The signal fixes the geometry that $f_{\theta}$ learns before any bits are assigned. The dominant supervised signal is \emph{pairwise}: a likelihood or contrastive loss over similar and dissimilar pairs, as in DPSH~\cite{Li16}, the Deep Hashing Network~\cite{Zhu16} and Deep Supervised Hashing~\cite{Liu16}. Such a loss pulls similar codes together in Hamming space and pushes dissimilar ones apart.

\iflong
Two refinements proved influential. Deep Cauchy Hashing~\cite{Cao18a} observes that pairwise losses concentrate poorly within the small Hamming radius at which hashtable lookup is efficient. It reshapes the objective with a Cauchy distribution, which pulls true neighbours inside a tight radius. Pairwise supervision also scales quadratically in the training set. Central Similarity Quantization~\cite{Yuan20} therefore substitutes a \emph{pointwise} signal: it assigns each class a well-separated ``hash centre'' towards which its items quantise. This gives linear-time supervision and globally separated codes. A complementary line addresses the scarcity of supervision rather than its form. HashGAN~\cite{Cao18b} synthesises training images with a pair-conditional generative adversarial network, which densifies sparse pairwise labels before the projection is learned.
\fi
\shortonly{Refinements reshape this signal. Deep Cauchy Hashing~\cite{Cao18a} concentrates true neighbours within a tight Hamming radius. Central Similarity Quantization~\cite{Yuan20} swaps quadratic pairwise supervision for linear-time pointwise ``hash centres''. HashGAN~\cite{Cao18b} densifies sparse labels with synthesised images.}

\iflong
\subsection{Unsupervised and self-supervised deep hashing}\label{sec:deep_unsupervised}

Where labels are unavailable, deep hashing must construct a surrogate training signal. Most unsupervised methods act on the projection axis by substituting a surrogate for supervision. Stochastic Generative Hashing~\cite{Dai17} frames codes generatively. It learns an encoder--decoder pair, so that the binary code compresses the input while permitting its reconstruction. It optimises this objective by a distributional gradient that sidesteps the binary constraint. DistillHash~\cite{Yang19} distils a confident set of similar and dissimilar pairs from the structure of a pretrained feature space. It then trains on these pseudo-labels, which imports the supervised pairwise machinery into the unsupervised setting. Twin-Bottleneck Hashing~\cite{Shen20} couples a continuous code that drives reconstruction with a binary code that builds a code-similarity graph. The two codes exchange information, so the binary bottleneck is shaped by the continuous one rather than truncated from it. Contrastive Information Bottleneck Hashing~\cite{Qiu21} brings the contrastive-learning paradigm to bear. It trains a stochastic binary encoder with a contrastive objective read through the information-bottleneck principle. The parameter-free Bi-half layer~\cite{Li21} is a distinct and instructive case, because it acts purely on the quantisation axis. It forces each bit towards a half-and-half (maximum-entropy) distribution across the batch. This maximises the information capacity of the code without any loss term, and the layer drops into an arbitrary projection backbone.

\subsection{Deep cross-modal hashing}\label{sec:deep_crossmodal}

The deep treatment of cross-modal retrieval is the direct analogue of the classical cross-modal methods of Section~\ref{sec:ch2_crossmodal_projection}. Each modality's projection is realised by its own network, and a shared Hamming space is enforced across them. Deep Cross-Modal Hashing~\cite{Jiang17} is the end-to-end counterpart of DPSH. It trains separate image and text networks under cross-modal pairwise supervision, so that paired items receive consistent codes. Self-Supervised Adversarial Hashing~\cite{Li18} strengthens the alignment with adversarial training. A self-supervised semantic network guides two modality-specific networks whose distributions are matched adversarially. This yields a more consistent shared quantisation space than a pairwise term alone provides.

\fi
\shortonly{Without labels, surrogates supervise the projection axis: generative reconstruction~\cite{Dai17}, distilled pseudo-pairs~\cite{Yang19}, a continuous--binary twin bottleneck~\cite{Shen20} and contrastive information-bottleneck codes~\cite{Qiu21}. The parameter-free Bi-half layer~\cite{Li21} acts instead on the quantisation axis; it drives each bit to a maximum-entropy split. Across modalities, each modality carries its own network into a shared Hamming space, by cross-modal pairwise supervision~\cite{Jiang17} or adversarial alignment~\cite{Li18}.}

\subsection{Resolved and open questions after deep hashing}\label{sec:deep_synthesis}

The lens permits a concise characterisation of the contribution of deep hashing readily. Recall the central hypothesis of this survey: relaxing the static, uniform and data-oblivious choices of LSH improves retrieval effectiveness. Deep hashing confirms this hypothesis jointly rather than one axis at a time. A data-dependent projection of essentially unbounded capacity, a data-dependent quantiser, and their coupling are optimised against a single criterion. This resolves the fourth classical deficit of Section~\ref{sec:classical_summary}, the scarcity of joint projection and quantisation learning. Deep hashing does not change the third stage of Definition~\ref{def:lens}, the organisation structure $\mathcal{I}$. Deep methods yield better codes, but they mostly still search them by Hamming ranking, a linear scan in code space, or a simple hashtable. The improved embedding does not of itself make the index more efficient. Production-scale systems therefore pair learned codes with the vector-quantisation and graph-based machinery taken up below. The resurgence of binary codes in retrieval-augmented serving~\cite{Yamada21} couples a deep projection and a binary quantiser with explicit re-ranking rather than Hamming search alone. Deep hashing has thus advanced the embedding and compression stages. It leaves the organisation stage largely untouched; the subsequent sections address that stage.

%% file: supp/auto_deep_binarisation.tex
%% Relocated from the CSUR main paper to the supplement (tab:deep_binarisation).
\begin{table}[!t]
\centering
\small
\begin{tabular}{@{}p{2.7cm} p{2.9cm} p{2.6cm} p{2.4cm}@{}}
\toprule
\textbf{Strategy} & \textbf{Representative methods} & \textbf{Forward $q$} & \textbf{Backward gradient} \\
\midrule
Relax + penalise & DPSH, DHN, DSH~\cite{Li16, Zhu16, Liu16} & $\tanh$ / identity $+\,\pm1$ penalty & exact (smooth surrogate) \\
Continuation & HashNet~\cite{Cao17} & $\tanh(\beta\mathbf{y})$, $\beta\!\to\!\infty$ & exact on each surrogate \\
Straight-through & GreedyHash~\cite{Su18} & $\mathrm{sgn}(\mathbf{y})$ (exact) & identity (biased) \\
Discrete optimisation & DSDH~\cite{Li17} & $\mathrm{sgn}$, exact codes & alternating; no relaxation \\
\bottomrule
\end{tabular}
\caption[Differentiable binarisation strategies in deep hashing]{Four strategies for placing a differentiable quantiser inside a deep network. Each strategy gives a distinct answer to the quantisation-axis question of where, and how sharply, to threshold while still permitting gradient-based training.}
\label{tab:deep_binarisation}
\end{table}

%% file: body/pq.tex
\shortonly{\label{sec:pq_additive}}

\longshort{Section~\ref{sec:ch2_quantisation} distinguished scalar from vector quantisation and set the latter aside. We noted only that vector quantisation often achieves lower reconstruction error at the cost of a query-time lookup table~\cite{Jegou11, He13}. We now take it up. The vector-quantisation family is the compression axis of Definition~\ref{def:lens} carried to its richer extreme. It therefore widens the second of the survey's guiding questions. The classical methods of Section~\ref{sec:ch2_quantisation} asked \emph{where to place the thresholds}. This family asks the more general question of \emph{how to partition the projected space into codes}. Scalar thresholding is the one-dimensional, binary special case of that question. Recall that binary hashing fixes the alphabet to $\mathcal{C} = \{0,1\}^{K}$ and compares codes by Hamming distance. This family instead lets $\mathcal{C}$ be a learned codebook of centroids and computes an \emph{asymmetric} distance through small lookup tables. It thereby resolves distances far more finely than $K$ bits of Hamming space can express. The associated cost is the lookup, and a body of work, reviewed below, is devoted to keeping it cheap. As the family develops, it does not remain on the compression axis. It reintroduces a learned \emph{projection} in its optimised variants and engages the \emph{organisation} axis in its indexed ones.}{Section~\ref{sec:ch2_quantisation} distinguished scalar from vector quantisation and set the latter aside~\cite{Jegou11, He13}. We now take it up. The vector-quantisation family is the compression axis of Definition~\ref{def:lens} carried to its richer extreme. It widens the survey's second question from \emph{where to place the thresholds} to \emph{how to partition the projected space into codes}; scalar thresholding is the one-dimensional, binary special case. Recall that binary hashing fixes the alphabet to $\mathcal{C} = \{0,1\}^{K}$ and compares codes by Hamming distance. This family instead lets $\mathcal{C}$ be a learned codebook of centroids and computes an \emph{asymmetric} distance through small lookup tables. These tables resolve distances far more finely than $K$ bits of Hamming space can express; the price is the lookup. As the family develops, it does not remain on the compression axis. It reintroduces a learned \emph{projection} in its optimised variants and engages the \emph{organisation} axis in its indexed ones.}

\subsection{From bits to codebooks: product quantisation}\label{sec:pq_pq}

A plain vector quantiser assigns each data-point to its nearest centroid in a codebook learned by $k$-means~\cite{Lloyd82}, representing $\mathbf{x}$ by a single index. Resolving neighbours finely, however, demands many centroids. A codebook of $2^{K}$ centroids is intractable to learn and store at the code lengths retrieval requires.

\iflong
Product quantisation (PQ)~\cite{Jegou11} resolves this with a Cartesian decomposition. It splits the $D$-dimensional space into $M$ disjoint subspaces,\footnote{Definition~\ref{def:lens} used the symbol $M$ for the dimensionality of the projected representation. We reuse it here in its conventional product-quantisation sense, the number of subspaces; the intended meaning is clear from context.} learns an independent $k$-means codebook (typically $256$ centroids, i.e.\ one byte) in each, and concatenates the $M$ sub-centroid indices. The effective codebook is then the Cartesian product of $256^{M}$ centroids drawn from $M$ byte-sized tables. PQ thus recovers cheaply a capacity that a flat quantiser could not reach. PQ computes distances \emph{asymmetrically}. The query is left uncompressed, and its distance to a database code is approximated by summing, over the $M$ subspaces, the precomputed distance from each query sub-vector to the relevant sub-centroid (asymmetric distance computation, ADC). Each summand is a constant-time lookup, so these distances are resolved far more finely than the integer-valued Hamming distances of binary codes. This finer resolution is the source of PQ's accuracy advantage. For non-exhaustive search, PQ is paired with a coarse quantiser and inverted lists (the IVFADC system of~\cite{Jegou11}). This configuration scaled the family to a billion vectors~\cite{Jegou11a} and first touched the organisation axis.

Formally, we write a data-point $\x \in \RE^{D}$ as the concatenation of $M$ sub-vectors, $\x = [\x^{1}, \ldots, \x^{M}]$ with each $\x^{m} \in \RE^{D/M}$. Every subspace carries its own sub-quantiser $q_{m}$, which maps $\x^{m}$ to the nearest centroid of a codebook $\mathcal{C}_{m} = \{\cc_{m,1}, \ldots, \cc_{m,k}\}$ of size $k$ (commonly $k = 256$). The product quantiser is the tuple of these assignments, $q(\x) = \big(i_{1}(\x), \ldots, i_{M}(\x)\big)$, where $i_{m}(\x)$ is the index of the centroid in $\mathcal{C}_{m}$ nearest to $\x^{m}$. The code therefore occupies $M\log_{2} k$ bits. Given a query $\y$, which is left uncompressed, the asymmetric approximation to the squared Euclidean distance is the sum of the per-subspace contributions:
\begin{equation}
\tilde{d}(\y, \x)^{2} = \sum_{m=1}^{M} \big\lVert \y^{m} - \cc_{m,\, i_{m}(\x)} \big\rVert_2^{2} .
\label{eqn:pq_adc}
\end{equation}
Each term is read from a table of $k$ entries, precomputed once per query for each subspace. The distance to every database code is therefore obtained by $M$ lookups and additions. This sum can take of order $k^{M}$ distinct values, far more than the $K+1$ values available to a $K$-bit Hamming distance. This finer resolution, rather than any difference in the projection, accounts for the accuracy of the family relative to binary hashing at equal code length.
\fi
\shortonly{Product quantisation (PQ)~\cite{Jegou11} resolves this with a Cartesian decomposition into $M$ subspaces, each with its own byte-sized $k$-means codebook; their product recovers a vast effective codebook cheaply. An asymmetric distance computation (ADC), summed by lookup over the subspaces against the uncompressed query, resolves neighbours far more finely than Hamming distance. This finer resolution is the source of the family's accuracy. Paired with a coarse quantiser and inverted lists (the IVFADC system of~\cite{Jegou11}, the IVF-PQ of the measured tables), PQ scaled to a billion vectors~\cite{Jegou11a} and first touched the organisation axis. Note that the symbol $M$ takes its conventional product-quantisation sense here, the number of subspaces.}

\subsection{Learning the projection before the codebook: optimised PQ}\label{sec:pq_opq}

Plain product quantisation splits the coordinate axes into subspaces as given. It thereby assumes implicitly that the variance of the data is balanced and aligned across that partition. When it is not, the per-subspace codebooks are allocated to poorly conditioned directions. The remedy lies on the projection axis. Optimised Product Quantisation~\cite{Ge13, Ge14} and, independently, Cartesian $k$-means~\cite{Norouzi13} learn an orthogonal rotation $\mathbf{R}$ that aligns the data to the subspace decomposition before quantising, minimising the distortion. In the terms of the lens, this is a learned projection $p(\mathbf{x}) = \mathbf{R}^{\intercal}\mathbf{x}$ composed with a product quantiser $q$. This is the same projection-then-quantise structure that organises the binary-hashing portion of the survey, so we may read the two literatures as addressing the same two problems.

\iflong
We find the rotation by minimising the total quantisation distortion jointly over the rotation and the codebooks:
\begin{equation}
\min_{\mathbf{R},\, \mathcal{C}} \; \sum_{i=1}^{N} \big\lVert \mathbf{R}^{\T}\x_{i} - q\big(\mathbf{R}^{\T}\x_{i}\big) \big\rVert_2^{2} \qquad \text{subject to} \qquad \mathbf{R}^{\T}\mathbf{R} = \mathbf{I},
\label{eqn:opq}
\end{equation}
where $q$ denotes the product quantiser applied in the rotated space. The problem is solved by alternating minimisation. With $\mathbf{R}$ held fixed, the codebooks are re-estimated by $k$-means on the rotated data. With the codebooks held fixed, the optimal rotation is obtained in closed form as the orthogonal Procrustes solution that aligns the data with their quantised reconstructions. This learned rotation is the quantisation-side analogue of the rotation that iterative quantisation~\cite{Gong11} learns on the binary-hashing side. Its appearance here is a further instance of the projection axis re-entering a family whose primary concern is quantisation.
\fi
\shortonly{The rotation is the quantisation-side analogue of the rotation that iterative quantisation learns on the binary-hashing side.}

\iflong
\subsection{Beyond independence: additive and composite quantisation}\label{sec:pq_additive}

PQ's Cartesian structure assumes the subspaces are independent, which caps how well it can model correlated dimensions. A second line relaxes this by making the code \emph{additive}: the data-point is approximated as a \emph{sum} of $M$ full-dimensional codewords drawn from $M$ codebooks, rather than as a concatenation of sub-centroids. Additive Quantisation~\cite{Babenko14} adopts this form directly. It gains expressiveness at the cost of an NP-hard encoding step, which later work rendered practical through an improved local-search encoder~\cite{Martinez16}. Composite Quantisation~\cite{Zhang14} constrains the inter-codebook cross terms to a constant, so the asymmetric distance remains a cheap sum of lookups; this trades model richness directly against table efficiency. Tree Quantisation~\cite{Babenko15} sits between PQ and full additivity, permitting codeword interactions only along the edges of a learned tree. A tension between more expressive codes and a cheaper distance table recurs across these methods. It is the compression-axis analogue of the bit-allocation question that drove the multi-threshold scalar quantisers of Section~\ref{sec:ch2_quantisation}.

\fi
\shortonly{A second line relaxes PQ's independence assumption by making the code \emph{additive}: each point is approximated as a sum of full-dimensional codewords (Additive~\cite{Babenko14}, Composite~\cite{Zhang14} and Tree~\cite{Babenko15} Quantisation). These methods trade richer codes against a cheaper distance table.}
\subsection{Organising the codes: inverted indices}\label{sec:pq_index}

The variants above improve the codebook. A parallel effort improves the structure over the codes, and so belongs to the organisation axis. The inverted multi-index~\cite{Babenko12} applies product quantisation to the \emph{coarse} quantiser itself. It subdivides the space far more finely than a flat inverted file at equal cost, shortening the candidate lists to be scanned. Locally Optimised Product Quantisation~\cite{Kalantidis14} makes the projection data-local. It learns a separate OPQ rotation and codebook within each coarse cell and encodes residuals, coupling a per-region projection to a per-region quantiser atop a multi-index. Packaged at scale in libraries such as FAISS~\cite{Johnson19}, these systems optimise all three of the lens's stages at once. They anticipate the co-design theme that the graph-based methods push furthest.

\subsection{Score-aware quantisation: ScaNN}\label{sec:pq_scann}

The methods so far all minimise reconstruction error, implicitly treating every direction of the residual as equally harmful. For maximum-inner-product search, the operation beneath modern embedding retrieval, this is the wrong objective. The residual component parallel to a database point dominates the error for exactly the high-scoring neighbours we most want to rank correctly. ScaNN's anisotropic vector quantisation~\cite{Guo20} reweights the loss to penalise that parallel component more heavily, making the quantiser itself \emph{score-aware}.

\iflong
We write the residual as $\mathbf{r} = \mathbf{x} - q(\mathbf{x})$ and decompose it into a part $\mathbf{r}_{\parallel}$ parallel to $\mathbf{x}$ and a part $\mathbf{r}_{\perp}$ orthogonal to it. The loss
\begin{equation}
\ell\big(\mathbf{x}, q(\mathbf{x})\big) = \eta\,\lVert \mathbf{r}_{\parallel} \rVert_2^{2} + \lVert \mathbf{r}_{\perp} \rVert_2^{2}, \qquad \eta > 1,
\label{eqn:scann}
\end{equation}
weights the parallel error more heavily than the orthogonal. This contribution sits purely on the compression axis, but it ties the quantiser to the downstream task rather than to generic fidelity. It underpins a system that performs strongly on contemporary ANN benchmarks.
\fi
\shortonly{This contribution sits purely on the compression axis, but it ties the quantiser to the downstream task rather than to generic fidelity. It underpins a system that performs strongly on contemporary ANN benchmarks.}

\subsection{Where the two axes meet: polysemous codes}\label{sec:pq_synthesis}

The polysemous code~\cite{Douze16} illustrates clearly that scalar-binary and vector-codebook quantisation are two settings of one axis, rather than rival paradigms. PQ codes are learned so that their centroid-index assignment \emph{also} behaves as a good binary code. A single code then offers the speed of a Hamming comparison with the accuracy of a codebook, through the binary-filter-then-fine-rerank arrangement that recurs at retrieval scale. Here it appears as Hamming filtering before ADC; in the modern binary-embedding resurgence it appears as a binary pass before full-precision rescoring. Table~S4 of the supplementary material summarises the family through the lens. The broader lesson is that the compression axis, pursued seriously, draws in a learned projection (optimised product quantisation), the organisation structure (the inverted multi-index), and finally the retrieval objective itself (ScaNN). This reinforces the claim that the three stages of Definition~\ref{def:lens} are ultimately to be co-designed, which the graph-based indexes of the next section take furthest.

\iflong
RaBitQ~\cite{Gao24} is a recent point on this axis. It quantises each vector to a single bit per dimension after a random rotation, in the manner of sign random projection, but with an unbiased distance estimator. The estimator carries a high-probability error bound decaying as $\mathcal{O}(1/\sqrt{D})$ and supports asymmetric re-ranking. A multi-bit extension trades that bound smoothly against space~\cite{Gao25}. In the lens's terms, RaBitQ is a randomised projection composed with a single-bit quantiser whose error is bounded with high probability rather than empirically. Its adoption into production libraries shows how the binary and codebook readings of the compression axis have converged. A one-bit scalar quantiser, in classical terminology, now carries much of the accuracy guarantee that motivated vector codebooks.
\fi
\shortonly{RaBitQ~\cite{Gao24} is a recent point on this axis: a randomised projection composed with a single-bit-per-dimension quantiser that carries a high-probability error bound and supports asymmetric re-ranking, with a space-tunable multi-bit extension. A one-bit scalar quantiser now carries much of the accuracy guarantee that motivated vector codebooks.}

\iflong\input{supp/auto_pq}\fi

%% file: supp/auto_pq.tex
%% Relocated from the CSUR main paper to the supplement (tab:pq).
\begin{table}[!t]
\centering
\small
\begin{tabular}{@{}p{2.9cm} p{2.5cm} p{2.9cm} p{2.5cm}@{}}
\toprule
\textbf{Method} & \textbf{Projection ($p$)} & \textbf{Code alphabet ($q$)} & \textbf{Organisation ($\mathcal{I}$)} \\
\midrule
Product quantisation~\cite{Jegou11} & \emph{axis-aligned split} & $M$ sub-codebooks, concatenated & IVFADC inverted lists \\
OPQ / ck-means~\cite{Ge14, Norouzi13} & learned rotation $\mathbf{R}$ & $M$ sub-codebooks & IVFADC \\
Additive / composite~\cite{Babenko14, Zhang14} & \emph{---} & sum of $M$ full-dim codewords & \emph{---} \\
Inverted multi-index~\cite{Babenko12} & \emph{---} & coarse product codebook & multi-index \\
LOPQ~\cite{Kalantidis14} & local rotation per cell & local sub-codebooks (residuals) & multi-index \\
ScaNN~\cite{Guo20} & \emph{(optional)} & anisotropic-loss codebook & partitioned + ADC \\
Polysemous~\cite{Douze16} & \emph{---} & PQ codes doubling as binary & Hamming filter $+$ ADC \\
\bottomrule
\end{tabular}
\caption[The product-quantisation family through the lens]{The product-quantisation family read through the PQO lens (Definition~\ref{def:lens}). The family principally acts on the compression axis ($q$), where real-valued codebooks with asymmetric lookup distance take the place of binary codes with Hamming distance; its optimised and indexed variants reintroduce a learned projection ($p$) and engage the organisation structure ($\mathcal{I}$). Em-dashes mark stages a method leaves untouched.}
\label{tab:pq}
\end{table}

%% file: body/graph.tex
The two preceding families concentrate their design effort on the embedding and compression stages of Definition~\ref{def:lens}. Graph-based indexes take the converse position. They leave those stages untouched and search raw or independently compressed vectors with exact distances. They concentrate instead on the organisation structure $\mathcal{I}$, an explicit graph whose edges encode proximity and whose traversal leads a query to its neighbourhood. Section~\ref{sec:lens_limits} anticipated this regime as the limit of the projection--quantisation perspective. The regime is significant in practice. On the recall--latency trade-offs that dominate benchmarking~\cite{Aumuller20, Simhadri22}, graph indexes are at present the strongest methods for in-memory search. We treat them because of the lens. They isolate the one axis the survey otherwise holds fixed, and that isolation renders the field's central trade-off intelligible.

\subsection{Greedy routing on proximity graphs}\label{sec:graph_routing}

The shared mechanism is simple. Each item is a vertex, and edges connect items that are near in the original metric. Search is greedy best-first traversal: the walk moves from an entry vertex to the neighbour closest to the query until no neighbour is closer. Both correctness and speed hinge on the edge set. Too few or poorly chosen edges trap the walk in local minima, while too many inflate memory and per-step cost. Nothing is projected or quantised. The method is entirely a statement about $\mathcal{I}$, and the search width that controls accuracy is the trade-off the organisation axis offers in place of code length.

\shortonly{Navigable small-world graphs~\cite{Malkov14} combine short-range links with long-range links. The short-range links make the graph locally dense, and the long-range links make it globally navigable. Greedy routing therefore reaches a query's neighbourhood in a number of hops that grows only logarithmically with database size. The search maintains a list of the $ef$ closest vertices found so far and halts when no unexplored candidate improves upon it. The width $ef$ trades distance computations against recall. We set out the routine and its candidate-list bookkeeping in Section~S6 of the supplementary material.}

\iflong
Navigable small-world graphs~\cite{Malkov14} gave the first influential answer. They combine short-range links that make the graph locally dense with long-range links that make it globally navigable. Greedy routing therefore reaches the query's neighbourhood in a number of hops that grows only logarithmically with database size.

The procedure is a single routine (Algorithm~\ref{alg:greedy_search}). The search maintains a list $W$ of the $ef$ closest vertices found so far. It repeatedly takes the most promising unexplored vertex, examines its neighbours, and admits any that improve upon $W$. It halts once the nearest unexplored candidate lies further from the query than the worst member of $W$. The width $ef$ is the trade-off the organisation axis offers in place of code length. A larger list explores more of the graph and returns more accurate neighbours at the cost of more distance computations. A smaller list returns sooner with lower recall.

\begin{algorithm}[!t]
\DontPrintSemicolon
\KwIn{query $\q$, entry vertex $v$, graph $G$, list width $ef$}
\KwOut{$ef$ approximate nearest neighbours of $\q$}
$C \gets \{v\}$ (min-heap on $d(\q,\cdot)$); \quad $W \gets \{v\}$ (max-heap on $d(\q,\cdot)$); \quad mark $v$ visited\;
\While{$C \neq \emptyset$}{
  $c \gets$ extract the nearest vertex from the min-heap $C$\; \tcp*{$\mathcal{O}(\log|C|)$}
  \If{$d(\q,c)$ exceeds the distance from $\q$ to the furthest vertex of $W$}{\textbf{break}\;}
  \ForEach{neighbour $e$ of $c$ in $G$ not yet visited}{
    mark $e$ visited\;
    \If{$d(\q,e)$ is below the furthest distance in $W$ \textbf{or} $|W| < ef$}{
      add $e$ to $C$ and to $W$\;
      \lIf{$|W| > ef$}{remove the furthest vertex of $W$ (the max-heap root)}
    }
  }
}
\Return{$W$}
\caption{Greedy search on a proximity graph}
\label{alg:greedy_search}
\end{algorithm}
\fi

\subsection{Hierarchy and pruning: HNSW and NSG}\label{sec:graph_hnsw}

Edge selection and graph topology are the organisation-axis analogue of the bit-allocation and threshold-placement questions of the other two axes. Both are budgeted decisions about which connections are worth keeping. The budget here is a memory and construction-cost ceiling rather than a code-length one.

\shortonly{Two refinements made proximity graphs the dominant in-memory index. The Hierarchical Navigable Small World graph (HNSW)~\cite{Malkov20} stacks navigable graphs into skip-list layers. Each element receives a maximum layer from an exponentially decaying random rule, so layer populations shrink geometrically and only a few vertices reach the top. A query descends from the sparse topmost layer and routes greedily within each layer before dropping to the next. This design yields robust logarithmic search and the de facto default index in modern libraries. The complementary concern is out-degree. The Navigating Spreading-out Graph (NSG)~\cite{Fu19} approximates a monotonic search network, pruning edges while preserving near-monotone paths. Its pruning rule descends from the occlusion criterion of FANNG~\cite{Harwood16}. Section~S6 of the supplementary material gives the layer rule and construction pipeline.}

\iflong
Two refinements made proximity graphs the dominant in-memory index. The Hierarchical Navigable Small World graph (HNSW)~\cite{Malkov20} stacks navigable graphs into skip-list layers. A sparse upper layer routes the query coarsely before handing off to denser layers below. This design yields robust logarithmic search and the de facto default index in modern libraries. The complementary concern is out-degree, since dense graphs are fast but memory-hungry. The Navigating Spreading-out Graph (NSG)~\cite{Fu19} approximates a monotonic search network. It approximates the monotonicity property rather than guaranteeing it, and so cuts out-degree and index size. Its pruning rule descends from the occlusion (relative-neighbourhood) criterion of FANNG~\cite{Harwood16}, which first framed edge selection as discarding links rendered redundant by a closer intermediate vertex. Exact construction is costly, so an approximate $k$-nearest-neighbour graph is usually built first, for example by EFANNA~\cite{Fu16}, and then refined.

The hierarchy follows a simple randomised rule: on insertion, an element is assigned a maximum layer
\begin{equation}
\ell = \big\lfloor -\ln(u)\, m_{L} \big\rfloor, \qquad u \sim \mathrm{Uniform}(0,1),
\label{eqn:hnsw_layer}
\end{equation}
where the normalising constant $m_{L}$ controls the expected number of layers. The canonical choice is $m_{L}=1/\ln M$, with $M$ the maximum out-degree. This choice yields an expected $\mathcal{O}(\log_{M} N)$ layers and makes each layer's population decrease geometrically, by a factor of $e^{-1/m_{L}}$, so that only a few vertices reach the top. A query descends from the single entry point in the topmost layer and routes greedily within each layer by Algorithm~\ref{alg:greedy_search} before dropping to the next. This descent yields logarithmic search at the cost of storing the multi-layer edge structure in memory.
\fi

\subsection{Scaling beyond memory: DiskANN and the return of the lens}\label{sec:graph_disk}

A pure graph over full-precision vectors is memory-bound, because it holds both the vectors and the edge lists in main memory. DiskANN~\cite{Subramanya19} overcomes this limit, and in doing so it reunites the three stages of the lens. Compression supplies cheap in-memory distance estimates, organisation supplies the routing, and a re-ranking pass restores the precision quantisation discards. At this scale, the graph family cannot ignore the projection--quantisation factorisation and must incorporate it.

\iflong
Concretely, DiskANN builds a single flat, long-range graph (Vamana) on solid-state storage. It keeps in memory only a product-quantised approximation, which gives fast but lossy distance estimates during traversal. It then fetches the full-precision vectors from disk to re-rank the candidates the walk surfaces. The result can search a billion points on a single machine. The system is an explicit co-design: compression ($q$, product quantisation) supplies the cheap in-memory distances, organisation ($\mathcal{I}$, the Vamana graph) supplies the routing, and a re-ranking pass restores the precision quantisation discards.
\fi

\subsection{Recent directions: hardware, freshness, and filtering}\label{sec:graph_recent}

The graph family has continued to develop along three axes of practical concern: hardware, freshness, and filtering. The first two refine the organisation axis without disturbing the lens. Filtering sits awkwardly within the PQO lens. A metadata predicate is neither a projection, a quantiser, nor a property of the proximity structure; it is a constraint on the candidate set. We regard it as a fourth concern, orthogonal to the three stages. It reminds us that the lens organises the geometry of similarity search rather than the full surface of a deployed retrieval system.

\shortonly{Representative work spans GPU-parallel search (CAGRA~\cite{Ootomo24}), disk-resident inverted indexes (SPANN~\cite{Chen21}) and their incrementally updatable successor for streaming corpora (SPFresh~\cite{Xu23}), and predicate-aware traversal for filtered retrieval (Filtered-DiskANN~\cite{Gollapudi23}, ACORN~\cite{Patel24}). Section~S6 of the supplementary material details each.}

\iflong
On \emph{hardware}, the massively parallel CAGRA~\cite{Ootomo24} adapts proximity-graph search to graphics processors. There the high memory bandwidth and thread count alter the trade-offs and deliver substantial throughput gains. On \emph{freshness}, recent work tackles directly the difficulty of updating a graph, a difficulty this survey has noted repeatedly. SPANN~\cite{Chen21} couples an in-memory navigating structure to disk-resident posting lists for billion-scale search. SPFresh~\cite{Xu23} adds incremental in-place insertion and deletion, supplying the updatable organisation streaming corpora require. On \emph{filtering}, retrieval-augmented applications routinely combine a similarity query with structured predicates such as a date range or access-control label. Recent work makes graph search predicate-aware. Filtered-DiskANN~\cite{Gollapudi23} builds label-aware edges, while ACORN~\cite{Patel24} uses a predicate-agnostic traversal that prunes during search. The former is more efficient when the predicates are known at construction time. The latter is more flexible when they are not.
\fi

\iflong\input{supp/auto_graph}\fi

\subsection{Advantages and resource costs of graph indexes}\label{sec:graph_lens}

The lens explains the empirical dominance of graph indexes. Binary hashing and product quantisation assume that good codes make the index trivial. Graphs assume that a good index makes elaborate codes unnecessary, and they search essentially raw vectors with exact distances. When memory is abundant and the corpus is static, the latter assumption is borne out. Greedy routing over a well-pruned graph then attains high recall at low latency without incurring any quantisation error. The cost falls precisely on the axis graph methods do not use. Holding both the full-precision vectors and the edge lists makes the graph the most memory-hungry of the three families and costly to construct. Because its edges encode global proximity, the graph is also awkward to update incrementally. Insertions and deletions can degrade navigability, so streaming and high-churn workloads remain a weakness. Binary codes occupy the opposite position, with the smallest footprint and the cheapest updates at the coarsest distance resolution. Product quantisation sits between the two, as Table~S6 of the supplementary material sets out. No family dominates along every axis. The choice among them therefore turns on which resource is scarcest, whether memory, accuracy, or update cost. The frontier lies in co-designing the three stages rather than in any single method. The binary codes with which the survey began have become resurgent precisely where memory is the binding constraint, namely retrieval-augmented serving at billion scale. There, compact codes increasingly appear as the compression layer beneath a graph rather than as a standalone index, or as a cheap filter re-ranked by a more accurate pass~\cite{Yamada21, Wang21}.

\iflong\input{supp/auto_bets}\fi

%% file: supp/auto_graph.tex
%% Relocated from the CSUR main paper to the supplement (tab:graph).
\begin{table}[!t]
\centering
\small
\begin{tabular}{@{}p{2.7cm} p{3.4cm} p{2.0cm} p{2.5cm}@{}}
\toprule
\textbf{Method} & \textbf{Structural idea} & \textbf{Hierarchy} & \textbf{Scale regime} \\
\midrule
NSW~\cite{Malkov14} & small-world: short $+$ long-range links & flat & in-memory, millions \\
HNSW~\cite{Malkov20} & navigable small-world $+$ skip layers & hierarchical & in-memory (default) \\
NSG~\cite{Fu19} & monotonic search net; edge pruning & flat, low degree & in-memory, large \\
FANNG~\cite{Harwood16} & occlusion / RNG edge rule & flat & precursor \\
DiskANN~\cite{Subramanya19} & Vamana graph $+$ PQ $+$ SSD re-rank & flat, SSD-resident & billion, single node \\
\bottomrule
\end{tabular}
\caption[Graph-based ANN indexes]{Graph-based indexes operate on the organisation axis $\mathcal{I}$ of Definition~\ref{def:lens} and leave projection and quantisation untouched. At billion scale, DiskANN reintroduces product quantisation to bound memory.}
\label{tab:graph}
\end{table}

%% file: supp/auto_bets.tex
%% Relocated from the CSUR main paper to the supplement (tab:bets).
\begin{table}[!t]
\centering
\small
\begin{tabular}{@{}p{2.9cm} p{2.6cm} p{2.6cm} p{2.6cm}@{}}
\toprule
 & \textbf{Binary hashing} & \textbf{Product quant.} & \textbf{Graph index} \\
\midrule
Principal axis & $p$ and $q$ & $q$ & $\mathcal{I}$ \\
Memory footprint & smallest (bits) & small (codes) & largest (vectors $+$ edges) \\
Distance resolution & coarse (Hamming) & fine (ADC) & exact (raw vectors) \\
Updates / streaming & easy & moderate & hard \\
Search over data & bucket / Hamming scan & list scan $+$ ADC & greedy traversal \\
\bottomrule
\end{tabular}
\caption[The three families and the principal axis of each]{The three dominant families of approximate nearest neighbour search, distinguished by the axis of Definition~\ref{def:lens} on which each principally acts, with the resource trade-offs that follow. No family dominates along every row. The choice depends on whether memory, accuracy, or update cost is the binding constraint.}
\label{tab:bets}
\end{table}

%% file: body/resurgence.tex
The classical methods of Sections~\ref{sec:ch2_quantisation} and~\ref{sec:ch2_projection} were developed when a data-point was a hand-crafted feature vector, such as the GIST descriptor~\cite{Oliva01} or a bag of visual words. In that setting, the difficulty of hashing lay in learning a projection that made the vector amenable to compact binary encoding. The deep hashing of Section~\ref{sec:deep} moved that projection into a network, though one still trained per task and per collection. The setting has since altered again. The projection stage of Definition~\ref{def:lens} is now usually supplied by a single large pretrained encoder, whose dense output captures semantic similarity beyond the reach of either. The need for compression has grown despite this stronger projection. Corpora have grown to billions of items, and embedding retrieval has become the substrate of retrieval-augmented generation~\cite{Lewis20}. The cost of storing and searching dense, full-precision vectors has therefore re-emerged as the binding constraint. This revives the compression questions that motivated the field at its inception. The consequence is a resurgence of compact codes, now applied to learned embeddings within vector database systems. We read this resurgence as a reassembly of the projection--quantisation pipeline from contemporary components.

\subsection{Compression returns as the binding constraint}\label{sec:res_constraint}

The shift also relocates the difficulty. The classical literature devoted its main effort to learning a separating projection. The modern setting inherits a projection of high quality and must instead confront the size of its output. A corpus of one billion $768$-dimensional single-precision embeddings occupies of the order of three terabytes before any index is added. This places exact search beyond commodity memory. The compression axis of Definition~\ref{def:lens} thus becomes decisive. The maturation of vector database systems has elevated quantisation to a documented, user-facing feature. This poses the question with which this survey began: how aggressively may a representation be compressed before retrieval quality is unacceptably degraded.
\iflong
The relevant systems include Milvus~\cite{Wang21}, its cloud-native successor~\cite{Guo22}, hybrid relational engines such as AnalyticDB-V~\cite{Wei20}, and the Postgres extension \texttt{pgvector}~\cite{Pgvector}. Each exposes quantisation as a user-facing setting. The reason is that the size of the stored embeddings dominates the operating cost of large retrieval systems.
\fi

\subsection{Binary and scalar quantisation of embeddings}\label{sec:res_quant}

Binary quantisation, the field's oldest technique, answers this question most directly. It thresholds each embedding dimension at zero, assigning a bit by its sign. It is therefore identical in form to the single-bit scalar quantiser of Section~\ref{sec:ch2_sbq_quant} and the sign random projection of Charikar~\cite{Charikar02}. The only difference is that the binarised projection is now a learned, semantically rich embedding rather than a random hyperplane. We measure the effect directly in Section~\ref{sec:ch4_measured} (Table~\ref{tab:emb_quant}). We find that the thirty-two-fold compression of the scanned representation is lossless once a re-ranking pass is retained. The operation is the classical single-bit quantiser applied to a far better projection than the field originally had.
\iflong
A single-precision dimension occupies thirty-two bits, so binarisation reduces storage thirty-two-fold and permits search in Hamming space. The milder scalar quantisation to eight-bit integers reduces storage fourfold while retaining a finer approximation of distances. Providers offer both as routine options and report modest quality loss~\cite{Cohere24, SBERT24}. Note that the more striking figures originate in industrial white papers rather than peer-reviewed evaluation, and should be read with caution.
\fi

\subsection{Compression followed by re-ranking}\label{sec:res_rerank}

Aggressive binary compression is made practical by a refinement met above in another guise. The binary codes serve only to generate a candidate set cheaply in Hamming space. The system then re-ranks the survivors with a more accurate distance computed from their eight-bit or full-precision vectors alone. This is the structure of the polysemous codes of Section~\ref{sec:pq_synthesis}, a Hamming filter preceding an asymmetric re-ranking, transposed to binary embeddings. The most complete peer-reviewed realisation is the Binary Passage Retriever~\cite{Yamada21}, which stores no continuous passage vectors at all. It generates candidates in Hamming space and re-ranks them by the inner product of the continuous query with the binary passage codes. In this way it reduces the index of the dense passage retriever~\cite{Karpukhin20} from approximately sixty-five gigabytes to two with no measurable loss. The stored-floats variant of the recipe underlies the binary-then-rescore pipelines documented for production embedding models~\cite{SBERT24}. The coupling also extends to vector quantisers: JPQ~\cite{Zhan21jpq} and RepCONC~\cite{Zhan22repconc} train product-quantisation codebooks jointly with the dual encoder against the ranking objective, coupling the projection and a vector quantiser end-to-end in the manner of Section~\ref{sec:deep}.
\iflong
Such a system relies on its organisation structure, whether flat, inverted, or graph-based, only for initial candidate generation.

The two-stage structure can be made precise. A passage is encoded into a continuous vector $\uu = f_{\theta}(\cdot) \in \RE^{K}$. During training, its binary code is obtained through a scaled hyperbolic tangent annealed towards the sign function, $\bb = \tanh(\beta\, \uu)$ with $\beta \to \infty$, in the manner of the continuation approach of Section~\ref{sec:deep_binarisation}. A query is encoded into $\mathbf{v} = f_{\phi}(\cdot)$ and binarised by the same annealed sign, $\bb_{q} = \tanh(\beta\,\mathbf{v})$. Candidate generation ranks passages by the Hamming distance between the binary query and passage codes, equivalently the binary inner product $\bb_{q}^{\T}\bb$; this is what makes the first pass cheap. The re-ranking loss scores the retained candidates by the inner product of the continuous query with the binary passage code, $\mathbf{v}^{\T}\bb$. The continuous passage representation $\uu$ exists only during training to derive $\bb$; it is never stored or scored at inference. Training combines the two stages as a sum of ranking losses,
\begin{equation}
\mathcal{L} = \mathcal{L}_{\mathrm{cand}}(\bb_{q}, \bb) + \mathcal{L}_{\mathrm{rerank}}(\mathbf{v}, \bb),
\label{eqn:bpr}
\end{equation}
where each term compares a positive passage against a set of negatives, so that the same encoder is accurate at both stages. This explicit two-stage objective, with the binary code as quantiser and the continuous representation as projection, allows the thirty-two-fold reduction in index size without a commensurate loss of accuracy.
\fi
\shortonly{We read this as the classical filter-then-verify recipe, with the binary code as quantiser and the continuous representation as projection, transposed onto learned embeddings. Section~S7 of the supplementary material gives the explicit two-stage training objective.}

\subsection{Adaptive-dimension embeddings}\label{sec:res_matryoshka}

A complementary line of work compresses on the projection axis rather than the quantisation axis. It learns embeddings whose dimensions are ordered by importance. Matryoshka representation learning~\cite{Kusupati22} trains a single embedding such that any leading prefix of its coordinates remains usable. A downstream system may therefore truncate the vector to a length suited to its memory and accuracy budget, without retraining. In effect, this is a learned analogue of the non-uniform bit allocation of Section~\ref{sec:ch2_quantisation}, since the most informative content is concentrated in the earliest dimensions.
\iflong
Later coordinates are discarded, just as less informative projected dimensions were once allocated fewer bits. Deployed embedding interfaces have adopted the technique and expose a parameter governing the dimensionality of the returned vector~\cite{OpenAI24}. The technique also composes naturally with the quantisation above. The two forms of compression act on different axes and therefore compound.
\fi

\subsection{Generative retrieval and semantic identifiers}\label{sec:res_generative}

A more radical proposal of the same period appears at first sight to lie outside the lens altogether. Generative retrieval~\cite{Tay22} dispenses with an explicit index. It instead trains a sequence model to emit, given a query, the identifier of a relevant document directly. The lens nonetheless places it within the factorisation. In the most developed systems~\cite{Rajput23}, the identifiers such models generate are the codes of a residual-quantising autoencoder. This is a particular quantiser $q$ of the \longshort{additive and residual kind reviewed in Section~\ref{sec:pq_additive}}{residual-quantisation kind introduced through the lens in Section~\ref{sec:lens_limits} (Table~\ref{tab:lens})}, and its codebook is the vocabulary of generation. Only the organisation stage differs: $\mathcal{I}$ is realised by an autoregressive decoder navigating the identifier space rather than by a lookup over stored codes.
\iflong
The differentiable search index~\cite{Tay22} and the neural corpus indexer~\cite{Wang22} are the canonical instances. The approach has been extended to recommendation as semantic identifiers~\cite{Rajput23}. Generative retrieval is thus a further instance of the factorisation, in which the quantisation axis reappears as the discrete vocabulary of a generative model. We refer the reader to the survey of Li et al.~\cite{Li25} for a treatment of the paradigm in its own right. We make only the narrower observation that its central representational device is a quantiser of a kind already considered.
\fi

\iflong\input{supp/auto_resurgence}\fi

\subsection{A note on multi-vector retrieval}\label{sec:res_multivector}

A parallel development represents each item by many embeddings, one per token, rather than by a single embedding. It scores a query by a late-interaction sum of per-token maxima rather than by a single inner product. ColBERT~\cite{Khattab20} is the canonical instance. Its distinctive move falls on the cardinality of the projection rather than on the quantiser or the organisation structure, and so lies largely outside the PQO lens. Its compression of those many vectors, however, returns squarely to the compression axis with the product-quantisation machinery of Section~\ref{sec:pq}.\shortonly{ A second parallel line is learned \emph{sparse} retrieval (SPLADE~\cite{Formal21}), in which a transformer predicts term weights scored on a classical inverted index. It likewise falls outside our scope, since we bound the survey to the dense compact codes on which the modern dense-retrieval stack depends.}
\iflong
ColBERT~\cite{Khattab20} is the canonical instance. Its revision~\cite{Santhanam22} contains the resulting storage by residual compression, a nearest-centroid index plus a quantised residual. We note this line chiefly in order to place it within the lens. The motivation has recently acquired a theoretical basis. A single-vector embedding is subject to a capacity limit set by its dimensionality. This limit bounds the document sets it can return as a top-$k$ result, irrespective of training~\cite{Weller25}. Late interaction multiplies the vectors per item, and is one of the few ways to exceed the limit. Multi-vector retrieval is thus orthogonal to the lens in its scoring model. It nonetheless depends on the lens for the compression without which it would not scale. A second parallel line is learned \emph{sparse} retrieval, in which a transformer predicts a sparse vector of term weights scored on a classical inverted index (SPLADE~\cite{Formal21}, and the COIL/uniCOIL family~\cite{Lin21}). It likewise falls outside scope, since it compresses a sparse lexical representation rather than a dense code and rests on inverted-index machinery. We note it only to bound the survey to the dense compact codes on which the modern dense-retrieval stack depends.
\fi

\subsection{Synthesis}\label{sec:res_synthesis}

\longshort{We regard the resurgence surveyed here, whose techniques are collected in the supplementary material, as a compact instance of the survey's overall argument. A learned encoder supplies the projection. Binary, scalar, or Matryoshka compression supplies the quantisation. A vector database, whether flat, inverted, or graph-based, supplies the organisation. Recall the lessons the classical literature established: the placement of thresholds matters, bits are most usefully allocated where the representation is most informative, and an aggressive compression may be made safe by a subsequent verification pass. These lessons return essentially unchanged, now upon learned embeddings and at a far greater scale. Their return follows from the constraint with which the section opened. When memory rather than computation is scarce, the compression axis can no longer be treated as a detail. Large-scale retrieval now raises the same questions that defined learning to hash. We take up the implications of this convergence, and the open problems it leaves, in the concluding discussion.}{We regard the resurgence surveyed here, whose techniques Table~S7 of the supplementary material collects, as a compact instance of the survey's overall argument. A learned encoder supplies the projection, binary, scalar, or Matryoshka compression the quantisation, and a vector database the organisation. The classical lessons return unchanged upon learned embeddings at far greater scale: threshold placement matters, bits belong where the representation is most informative, and aggressive compression is made safe by a verification pass. We take up the implications of this convergence, and the open problems it leaves, in the concluding discussion.}

%% file: supp/auto_resurgence.tex
%% Relocated from the CSUR main paper to the supplement (tab:resurgence).
\begin{table}[!t]
\centering
\small
\begin{tabular}{@{}p{3.2cm} p{4.0cm} p{3.4cm}@{}}
\toprule
\textbf{Technique} & \textbf{Reading through the lens} & \textbf{Status of evidence} \\
\midrule
Binary quantisation of embeddings & single-bit scalar $q$ on a learned projection & peer-reviewed precedent~\cite{Yamada21}; industry feature~\cite{Cohere24, Pgvector} \\
Scalar (8-bit) quantisation & low-bit scalar $q$ & industry~\cite{Cohere24, SBERT24} \\
Compression $+$ re-ranking & cheap $q$ filter, accurate verification & peer-reviewed~\cite{Yamada21} \\
Matryoshka embeddings & learned ordered (non-uniform) projection $p$ & peer-reviewed~\cite{Kusupati22}; deployed~\cite{OpenAI24} \\
Semantic identifiers & residual-quantised $q$; decoder as $\mathcal{I}$ & peer-reviewed~\cite{Rajput23} \\
\bottomrule
\end{tabular}
\caption[The modern resurgence read through the lens]{Contemporary techniques for compact retrieval, read through the PQO lens of Definition~\ref{def:lens}. The final column notes whether the supporting evidence is peer-reviewed or originates in industrial documentation. Each technique instantiates an axis the survey has already examined in a classical setting.}
\label{tab:resurgence}
\end{table}

%% file: supp/methodology_metrics.tex
%% Full definitions of the evaluation metrics (F-beta, AUPRC, mAP, nDCG),
%% relocated from the CSUR main paper Evaluation Metrics section.
We summarise retrieval effectiveness with standard IR measures. An unranked retrieved set is formed either by hash-table collisions or by thresholding a Hamming radius. For such a set, we micro-average precision $P = TP/(TP+FP)$ and recall $R = TP/(TP+FN)$ over the queries against the ground-truth neighbour matrix $\mathbf{S} \in \{0,1\}^{N_{trd}\times N_{trd}}$ (Section~\ref{sec:ch4_groundtruth}). We combine the two by the $F_{\beta}$-measure~\cite{Rijsbergen79},
\begin{equation}
F_{\beta} = \frac{(1+\beta^{2})\,P\,R}{\beta^{2}P+R}.
\label{eqn:ch4_fbeta}
\end{equation}
Sweeping the operating point, for instance the Hamming radius $d=1,\ldots,d_{\max}$ with $d_{\max}\le K$, traces a precision--recall curve. The area under this curve (AUPRC)\label{sec:ch4_auprc} gives a single-number summary over recall,
\begin{equation}
AUPRC = \int_{0}^{1} P(R)\, dR = \sum_{d=1}^{d_{\max}} P(d)\,\Delta R(d),
\label{eqn:ch4_auprc}
\end{equation}
with $P(d)$ the precision at radius $d$ and $\Delta R(d)$ the recall increment from $d-1$ to $d$.\footnote{The displayed sum is the rectangle (interpolated-precision) approximation; a trapezoidal estimate (e.g., \texttt{trapz} in MATLAB) gives a finer one.} The complementary ranking summary is mean average precision (mAP)\label{sec:ch4_map}. Writing $P_{\mathbf{q}}(r)$ for the precision at rank $r$, $\delta(r)$ for the relevance indicator, and $L$ for the number of items relevant to $\mathbf{q}$, we define
\begin{equation}
AP(\mathbf{q})=\frac{1}{L}\sum_{r=1}^{N} P_{\mathbf{q}}(r)\,\delta(r), \qquad mAP = \frac{1}{|Q|}\sum_{i=1}^{|Q|} AP(\mathbf{q}_{i}).
\label{eqn:ch4_map_metric}
\end{equation}
The two summaries differ chiefly in their averaging. Note that mAP is approximately the per-query average of AUPRC~\cite{Turpin06}. The two therefore agree when the number of relevant items per query is balanced and diverge under skew. Intuitively, AUPRC is a micro-average and favours queries with many relevant items, whereas mAP is a macro-average and weights queries equally. The choice is therefore application-dependent~\cite{Sebastiani02}: system-oriented tasks tend to the former and user-oriented tasks to the latter. We report mean average precision under the class-based ground truth of Section~\ref{sec:ch4_class_ground} for the supervised CIFAR-10 study of Section~\ref{sec:ch4_measured}. The recall, throughput and memory measures defined next are those of the metric and at-scale studies.

Where relevance is graded rather than binary, as in the neural-retrieval benchmarks, we summarise effectiveness instead by the \emph{normalised discounted cumulative gain} (nDCG@$k$). This measure discounts each relevant item by the logarithm of its rank and normalises by the ideal ranking,
\begin{equation}
\text{DCG@}k = \sum_{r=1}^{k} \frac{2^{\,\text{rel}_{r}}-1}{\log_{2}(r+1)}, \qquad \text{nDCG@}k = \frac{\text{DCG@}k}{\text{IDCG@}k},
\label{eqn:ch4_ndcg}
\end{equation}
with $\text{rel}_{r}$ the relevance grade of the item at rank $r$ and $\text{IDCG@}k$ the discounted cumulative gain of the grade-sorted ideal ranking. It is the standard measure on the BEIR and MTEB embedding benchmarks~\cite{Thakur21, Muennighoff23}. We report it for the embedding-compression study of Section~\ref{sec:ch4_measured}.

%% file: supp/scale_metrics.tex
%% Formal definitions and protocols for at-scale evaluation (k-recall@k, QPS,
%% bytes-per-vector; ANN-Benchmarks and billion-scale protocols). Summarised in
%% the main paper; rendered by the arXiv-long build and the supplement.
The protocol described above measures the quality of a ranking. It is the appropriate instrument for the binary-hashing methods with which the field began. For those methods the code length is fixed, and the question is how well the codes preserve neighbourhood structure. The methods that have since come to dominate large-scale retrieval are most usefully compared along the trade-off between accuracy and the resources that accuracy consumes, rather than at a fixed code length. A complementary body of evaluation practice has developed to measure exactly this trade-off. We summarise it here, since the later sections of the survey appeal to it repeatedly.

\paragraph{Metrics.} We report the accuracy of an approximate search by \emph{$k$-recall@$k$}, the fraction of the true $k$ nearest neighbours recovered among the $k$ items returned:
\begin{equation}
k\text{-recall@}k = \frac{1}{k}\,\big|\, \mathcal{N}_{k}(\q) \cap \tilde{\mathcal{N}}_{k}(\q) \,\big|,
\label{eqn:recall_at_k}
\end{equation}
where $\mathcal{N}_{k}(\q)$ is the set of exact $k$ nearest neighbours of the query $\q$ and $\tilde{\mathcal{N}}_{k}(\q)$ the set of $k$ items returned, both of cardinality $k$. Normalisation by $k$ therefore coincides with division by $|\mathcal{N}_{k}(\q)|$, which is the $k$-recall@$k$ convention of~\cite{Aumuller20}. We report recall against a measure of cost, most commonly the number of queries answered per second (the throughput, or QPS) on fixed hardware. Varying a method's accuracy parameter, such as the list width $ef$ of Algorithm~\ref{alg:greedy_search} or the number of inverted lists probed, traces out a recall--throughput curve. One method dominates another when its curve lies above and to the right. The third axis of comparison is the \emph{memory footprint}, reported as the number of bytes stored per vector. The survey's motivation identified this axis as decisive. A flat index of single-precision vectors costs $4D$ bytes per item, a product-quantised index costs $M\log_{2}k / 8$ bytes, and a binary code costs $K/8$ bytes. Compact codes recover their advantage over graphs along this axis.

\paragraph{Datasets and protocols.} The classical protocol used image-feature collections of up to a million items (Section~\ref{sec:ch4_dataset_unimodal}). The modern benchmarks are larger and increasingly drawn from learned embeddings. Standard unimodal sets include the billion-vector SIFT and Deep collections (BIGANN and Deep1B), together with the GIST and GloVe sets used by the ANN-Benchmarks suite~\cite{Aumuller20}. For text retrieval over neural embeddings, the standard sets include the passage collections that underlie open-domain question answering. They also include the heterogeneous zero-shot retrieval benchmark BEIR~\cite{Thakur21} and the massive text-embedding benchmark MTEB~\cite{Muennighoff23}. These two benchmarks have become standard for evaluating the learned embeddings on which modern retrieval depends. The ANN-Benchmarks methodology~\cite{Aumuller20} fixes the dataset, the hardware, and the recall definition, and compares methods purely on their recall--throughput curves. This methodology has become the de facto standard for in-memory evaluation. The billion-scale challenge of~\cite{Simhadri22} extends it to the regime in which the index no longer fits in memory. In that regime the bytes-per-vector and build-time axes become binding.

\paragraph{Reading the axes.} These measurements map cleanly onto the lens. Recall at a fixed operating point reflects the combined quality of the projection and the quantiser. The position of the recall--throughput curve reflects the efficiency of the organisation structure. The bytes-per-vector figure is a direct measure of the compression axis. The classical ranking metrics of the preceding subsections and the operating-point metrics introduced here are therefore complementary: they project the same system onto different axes. A complete account of a modern method reports both.

%% file: supp/meas_empirical.tex
%% Representative cross-family comparison drawn from the published benchmarks
%% (literature-derived, not directly comparable across rows). Abbreviated out of
%% the CSUR main paper, where the measured Tables of Section~\ref{sec:ch4_measured}
%% carry the argument; retained here and in the arXiv-long build for completeness.
\paragraph{A representative comparison.} Table~\ref{tab:empirical} makes these trade-offs concrete. For each family we report the memory footprint, which follows by arithmetic for a $768$-dimensional embedding, alongside a representative recall and throughput. We draw these figures from the public benchmarks and the methods' original evaluations. We deliberately do not force the operating points onto a common task. The in-memory methods are quoted on the GloVe-100-angular benchmark of ANN-Benchmarks~\cite{Aumuller20}. DiskANN is quoted at billion scale on SIFT1B~\cite{Subramanya19}, and the binary-embedding designs are quoted on open-domain question answering~\cite{Yamada21}. The rows therefore convey the characteristic position of each family rather than a ranking on a single dataset. The pattern nonetheless follows the lens's prediction. Graph indexes attain the highest recall and throughput at the cost of the largest footprint. The quantisation families trade a controlled loss of recall for an order-of-magnitude reduction in memory. The binary and re-ranking designs occupy the extreme low-memory corner and recover accuracy through a continuous re-ranking pass~\cite{Yamada21, Gao24}.

\begin{table}[!t]
\centering
\small
\caption[Representative memory and accuracy by family]{Representative comparison by family. Bytes per vector follow by arithmetic for a $768$-dimensional embedding ($D=768$): a flat index costs $4D$, and a product or binary code costs roughly $M$ or $K/8$ bytes. The recall and throughput figures are representative operating points from the cited sources and are not directly comparable across rows: the in-memory methods are quoted on GloVe-100-angular (ANN-Benchmarks~\cite{Aumuller20}), DiskANN at billion scale on SIFT1B~\cite{Subramanya19}, and the binary-embedding designs on open-domain question answering~\cite{Yamada21}; consult the live benchmarks~\cite{Aumuller20, Simhadri22} for current figures. $^{\dagger}$~recall attained with a continuous or full-precision re-ranking pass; $^{\ddagger}$~DiskANN figures are $1$-recall@$1$ at $>$5000 QPS on a 16-core node~\cite{Subramanya19}, not the $10$-recall@$10$ of the in-memory rows.}
\label{tab:empirical}
\begin{tabular}{@{}p{3.0cm} p{3.0cm} p{1.8cm} p{1.9cm}@{}}
\toprule
\textbf{Family} & \textbf{Bytes / vector} & \textbf{Recall} & \textbf{Throughput} \\
\midrule
Exact (flat) & $4D \approx 3072$ & $1.00$ & brute force \\
HNSW~\cite{Malkov20} & $4D +$ edges $\approx 3.3$\,K & $\approx 0.99$ & very high \\
IVF-PQ~\cite{Jegou11} & $\approx 96$--$192$ & $\approx 0.85^{\dagger}$ & high \\
ScaNN~\cite{Guo20} & $\approx 96$--$192$ & $\approx 0.99$ & very high \\
DiskANN~\cite{Subramanya19} & $\approx 96$ RAM, $4D$ disk & $0.95^{\ddagger}$ & $>5000$ QPS$^{\ddagger}$ \\
Binary $+$ re-rank~\cite{Yamada21} & $K/8 \approx 96$ & no loss$^{\dagger}$ & high \\
RaBitQ~\cite{Gao24} & $D/8 \approx 96$ & $\approx 0.99^{\dagger}$ & high \\
\bottomrule
\end{tabular}
\end{table}

%% file: supp/meas_cifar_diag.tex
%% Per-bit diagnostic of the supervised CIFAR-10 codes (information vs. separation).
%% Abbreviated out of the CSUR main paper; retained here and in the arXiv-long build.
\begin{figure}[!t]
\centering
\includegraphics[width=\textwidth]{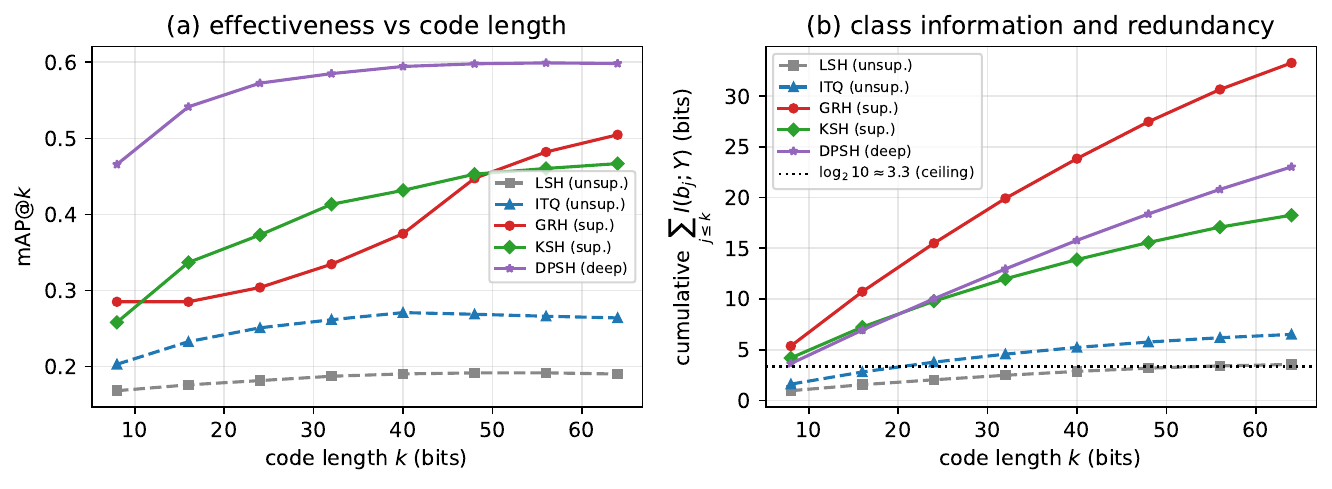}
\caption[Diagnostics of the supervised CIFAR-10 codes]{Diagnostics of the supervised CIFAR-10 codes, computed on a single $64$-bit code per method whose bits are ordered most-class-informative first, so that a prefix of length $k$ retains the $k$ most informative bits; these prefixes probe how information accumulates within one code, and they differ from the separately trained per-length codes of Table~\ref{tab:cifar_measured}, whose dedicated short codes score higher. (a)~Mean average precision of the leading $k$ bits. (b)~Cumulative per-bit mutual information with the class, $\sum_{j\le k} I(b_{j};Y)$, against the $\log_{2}10 \approx 3.3$-bit information ceiling (dotted); the excess above the ceiling measures the redundancy ($E_{4}$) of a code.}
\label{fig:cifar_diagnostics}
\end{figure}

The diagnostic of Figure~\ref{fig:cifar_diagnostics} exhibits a mechanism that the effective-hashcode properties of Section~\ref{sec:ch2_effective_hashcodes} anticipate. Summed across the sixty-four bits, the per-bit mutual information with the class reaches some $33$ bits for GRH, against an information ceiling of $\log_{2}10 \approx 3.3$ bits. This tenfold excess measures the redundancy ($E_{4}$) of its codes. The sequential residue of KSH learns each bit against the errors of its predecessors; this decorrelates the bits and roughly halves the figure to some $18$ bits. The deep code occupies the middle of this range, at some $23$ bits. It nonetheless attains both the highest precision and the largest separation between the same- and different-class Hamming distributions ($d' = 1.65$, against $1.50$ for GRH and $1.45$ for KSH). Redundancy and separation therefore vary independently: a stronger projection extracts a larger margin from fewer redundant bits. On these representations the redundancy proves margin-amplifying rather than harmful. Repeating the class across many bits stretches the Hamming distance between the classes, so the precision of these methods rises with the code length. We therefore read the redundancy as a trade of information efficiency for separation, with the ratio determined by the projection, rather than as wasted capacity.

%% file: supp/meas_cifar_pareto.tex
%% Supervised codes against float and product quantisation at matched memory
%% (CIFAR-10). A corollary of the supervised measurement; abbreviated out of the
%% CSUR main paper and retained here and in the arXiv-long build.
\paragraph{Supervised compression.} The comparison of Table~\ref{tab:cifar_measured} fixes the code length and asks how well each method ranks. The system poses a sharper question: how do the codes fare against the wider alternatives at a matched memory budget? Table~\ref{tab:cifar_pareto} sets the same supervised codes against full-precision float and product quantisation on the class task. The outcome is the supervised counterpart of the resurgence regime that follows. The ImageNet-pretrained features are not tuned to the CIFAR classes. The full-precision float ranking therefore attains a class-mAP of only $0.223$ at two kilobytes per vector. Product quantisation attains a little more at sixty-four bytes, and the unsupervised binary codes sit near the same level at eight bytes. The supervised codes fold the class structure into the projection and more than double this figure. GRH reaches $0.508$ and KSH reaches $0.473$ at those same eight bytes, and the deep hash head is higher still at $0.600$. This two-hundred-and-fifty-six-fold compression improves, rather than degrades, the semantic ranking. The standard deviations over three random splits are a few thousandths for every method, so the gap is not an artefact of a particular partition. Recall that the unsupervised Quora corpus of Table~\ref{tab:quora_pareto} offered a supervised method no signal it could use. Here the labels are present, and supervision improves quality at matched memory as it does at matched code length.

\begin{table}[!t]
\centering
\small
\caption[Supervised codes against float and product quantisation at matched memory]{Semantic retrieval on CIFAR-10 ($512$-dimensional ResNet-18 features, $1{,}000$ queries against a $59{,}000$-image database, class-based relevance, class-mAP over the full ranking, mean $\pm$ standard deviation over three random query/database splits) for full-precision float, product quantisation, and $64$-bit binary codes. The pretrained features are not tuned to the CIFAR classes, so the supervised binary codes (GRH, KSH) at eight bytes more than double the class-mAP of the $2{,}048$-byte float, and a deep hash head on the same features (DPSH) is higher still; this is the supervised counterpart of the resurgence's compression result. The binary Hamming search is also the fastest of the methods.}
\label{tab:cifar_pareto}
\begin{tabular}{@{}l r r@{}}
\toprule
\textbf{Method} & \textbf{Bytes / vec} & \textbf{class-mAP} \\
\midrule
Exact (flat) & $2048$ & $0.223 \pm 0.002$ \\
IVF-PQ~\cite{Jegou11} & $64$ & $0.241 \pm 0.002$ \\
LSH (unsupervised) & $8$ & $0.188 \pm 0.003$ \\
ITQ (unsupervised)~\cite{Gong11} & $8$ & $0.265 \pm 0.002$ \\
GRH (supervised)~\cite{Moran15c} & $8$ & $0.508 \pm 0.004$ \\
KSH (supervised)~\cite{Liu12} & $8$ & $0.473 \pm 0.006$ \\
DPSH (deep)~\cite{Li16} & $8$ & $0.600 \pm 0.002$ \\
\bottomrule
\end{tabular}
\end{table}

%% file: supp/emb_generality.tex
%% One-bit compression across seven embedders (tab:emb_generality).
%% Referenced from the main paper's generality paragraph; rendered by the
%% arXiv-long build and the supplement.
\begin{table}[!t]
\centering
\small
\caption[One-bit compression across seven embedders]{nDCG@$10$ retained by a one-bit code as a percentage of the floating-point ranking, for seven embedders over three BEIR corpora (scifact, nfcorpus, arguana), via the released \textsc{BitBudget} benchmark; the floating-point figures therefore differ slightly from Table~\ref{tab:emb_quant}, which averages over four corpora including fiqa. The two one-bit columns threshold each coordinate at zero and at its mean respectively; the final column re-ranks the top-$100$ Hamming candidates with full-precision vectors. We analyse the e5-base-v2 exception in the generality discussion of Section~\ref{sec:ch4_measured}.}
\label{tab:emb_generality}
\begin{tabular}{@{}l r r c c r@{}}
\toprule
\textbf{Embedder} & \textbf{dim} & \textbf{float} & \multicolumn{2}{c}{\textbf{1-bit} (\% float)} & \textbf{1-bit$+$re-rank} \\
\cmidrule(lr){4-5}
 & & nDCG@10 & zero $\theta$ & mean $\theta$ & \% float \\
\midrule
text-embedding-3-large & $3072$ & $0.543$ & $99$ & $97$ & $100$ \\
text-embedding-3-small & $1536$ & $0.508$ & $96$ & $96$ & $100$ \\
mxbai-embed-large & $1024$ & $0.528$ & $93$ & $94$ & $100$ \\
bge-base-en-v1.5 & $768$ & $0.523$ & $89$ & $90$ & $100$ \\
gte-base & $768$ & $0.517$ & $87$ & $93$ & $100$ \\
all-MiniLM-L6-v2 & $384$ & $0.444$ & $89$ & $87$ & $100$ \\
\midrule
e5-base-v2 \emph{(exception)} & $768$ & $0.461$ & $53$ & $86$ & $90$ \\
\bottomrule
\end{tabular}
\end{table}

%% file: supp/emb_axis.tex
%% Projection-vs-quantisation at matched bytes (tab:emb_axis) + its discussion.
%% Relocated from the CSUR main paper; the finding is stated there in prose.
The same embeddings answer a question that the lens poses but the literature rarely tests directly. At a fixed memory budget, should the bytes go to the projection axis, by reducing dimensions, or to the quantisation axis, by reducing bits? In Table~\ref{tab:emb_axis} we compare the two at three matched budgets per embedder. We reduce dimensions by PCA or prefix-truncation, and we reduce bits by uniform scalar quantisation. The Matryoshka-trained embedding makes the truncation side fair. Its coordinates are ordered by importance during training, so prefix-truncation is the operation it is built to support. Truncating MiniLM is naive by contrast, because its coordinates carry no such ordering. The quantisation axis dominates at every budget for both embedders by a wide margin. A full-dimensional one-bit code preserves far more of the ranking than a reduced-dimensional floating-point vector of the same size. Intuitively, the discriminative information is spread thinly across all the coordinates rather than concentrated in a few. The Matryoshka ordering closes much of the gap and lifts truncation from its near-useless naive figures to within reach of PCA. This lift quantifies the benefit of the learned ordering. The conclusion is unchanged: for these representations, even under fair truncation, reducing bits preserves more ranking quality than reducing dimensions.

\begin{table}[!t]
\centering
\small
\caption[Projection versus quantisation at matched bytes]{Retrieval quality (nDCG@$10$, mean over scifact, nfcorpus, arguana and fiqa) when a fixed byte budget is allocated to the projection axis (reducing dimensions, by PCA or by prefix-truncation, at full precision) against the quantisation axis (all dimensions at the stated bit depth). Truncating the Matryoshka-trained mxbai embedding is the operation it is designed to support, so its truncation column is a fair baseline; truncating MiniLM, whose coordinates carry no learned ordering, is naive. The quantisation axis is the most byte-efficient for both embedders, and it remains so even against the fair Matryoshka baseline.}
\label{tab:emb_axis}
\begin{tabular}{@{}l r r r@{}}
\toprule
\textbf{Budget} & \textbf{PCA} & \textbf{truncate} & \textbf{quantise} \\
\midrule
\multicolumn{4}{@{}l}{\emph{MiniLM-L6 ($384$-d), naive prefix-truncation}}\\
$192$ bytes & $0.285$ & $0.217$ & $0.424$ \\
$96$ bytes & $0.181$ & $0.107$ & $0.400$ \\
$48$ bytes & $0.090$ & $0.037$ & $0.404$ \\
\midrule
\multicolumn{4}{@{}l}{\emph{mxbai-embed-large ($1024$-d), Matryoshka truncation}}\\
$512$ bytes & $0.418$ & $0.348$ & $0.507$ \\
$256$ bytes & $0.349$ & $0.215$ & $0.483$ \\
$128$ bytes & $0.250$ & $0.101$ & $0.488$ \\
\bottomrule
\end{tabular}
\end{table}

%% file: supp/meas_quora_curves.tex
%% Full recall--throughput and recall--memory operating curves behind the Quora
%% comparison (Table~\ref{tab:quora_pareto}). Abbreviated out of the CSUR main
%% paper; retained here and in the arXiv-long build.
\begin{figure}[!t]
\centering
\includegraphics[width=\textwidth]{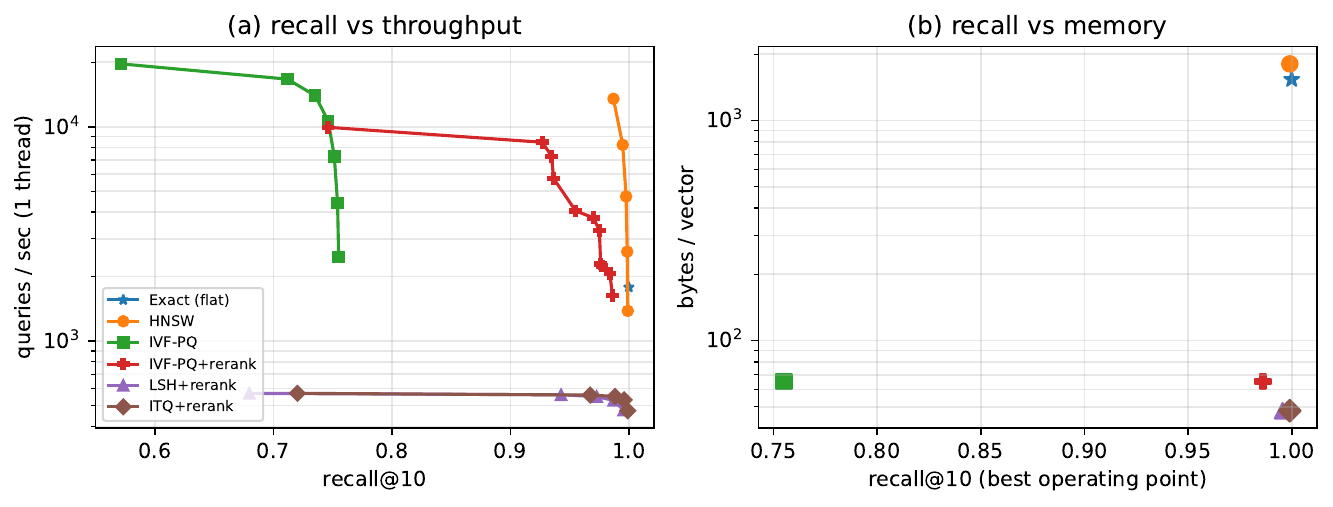}
\caption[Recall--throughput and recall--memory frontiers on Quora]{The full operating curves behind Table~\ref{tab:quora_pareto} (Quora, MiniLM-L6); we sweep each method over its accuracy parameter and reduce it to its Pareto-optimal points, under the protocol and environment of the scope note. \textbf{(a)}~Recall against single-thread throughput: the graph index (HNSW) holds the high-recall, high-throughput corner; the inverted-file product quantiser saturates near recall $0.76$ unless it is granted the full-precision re-ranking pass (IVF-PQ$\,+\,$rerank, \longshort{Table~\ref{tab:rerank_control}}{Table~S11 of the supplementary material}), which carries it to $0.99$ at intermediate throughput; the binary codes with re-ranking reach the highest recall of the compact codes at lower throughput. \textbf{(b)}~Each method at its best-recall operating point against its memory footprint: the binary codes occupy the low-memory, high-recall corner of the frontier at $48$ bytes per vector, with the re-ranked quantiser at $65$ bytes and recall $0.986$.}
\label{fig:quora_curves}
\end{figure}

%% file: supp/rerank_control.tex
%% Matched-shortlist re-ranking control (tab:rerank_control). The evidence table
%% behind the main paper's claim that the binary recall ordering survives
%% granting IVF-PQ the identical re-ranking pass. Rendered by the arXiv-long
%% build and the supplement. Data: benchmark/results/rerun/*_points.csv
%% (thread-pinned re-run; IVF-PQ at nprobe=128, the largest swept).
\begin{table}[!t]
\centering
\small
\caption[Matched-shortlist re-ranking control on Quora]{Recall@$10$ on the Quora panels of Table~\ref{tab:quora_pareto} when every method receives the identical full-precision re-ranking pass over a shortlist of the stated length. IVF-PQ is searched at $\mathit{nprobe}{=}128$, the largest value swept; byte budgets per panel are those of Table~\ref{tab:quora_pareto}. Protocol and environment are those of the scope note. At shortlist length $10$ with recall@$10$, re-ranking is a set-preserving no-op, so those cells measure candidate generation alone.}
\label{tab:rerank_control}
\begin{tabular}{@{}l ccccc@{}}
\toprule
\textbf{Shortlist length} & $10$ & $50$ & $100$ & $200$ & $500$ \\
\midrule
\multicolumn{6}{@{}l}{\emph{Quora, MiniLM-L6 ($384$-d)}}\\
IVF-PQ $+$ rerank & $0.755$ & $0.960$ & $0.977$ & $0.984$ & $0.986$ \\
LSH $+$ rerank & $0.680$ & $0.943$ & $0.973$ & $0.987$ & $0.996$ \\
ITQ $+$ rerank & $0.720$ & $0.967$ & $0.988$ & $0.996$ & $0.999$ \\
\midrule
\multicolumn{6}{@{}l}{\emph{Quora, mxbai-embed-large ($1024$-d)}}\\
IVF-PQ $+$ rerank & $0.779$ & $0.958$ & $0.973$ & $0.978$ & $0.979$ \\
LSH $+$ rerank & $0.742$ & $0.977$ & $0.993$ & $0.998$ & $0.999$ \\
ITQ $+$ rerank & $0.784$ & $0.989$ & $0.997$ & $0.999$ & $1.000$ \\
\bottomrule
\end{tabular}
\end{table}

%% file: supp/at_scale.tex
%% At-scale validation (90M-vector BigANN slice + 8.84M-passage MS MARCO).
%% Relocated from the CSUR main paper; the result is summarised there.
The measurements above are deliberately of workstation scale, so that every number is reproducible without special hardware. To check that the lens's crossovers do not dissolve at the scale that motivates modern retrieval, we repeated two of them several orders of magnitude larger, on a single cloud machine; the instance configuration and the scale-adjusted index parameters for these runs are recorded in the released harness rather than reproduced here, and their throughputs are not comparable with the workstation columns of the main paper. The first is a ninety-million-vector slice of the BigANN \texttt{SIFT1B} collection~\cite{Jegou11a}, with exact ground truth by brute force. It places one method per family on the recall--memory--throughput frontier (Table~\ref{tab:scale_bigann}). The second is the full MS~MARCO passage corpus~\cite{Nguyen16}, $8.84$ million passages embedded with MiniLM-L6. On this corpus we apply the same compression operators as in the resurgence regime and measure nDCG@10 against the released relevance judgements (binary for the MS~MARCO development queries) (Table~\ref{tab:scale_msmarco}).

\begin{table}[!t]
\centering
\caption[Recall, memory and throughput at ninety million vectors]{One method per family on a ninety-million-vector slice of BigANN \texttt{SIFT1B} ($128$-d, exact ground truth). The lens crossover holds at $10^{8}$: the graph attains high throughput at full-precision memory, the inverted product quantiser reduces the footprint by an order of magnitude at lower recall at lower recall, and the compact binary code has the smallest footprint of all. We use a single machine and a single thread for the query timing.}
\label{tab:scale_bigann}
\begin{tabular}{@{}l c c r@{}}
\toprule
\textbf{Method} & \textbf{Bytes / vec} & \textbf{Recall@10} & \textbf{QPS} \\
\midrule
Exact (flat) & $512$ & $1.000$ & $6$ \\
HNSW~\cite{Malkov20} & $704$ & $0.966$ & $17{,}448$ \\
IVF-PQ~\cite{Jegou11} & $64$ & $0.829$ & $434$ \\
ITQ $+$ rerank~\cite{Gong11} & $16$ & $0.430$ & $120$ \\
\bottomrule
\end{tabular}
\end{table}

\begin{table}[!t]
\centering
\caption[Compression at RAG scale on MS MARCO]{The resurgence-regime compression operators on the full MS~MARCO passage corpus ($8.84$M passages, MiniLM-L6, nDCG@10 against the released relevance judgements). A one-bit code with re-ranking is lossless at thirty-two-fold compression at this scale. This reproduces the workstation result on a corpus two orders of magnitude larger.}
\label{tab:scale_msmarco}
\begin{tabular}{@{}l c c c@{}}
\toprule
\textbf{Method} & \textbf{Bytes / vec} & \textbf{nDCG@10} & \textbf{\% of float} \\
\midrule
float32 & $1536$ & $0.3653$ & $100$ \\
int8 & $384$ & $0.3661$ & $100$ \\
binary & $48$ & $0.3412$ & $93$ \\
binary $+$ rerank & $48$ & $0.3652$ & $\mathbf{100}$ \\
RaBitQ~\cite{Gao24} & $48$ & $0.3408$ & $93$ \\
\bottomrule
\end{tabular}
\end{table}

Two features of these results are worth drawing out. On MS~MARCO the picture is unchanged from the workstation corpora. A one-bit code with a re-ranking pass remains lossless at thirty-two-fold compression, now on $8.84$ million passages. The quantisation axis again retains more quality per byte than the dimensions it replaces. On BigANN the crossover is equally clear but more instructive. Brute-force search, exact by construction, manages only six queries per second at ninety million vectors and is too slow to be usable. The graph restores interactive throughput by storing the vectors in full. The product quantiser trades an order of magnitude of memory for a portion of the recall. The compact binary code degrades at this scale: its recall falls to $0.43$ here, well below the level it sustains at workstation scale. This happens because a fixed-length code shortlists a fixed number of candidates by Hamming distance. At ninety million vectors the shortlist therefore captures a vanishing fraction of the corpus, and the true neighbour increasingly falls outside it before re-ranking can recover it. The failure arises in the organisation stage, since the same bits, navigated by a graph rather than scanned in a flat shortlist, need not lose those neighbours. The organisation axis, and indexes built natively on compact codes, are intended to close exactly this gap.

%% file: supp/llm_weights.tex
%% Weight quantisation read through the lens (the full correspondences).
%% Summarised in the CSUR main paper (Section on model serving); rendered in full
%% by the arXiv-long build and the supplement.
\paragraph{Weights and the product-quantisation family.} The compression of the model's weights, the other large serving cost, exhibits the same correspondences still more sharply. Beyond the standard post-training scalar quantisers (GPTQ~\cite{Frantar23}, AWQ~\cite{Lin24}), we name two developments. First, the most accurate low-bit weight quantisers replace scalar rounding with a learned codebook. The additive quantisation of weight blocks in AQLM~\cite{Egiazarian24} and the lattice and incoherence-processed codebooks of QuIP and QuIP\#~\cite{Chee23, Tseng24} are the additive and product quantisation of Section~\ref{sec:pq}, with a weight matrix in place of a database of descriptors. AQLM indeed shares an author with the additive-quantisation line in vision. Second, these and other methods apply a rotation before quantising (QuaRot~\cite{Ashkboos24}, SpinQuant~\cite{Liu24spin}). The rotation spreads the quantisation error evenly across coordinates and dissolves the outliers that a uniform grid handles poorly. This rotation is the random rotation of RaBitQ~\cite{Gao24}, used for the same purpose and justified by the same isotropy argument. Outlier-aware mixed precision (LLM.int8()~\cite{Dettmers22}, SmoothQuant~\cite{Xiao23}) plays, for model serving, the role that re-ranking and non-uniform bit allocation play in retrieval. It plays the role that re-ranking and non-uniform bit allocation play in retrieval.

%% file: supp/eval_pitfalls.tex
%% Evaluation pitfalls and recommended practice, in full. Summarised in the
%% main paper's conclusion; rendered by the arXiv-long build and the supplement.
We also observed recurring \emph{evaluation pitfalls}. First, inconsistent groundtruth definitions (class labels versus metric $\epsilon$-balls) hindered comparability. Second, widely adopted split protocols risked overfitting by reusing one database for training and evaluation. Third, reporting alternated between mAP and AUPRC without clarifying when each was appropriate.

\paragraph{Recommended evaluation practice.} For reproducibility and comparability, we recommend three practices: (1) held-out test queries and a disjoint test database (Section~\ref{sec:ch4_improved_splits}); (2) reporting both AUPRC and mAP with clear groundtruth definitions; and (3) releasing code, splits, and preprocessing details. For the contemporary setting we would add recall against latency and memory footprint per item. These are the axes along which the dominant modern methods are best compared.